\documentclass[a4paper,11pt]{article}
\pdfoutput=1  

\usepackage{jheppub}  

\usepackage[T1]{fontenc}  
\usepackage{bm}
\usepackage{graphics}
\usepackage{hyperref}
\usepackage{slashed}
\usepackage{mathrsfs}
\usepackage[normalem]{ulem}
\usepackage{cancel}

\newcommand{\dd}{\mathrm d} 
\newcommand{\leri}[1]{\left(#1\right)} 
\newcommand{\ord}[1]{{\mathcal O}(#1)}

\newcommand{\Lyby}{\log(\bar y/y)}
\newcommand{\LLyby}{\log^2(\bar y/y)}
\newcommand{\bnslash}{\bar{n}\!\!\!\slash}
\newcommand{\cP}{{\mathcal P}}
\newcommand{\bn}{\bar{n}}
\newcommand{\bnP}{\overline {\mathcal P}}
\newcommand{\Jme}{\!\mathscr J}
\newcommand{\Scme}{\mathscr S}
\newcommand{\msb}{{\overline{\mathrm{MS}}}}
\newcommand{\eps}{\varepsilon}
\newcommand{\nn}{\nonumber}

\DeclareRobustCommand{\eq}[1]{eq.~\eqref{eq:#1}}
\DeclareRobustCommand{\eqs}[2]{eqs.~\eqref{eq:#1} and \eqref{eq:#2}}
\DeclareRobustCommand{\eqss}[2]{eqs.~\eqref{eq:#1}\,--\,\eqref{eq:#2}}
\DeclareRobustCommand{\fig}[1]{figure~\ref{fig:#1}}

\DeclareRobustCommand{\sec}[1]{section~\ref{sec:#1}}

\DeclareRobustCommand{\app}[1]{appendix~\ref{app:#1}}

\DeclareRobustCommand{\rcite}[1]{ref.\,\cite{#1}}
\DeclareRobustCommand{\rcites}[1]{refs.\,\cite{#1}}

\title{\boldmath 
Two-Loop Massive Quark Jet Functions in SCET
}

\preprint{
\begin{flushright}
UWThPh-2019-13\\ 
MITP/19-025
\end{flushright}
}

\author[a,b]{Andr\'e H. Hoang}
\author[a]{Christopher Lepenik}
\author[c]{Maximilian Stahlhofen}

\affiliation[a]{University of Vienna, Faculty of Physics,\\Boltzmanngasse 5, A-1090 Wien, Austria}
\affiliation[b]{Erwin Schr\"odinger International Institute for Mathematical Physics,\\
University of Vienna, Boltzmanngasse 9, A-1090 Wien, Austria}
\affiliation[c]{PRISMA Cluster of Excellence, Johannes Gutenberg University, D-55128 Mainz, Germany}

\emailAdd{andre.hoang@univie.ac.at}
\emailAdd{christopher.lepenik@univie.ac.at}
\emailAdd{mastahlh@uni-mainz.de}

\abstract{
We calculate the $\ord{\alpha_s^2}$ corrections to the primary massive quark jet functions in Soft-Collinear Effective Theory (SCET). They are an important ingredient in factorized predictions for inclusive jet mass cross sections initiated by massive quarks emerging from a hard interaction with smooth quark mass dependence.
Due to the effects coming from the secondary production of massive quark-antiquark pairs there are two options to define the SCET jet function, which we call universal and mass mode jet functions.
They are related to whether or not a soft mass mode (zero) bin subtraction 
is applied for the secondary massive quark contributions and differ in particular
concerning the infrared behavior for vanishing quark mass.
We advocate that a useful alternative to the common zero-bin subtraction concept is to define the SCET jet functions through subtractions related to collinear-soft matrix elements. This avoids the need to impose additional power counting arguments as required for zero-bin subtractions. We demonstrate how the two SCET jet function definitions may be used in the context of two recently developed factorization approaches to treat secondary massive quark effects. We clarify the relation between these approaches and in which way they are equivalent.
Our two-loop calculation involves interesting technical subtleties related to spurious rapidity divergences and infrared regularization in the presence of massive quarks.
}

\begin{document}
\maketitle
\flushbottom

\section{Introduction}
\label{sec:intro}

Gaining higher precision in theoretical predictions for the production of massive quarks in hard particle collisions represents an important field of research in the context of the LHC as well as future colliders (see e.g.\ \rcites{Bendavid:2018nar,Abramowicz:2018rjq,Azzi:2019yne}). Factorized predictions are of special relevance since they provide a separation of physical effects from different momentum scales for cases where the scale hierarchies are large, such as for kinematic edges or endpoint regions or when there are hierarchies between particle masses and dynamical scales. Once factorization is established for a given observable, it can be written as a product or a convolution of functions encoding the dynamics of particular phase space regions. The resulting factorization formulae (or theorems) provide an approximation in the limit where one can expand in the ratios of the hierarchical scales. This allows to sum large logarithms related to these scales and in addition may provide the basis for a field theoretic treatment of low-energy hadronization effects for processes dominated by the strong interaction. As such, factorization theorems provide important pieces of information that are not accessible directly in corresponding calculations obtained in fixed-order perturbation theory within full QCD. 
Factorization also entails that the various functions occurring in the factorization theorems frequently have a universal character and can be applied for different processes.

For the description of the dynamics related to energetic radiation that is emitted collinearly to a quark produced at high energy, the so-called quark jet function is an essential ingredient in factorization theorems for a variety of processes where the inclusive invariant mass of all the collinear radiation (including the energetic quark) enters. The quark jet function was first introduced in the context of the factorization framework of the Soft-Collinear Effective Theory (SCET) \cite{Bauer:2000yr,Bauer:2001yt,Bauer:2001ct} for the theoretical description of inclusive semileptonic or radiative $B$ meson decays in the kinematic regions where the produced hadrons form a jet. For the case of massless quarks the jet function is defined as a non-local correlation function of two SCET massless quark jet fields. It was calculated at ${\cal O}(\alpha_s)$ and ${\cal O}(\alpha_s^2)$ in \rcites{Lunghi:2002ju} and \cite{Becher:2006qw}, respectively, and recently also the $\ord{\alpha_s^3}$ corrections became available in \rcite{Bruser:2018rad}. These results are applicable for the treatment of light quarks but can also be applied for massive quarks as long as the jet invariant mass $\sqrt{p^2}$ is much larger than the quark mass $m$, i.e.\ $p^2\gg m^2$, and the associated mass corrections are negligible.

When quark mass effects are considered, it is useful to distinguish two types of mass effects. One is related to the quark produced by the hard interaction and which initiates the jet, called \textit{primary}, and the other is related to quark-antiquark vacuum polarization effects, called \textit{secondary}. Interestingly, primary as well as secondary quark mass effects introduce new kinds of subtleties.

For an inclusive jet initiated by a primary heavy quark and with invariant mass $p^2$, two kinematic regions where quark mass effects are important emerge:
$p^2\sim p^2-m^2\sim m^2$ and $p^2-m^2\ll m^2\approx p^2$. The region $p^2\sim p^2-m^2\sim m^2$, called SCET region, is relevant for processes where the jet invariant mass is close to the primary quark mass $m$ but is also allowed to fluctuate at the same level. The corresponding jet function(s) can be formulated with SCET massive collinear quark jet fields and require that the massless quark SCET formalism is extended to account for mass effects modifying the propagation and interactions of collinear quarks \cite{Leibovich:2003jd,Rothstein:2003wh,Fleming:2007qr}. Jet functions of this kind are therefore called SCET jet functions. In the context of bottom and charm quark production in hard collisions the SCET regime $p^2\sim p^2-m^2\sim m^2$ is essentially the only one that can ever arise in practical applications where quark mass effects are important due to the sizable momentum smearing coming from non-perturbative effects~\cite{Dehnadi:2016snl}.
The region $p^2-m^2\ll m^2\approx p^2$, also called bHQET limit or heavy quark limit, is relevant for processes where the jet invariant mass is very close to the primary quark mass $m$ and only allowed to fluctuate at a scale much smaller than $m$. Here, the SCET description does not provide a full separation of all dynamical effects because $\hat s\equiv (p^2-m^2)/m\ll m\approx \sqrt{p^2}$ emerges as a new relevant scale. This separation is achieved in the context of boosted Heavy Quark Effective Theory (bHQET) where the corresponding collinear SCET sector is matched on a theory containing a super-heavy quark with its small offshell field component being integrated out \cite{Fleming:2007qr,Fleming:2007xt}. Jet functions of this kind are therefore called bHQET jet functions. In practice, the bHQET limit $p^2-m^2\ll m^2\approx p^2$ can only arise in the context of top quark production due to the large size of the top quark mass. Here, scales below the top quark mass may be resolved and not smeared out by hadronization effects \cite{Dehnadi:2016snl,Dehnadi:unpub}.\footnote{For similar discussions in the context of groomed jets see e.g.\ \rcites{Hoang:2017kmk,Makris:2018npl,Lee:2019lge}.} The bHQET regime is particularly important for the theoretical understanding of top quark mass determinations from observables tied to kinematic regions with top invariant masses close to the reconstructed top quark resonance \cite{Butenschoen:2016lpz,Hoang:2017kmk,Hoang:2018zrp}. The ${\cal O}(\alpha_s)$ primary massive quark SCET and the bHQET jet functions were calculated in \rcite{Fleming:2007xt}. The ${\cal O}(\alpha_s^2)$ corrections to the bHQET jet function were computed in \rcite{Jain:2008gb}.

Secondary massive quark effects in a jet function occur at ${\cal O}(\alpha_s^2)$ (NNLO) in the fixed-order expansion and thus only become relevant in high precision predictions. Their treatment in a factorization approach that separates collinear and soft quantum fluctuations, however, inevitably leads to $\mathrm{SCET_{II}}$-type rapidity singularities and associated large logarithms in factorization theorems for physical cross sections \cite{Gritschacher:2013pha}. These large logarithms contribute at the next-to leading logarithmic (NLL) order. They arise from modes of equal virtuality (i.e.\ invariant mass $k^2\sim k_+ k_-\sim m^2$) but widely separated rapidity (i.e.\ the size of the ratio $k_+/k_-$), see e.g.\ \rcite{Chiu:2012ir}. The treatment of these rapidity singularities in the context of jet functions is tied to the definition of jet functions being infrared finite (matching) functions describing collinear fluctuations with virtuality of order $p^2$. Traditionally the required infrared (IR) subtractions are associated to so-called zero-bin subtractions \cite{Manohar:2006nz} (that can be understood from the point of view of the SCET label formalism \cite{Bauer:2001ct}, where the collinear label momentum zero is removed as it may not be power-counted as containing collinear modes) or are simply ignored \cite{Becher:2006qw} (since frequently they are related to scaleless integrals that vanish identically in dimensional regularization). However, when secondary massive quark effects (or, conceptually equivalent, the exchange of massive gauge bosons \cite{Gritschacher:2013pha}) are considered, these subtractions are associated to non-vanishing contributions that need to be computed and specified in a well-defined and systematic fashion. Here, the concept of a zero-bin subtraction gets tied up with the concept of a soft mass mode bin subtraction \cite{Chiu:2009yx,Gritschacher:2013pha,Pietrulewicz:2014qza}. The ${\cal O}(\alpha_s^2)$ secondary massive quark corrections for the SCET jet function for primary massless quarks with imposing zero-bin as well as soft mass mode bin subtractions were calculated in \rcite{Pietrulewicz:2014qza}. However, the difference between both types of subtractions makes the treatment of secondary quark mass effects subtle since the mass mode subtractions may or may not be associated to lower virtuality modes that are, depending on the application, located in separated phase space regions (i.e.\ having different power counting).

We propose that an alternative and in fact more transparent view is to associate the subtractions needed to define the jet functions to so-called collinear-soft matrix elements \cite{Bauer:2011uc,Procura:2014cba}, an idea that was already explored in different contexts in \rcites{Lee:2006nr,Idilbi:2007ff}. This leads to jet function results that agree with the results based on the zero-bin and mass mode bin  subtraction concepts, but avoids the need to impose power counting arguments to determine whether a bin subtraction for a given diagram is relevant. Interestingly, there are two alternative ways to define the subtractions which entail two options to define jet functions, both of which differ analytically once secondary quark mass effects at ${\cal O}(\alpha_s^2)$ are accounted for. Both options differ in whether or not the collinear-soft matrix element used for subtractions accounts for secondary massive quark corrections. The former option leads to what we call ``universal'' jet	functions which merge into the well known massless quark jet functions in the limit $m\to 0$ and which were already introduced in \rcites{Pietrulewicz:2014qza,Hoang:2015iva}. The latter option leads to what we call ``mass mode'' jet functions, which are divergent in the limit $m\to 0$ and contain (universal) rapidity singularities. The mass mode jet functions are analogous to a corresponding definition for invariant mass dependent beam functions introduced in \rcite{Pietrulewicz:2017gxc}. Both types of jet functions are useful to gain a transparent view on the treatment of secondary massive quark effects and can be employed in practical applications. They are also useful to better understand the relation between the secondary massive quark factorization approaches provided in \rcites{Gritschacher:2013pha,Pietrulewicz:2014qza,Hoang:2015iva} and in \rcite{Pietrulewicz:2017gxc} which lead to equivalent results. We stress that the concepts of the ``universal'' and the ``mass mode'' jet functions also applies to the analogous situation when the effects of massive (e.g.\ electroweak) gauge bosons are accounted for. This is because the structure of the relevant dynamical modes is equivalent~\cite{Gritschacher:2013pha}.
For the exchange of massive gauge bosons, however, the mass effects arise already at ${\cal O}(\alpha_{em})$, $\alpha_{em}$ being the electromagnetic coupling, which will be explored elsewhere. Furthermore, the same concepts can also be applied for invariant mass dependent beam functions \cite{Samitz:thesis}.

The main aim of this paper is to present the ${\cal O}(\alpha_s^2)$ results (and some details of their computation) for the primary massive quark SCET jet functions.
They represent important ingredients in factorization theorems for massive quark production within an inclusive jet where the massive quark is produced by the hard interaction and where the jet masses are similar in size to the mass of the quark.
To illustrate the application of the universal and the mass mode jet functions
we also discuss their role in the equivalent factorization approaches of \rcites{Gritschacher:2013pha,Pietrulewicz:2014qza} and \cite{Pietrulewicz:2017gxc} using  as an example the double differential hemisphere mass distribution in $e^+ e^-$ annihilation (for the hard production of a boosted massive quark-antiquark pair). This discussion also clarifies the relation between both approaches and furthermore emphasizes that a smooth dependence on the value of the quark mass $m$ (in the sense that assumptions on the hierarchy of $m$ with respect to the other physical scales are not mandatory) can be achieved with them.
The latter issue was already fully accounted for in \rcites{Gritschacher:2013pha,Pietrulewicz:2014qza,Hoang:2015iva} but not explored in \rcite{Pietrulewicz:2017gxc}. On the other hand, in \rcite{Pietrulewicz:2017gxc} a more systematic representation of the building blocks to sum virtuality and rapidity logarithms on an equal footing was provided.

The content of the paper is organized as follows. In \sec{notation} we set up our notation and provide the definition of the universal and the mass mode SCET jet functions in the context of subtractions based on collinear-soft matrix elements. In \sec{results} we present our results for both SCET jet functions for massive primary quarks at ${\cal O}(\alpha_s^2)$ and show that (and in which way) the results are consistent with respect to available ${\cal O}(\alpha_s^2)$ results in the limit of massless quarks \cite{Becher:2006qw} and the bHQET limit \cite{Jain:2008gb}. The practical use of the jet functions in the context of the factorization approaches of \rcites{Gritschacher:2013pha,Pietrulewicz:2014qza,Hoang:2015iva} and of \rcite{Pietrulewicz:2017gxc} is discussed in \sec{consistency} for the example of the double differential hemisphere mass distribution in $e^+ e^-$ annihilation. The discussion should be sufficient to illustrate how the jet functions can be used for other processes in the context of QCD as well as the electroweak theory. Here, we also show how both approaches, considered together, are useful to gain a clearer view on the summation of virtuality and rapidity logarithms, the structure of collinear and soft mass mode corrections and on how a smooth dependence on the quark mass $m$ can be achieved in practical applications. Section \ref{sec:calculations} contains details of the calculations for the ${\cal O}(\alpha_s^2)$ corrections to the massive primary quark SCET jet functions. This section is self-contained and focuses on computational details because the calculations are involved and require specific methods for different types of diagrams that may be useful for similar future work. The readers not interested in these technical details may skip this section except for \sec{commentssummary} which summarizes the results for the jet-function and collinear-soft matrix elements with different infrared regulators. In \sec{analysis} we provide a brief numerical analysis of the universal primary massive quark SCET jet function at ${\cal O}(\alpha_s^2)$ in comparison to the corresponding known result in the massless quark and the bHQET limits. Finally, in \sec{conclusions} we conclude.
The paper also contains a number of appendices. 
In \app{OtherJet} we provide the expressions for the universal and the mass mode jet functions at ${\cal O}(\alpha_s^2)$ for massless primary quarks for the sake of future use. The latter can be extracted from \rcite{Pietrulewicz:2014qza} but was not given there. The massless limit of the non-distributional corrections of the primary massive SCET jet function is provided in \app{Gmto0}, and a number of virtuality and rapidity anomalous dimensions used in the discussion of \sec{consistency} are collected in \app{AnoDim}. Finally, in \app{MI} we provide the results for the two-loop master integrals needed for our calculations.

\section{Jet Function Definitions and Notation}
\label{sec:notation}

Quark jet functions in SCET are based on the vacuum correlator of two SCET jet fields, encoding the inclusive collinear dynamics of quark fields coherently accounting for the collinear gluon radiation from all other color sources of a process. 
{\it We work in QCD with $n_\ell$ massless quark flavors and one quark flavor $Q$ with mass $m$.}
The SCET jet-function matrix element (ME) for a jet initiated by the primary quark $f$ in light-like direction $n^\mu$ ($\vec{n}^2=1$) is given by~\cite{Bauer:2001yt,Fleming:2007qr}\footnote{
We note that in this paper we write jet functions and jet-function MEs as functions of the invariant mass $p^2\ge m^2$ in order to make all dependence on the quark mass $m$ explicit. 
}
\begin{align} 
\label{eq:JmeSCET}
\Jme_f^{(n_\ell+1)}(p^2, m^2)
&= \frac{(2\pi)^3}{N_c\, p^-}\,
   \mathrm{Tr}\,
   \Big\langle 0 \Big|  \frac{\bnslash}{2} \chi^n_f (0) \,
   \delta( p^+  \!+  \hat{p}^+)\, \delta(p^- \!+ \bnP_n)\, \delta^{(2)}(p_\perp \!+\! \cP_{n\perp})\, \overline{\chi}^n_f(0) 
   \Big| 0 \Big\rangle
\\
&= \delta(p^2-m^2) + \ord{\alpha_s}
   \,.\nn
\end{align}
We decompose four-vectors in lightcone components according to $p^\mu = p^- n^\mu/2 + p^+ \bn^\mu/2 + p_\perp^\mu$ with $n^2=\bn^2=0$, $\bar n\cdot n=2$, and $n\cdot p_\perp =\bn\cdot p_\perp=0$.
The SCET label momentum operators $\bnP_n$ and $\cP_{n\perp}$ yield the sum of the large minus and perpendicular light-cone momentum labels of the fields on their right, respectively~\cite{Bauer:2001ct}, while  $\hat p^+$ is the momentum operator of the small plus momentum.
The trace is over color ($N_c = 3$) and spinor indices.
The SCET jet fields $\chi_f^n(x)$ are defined by
\begin{equation}
\label{eq:jetfielddef1}
 \chi_f^n(x)=W^\dagger_n(x)\xi^n_f(x)\,,
\end{equation}
where $\xi_f^n(x)$ is the SCET primary quark field and the $n$-collinear Wilson line can be written as
\begin{equation}
W_n(x)=\sum_\mathrm{perms}\exp\bigg(\frac{-g}{\bar n\cdot \mathcal P}\,\bar n\cdot A_n(x)\bigg)
\,.
\end{equation}
The Wilson lines in \eq{JmeSCET}  encode the universal coherent dynamics of collinear gluons coming from other color sources, which are boosted in direction $\vec{\bn}=- \vec{n}$ with respect to the direction of motion of the primary quark. 
They entail that the jet-function ME is gauge-invariant with respect to collinear gauge transformations \cite{Bauer:2000yr}.

After BPS field redefinition the collinear sectors of the (leading-order) SCET Lagrangian are equivalent to boosted versions of the QCD Lagrangian~\cite{Bauer:2000yr,Bauer:2001yt}.
We can thus express the $n$-collinear jet-function ME also in terms of QCD quark fields as~\cite{Becher:2006qw}
\begin{align}
\Jme_f^{(n_\ell+1)} &(p^2, m^2) =
\frac{1}{4 \pi N_c \,(\bar n\cdot p)}
\int\! \dd^4x \; e^{-ip\cdot x}\,
\mathrm{Tr}\, 
\Big\langle 0 \Big| \frac{\bnslash}{2}
W^\dagger(0)\, \psi_f(0)\,\overline{\psi}_f(x)\, W(x)
\Big| 0 \Big\rangle
 \notag \\[2mm]
&= \mathrm{Im}\bigg[
\frac{i}{2 \pi N_c \,(\bar n\cdot p)}
\int\! \dd^4x \; e^{-ip\cdot x}\,
\mathrm{Tr}\, 
\Big\langle 0 \Big|  \mathrm{T} \bigg\{  \frac{\bnslash}{2}
W^\dagger(0)\, \psi_f(0)\,\overline{\psi}_f(x)\, W(x)
\bigg\}
\Big| 0 \Big\rangle
\bigg] \,,
\label{eq:JmeQCD}
\end{align}
with the corresponding QCD definition for the $n$-collinear Wilson line
\begin{align}
W(x) &= \mathrm{P}\, \exp \biggl[i\, g \int_{-\infty}^0 \!\!  \dd s \, \bn \!\cdot\! 
A(x+s \bn)  \biggr]
\,.
\end{align}
In \eq{JmeQCD} we have reexpressed the jet-function ME as the imaginary part of a jet field correlator.
Taking the imaginary part is equivalent to half the discontinuity w.r.t.\ $s=p^2-m^2 > 0$.
In practice, for the computation of the two-loop corrections it is more convenient to work with \eq{JmeQCD} rather than with the corresponding SCET expression of \eq{JmeSCET}, because the QCD Feynman rules are simpler than the SCET Feynman rules and lead to less diagrams.

In \eqs{JmeSCET}{JmeQCD} the quark flavor index $f$ stands for the flavor of the incoming primary quark and can either be one of the $n_\ell$ massless quarks, generically referred to as $q$, or the massive quark $Q$. The argument $m$ represents the dependence on the mass of quark $Q$ and arises either from secondary $Q$ effects (due to gluon splitting) if the primary quark is a massless flavor $q$ or from primary and secondary $Q$ effects. The superscript $(n_\ell+1)$ in \eqs{JmeSCET}{JmeQCD} (and for ME definitions and factorization functions used throughout this paper) indicates that we consider the MEs in the context of QCD with $n_\ell$ massless quarks and the massive quark $Q$. We stress, however, that here and throughout this work the superscript $(n_\ell+1)$ (or $n_\ell$) on MEs or factorization functions does not automatically imply that also the corresponding flavor scheme for the $\overline{\rm MS}$ strong coupling is used. The latter can in principle be chosen independently, particularly for renormalization scales $\mu$ close to the quark mass $m$. We therefore specify explicitly the flavor scheme of $\alpha_s$ when we quote results parametrized by either $\alpha_s^{(n_\ell)}$ or  $\alpha_s^{(n_\ell+1)}$.

It is well known \cite{Fleming:2007xt,Fleming:2007qr} that the jet-function ME $\Jme_f^{(n_\ell+1)}(p^2,m^2)$ defined in \eqs{JmeSCET}{JmeQCD} is per se not infrared-finite and commonly said to implicitly require as part of its definition zero-bin subtractions \cite{Manohar:2006nz} to avoid double counting concerning the collinear overlap with softer or lower virtuality regions described by other functions in factorization theorems where the jet function appears. {\it Alternatively}, one can consider the jet function as an IR-finite matching function that describes fluctuations with virtualities $p^2-m^2\sim p^2 \gtrsim m^2$ and that is defined by the jet-function ME $\Jme_f^{(n_\ell+1)}(p^2,m^2)$ with {\it additional} (i.e.\ independent) ME subtractions related to (soft and possibly collinear) modes at smaller virtualities.
So far in the literature the required subtractions, when they contributed non-vanishing terms, were implemented mostly via the zero-bin \cite{Fleming:2007xt,Fleming:2007qr} or soft mass mode bin \cite{Gritschacher:2013pha,Pietrulewicz:2014qza,Gritschacher:2013tza} prescriptions that are imposed by hand. However, from the point of view of being matching functions, the jet functions can also be defined in a conceptually more systematic way by explicitly dividing out fluctuations with momenta $n\cdot k=k^+\sim p^2/Q$ and virtualities $k^2\ll p^2-m^2$ 
in terms of a so-called collinear-soft ME. 
This also has the advantage that no additional power counting arguments are mandatory to identify which contributions have to be accounted for (and which not), as it is necessary when implementing a bin subtraction.

In case that all $(n_\ell+1)$ quarks are massless this results in the jet function definition
\begin{align}
\label{eq:jm0def}
J_f^{(n_\ell+1)}(p^2,0) \equiv {}& \int\dd\ell\, \Jme_f^{(n_\ell+1)}(p^2-(\bar n\cdot p)\ell,0)\Bigl(\Scme^{(n_\ell+1)}(\ell,0)\Bigr)^{-1}
\,.
\end{align}
Here $\Scme^{(n_\ell+1)}(\ell,m)$ is the collinear-soft ME of our $(n_\ell+1)$ flavor theory (with $n_\ell$ massless quarks $q$ and one quark $Q$ with mass $m$) and defined as\footnote{Note that the collinear-soft matrix element is universal in the sense that it only depends on the color representations of the partons involved as noted in \rcite{Bauer:2011uc}. Therefore the result is equivalent to the one given in eq.\ (B.51) of \rcite{Pietrulewicz:2017gxc} where massive quark effects in exclusive Drell-Yan were considered.}
\begin{equation}
\label{eq:Sdef1}
 \Scme^{(n_\ell+1)}(\ell,m)=\frac{1}{N_c}\mathrm{Tr}\left\langle 0\left|\overline{\mathrm{T}}\bigl[V_n^{(0)\dagger}(0)X_n^{(0)}(0)\bigr]\,\delta(\ell-\hat{p}^+)\,\mathrm T\bigl[X_n^{(0)\dagger}(0)V_n^{(0)}(0)\bigr]\right|0\right\rangle\,,
\end{equation}
with the collinear-soft Wilson lines~\cite{Bauer:2011uc,Procura:2014cba}
\begin{align}
X_n(x)={}&\sum_\mathrm{perms}\exp\biggl(\frac{-g}{n\cdot \mathcal P}\frac{\nu^{\eta/2}}{|\bar n\cdot \mathcal P|^{\eta/2}}\, n\cdot A_{cs}(x)\biggr)\,,\nonumber\\
V_n(x)={}&\sum_\mathrm{perms}\exp\biggl(\frac{-g}{\bar n\cdot \mathcal P}\frac{\nu^{\eta/2}}{|\bar n\cdot \mathcal P|^{\eta/2}}\, \bar n\cdot A_{cs}(x)\biggr)\,,
\label{eq:XVdef}
\end{align}
including the symmetric analytic rapidity regularization of \rcite{Chiu:2012ir}.
The inverse of the collinear-soft ME is defined by the convolution
\begin{equation}
\label{eq:SSinv}
\int\dd\ell'\, \Scme^{(n_\ell+1)}(\ell-\ell',m)\Bigl(\Scme^{(n_\ell+1)}(\ell',m)\Bigr)^{-1}=\delta(\ell)\,,
\end{equation}
and we use the analogue relations for defining the inverse of all (factorization) functions depending on dynamical variables.
The combination of MEs in \eq{jm0def} yields the well known IR finite massless quark jet function at one~\cite{Lunghi:2002ju}, two~\cite{Becher:2006qw} and three loops~\cite{Bruser:2018rad}. 
The analogous definition (with the appropriate color representation for the Wilson lines) can be applied also for the gluon jet function~\cite{Becher:2009th,Becher:2010pd,Banerjee:2018ozf}.
For $(n_\ell+1)$ massless quarks the subtraction of the low virtuality modes is from the purely computational point of view, however, not strictly mandatory for computations of the jet function when dimensional regularization for UV and IR divergences is used because then $\Scme^{(n_\ell+1)}(\ell,0)$ yields vanishing scaleless integrals beyond tree-level.

On the other hand, when considering the quark $Q$ having a finite mass $m$, $\Scme^{(n_\ell+1)}(\ell,m)$ is not anymore scaleless and yields non-vanishing ${\cal O}(\alpha_s^2 C_F T_F)$
contributions in dimensional regularization due to mass mode fluctuations with virtualities $k^2\sim m^2$. We note that the mass mode contributions in the collinear-soft ME $\Scme^{(n_\ell+1)}(\ell,m)$ are only related to secondary effects of the massive quark $Q$ due to closed $Q$ loops, so that it is the same for the jet functions for a massless and a massive primary quark. 

Interestingly, secondary massive quark effects do not lead to IR divergences (as long as one does not take the massless limit) that must always be subtracted. It is therefore not mandatory to use the massive quark analogue of \eq{jm0def} to define the jet function. This is also related to the emergence of rapidity singularities, which one may not associate to be of either UV or IR type and leads to two options to define the SCET jet function in the context of QCD with $n_\ell$ massless flavors and one quark $Q$ with mass $m$.

The first option is just the analogue of \eq{jm0def} but with the $(n_\ell+1)$st quark $Q$ having mass $m$,
\begin{equation}
\label{eq:jmufdef}
J_f^{\mathrm{uf},(n_\ell+1)}(p^2,m^2)\equiv\int\dd\ell\, \Jme_f^{(n_\ell+1)}(p^2-(\bar n\cdot p)\ell,m^2)
\Bigl(\Scme^{(n_\ell+1)}(\ell,m^2)\Bigr)^{-1}\,.
\end{equation}
The primary jet-initiating quark flavor $f$ can either be one of the $n_\ell$ massless quarks $q$ or the massive quark $Q$. 
Here the collinear-soft ME is calculated in the full $(n_\ell+1)$ flavor theory and thus encodes in particular the secondary effects coming from the massive quark $Q$.
This jet function yields by construction the massless quark jet function $J_f^{(n_\ell+1)}(p^2,0)$ of \eq{jm0def} in the limit $m\to 0$. $J_f^{\mathrm{uf},(n_\ell+1)}(p^2,m^2)$ and $J_f^{(n_\ell+1)}(p^2,0)$ have the same anomalous dimension and are furthermore free of rapidity divergences as well as of any large rapidity logarithms. For $J_f^{\mathrm{uf},(n_\ell+1)}(p^2,m^2)$ both cancel between $\Jme_f^{(n_\ell+1)}$ and $\Scme^{(n_\ell+1)}$.
Upon renormalizing $J_f^{\mathrm{uf},(n_\ell+1)}$ the only dependence on a renormalization scale is that on the virtuality scale $\mu$.
Following the philosophy of the SCET jet function being a matching function describing fluctuations with virtuality $p^2-m^2 \sim p^2$, the jet function of \eq{jmufdef} is the mandatory definition for the case where $m^2\ll p^2$ where the quark mass $m$ is an IR scale and mass effects represent power corrections. In that regime, however, one may as well use the massless quark jet function $J_f^{(n_\ell+1)}(p^2,0)$.
The jet function $J_f^{\mathrm{uf},(n_\ell+1)}(p^2,m^2)$ can, however, also be employed for $p^2\sim p^2-m^2\sim m^2$ following the approach of \rcites{Gritschacher:2013pha,Pietrulewicz:2014qza} (originally developed for massless primary quark jet functions) where universal hard, jet and soft (and in principle also invariant mass dependent beam) functions can be defined valid for {\it any} value of $m^2$, with respect to the other dynamic scales and the renormalization scale $\mu$. This approach makes the $\mu$-evolution in $n_\ell$ or $(n_\ell+1)$ flavor schemes to sum virtuality logarithms particularly transparent and allows to formulate factorization theorems that smoothly interpolate all possible hierarchies as far as the value of $m$ with respect to the other kinematic scales is concerned. 
The approach of \rcites{Gritschacher:2013pha,Pietrulewicz:2014qza} also 
entails that rapidity singularities (and the respective resummation of logarithms) only arise within the mass threshold matching conditions for the individual hard, jet (as well as beam) and soft factorization functions. Since these threshold matching conditions consist of a few universal collinear and soft mass mode factors, there can be cancellations among them in the context of a factorization theorem related to consistency relations. So mass mode factors can arise in the matching conditions that are physically not relevant for a particular observable and eventually cancel in the corresponding factorization theorem.
We call this the ``universal factorization function'' approach and therefore give the jet function definition in \eq{jmufdef} the superscript ``uf''.
In the following we call it the ``universal jet function''.

The second option is to define the jet function in the presence of a massive quark by
\begin{equation}
\label{eq:jmmfdef}
J_f^{\mathrm{mf},(n_\ell+1)}(p^2,m^2)=\int\dd\ell\, \Jme_f^{(n_\ell+1)}(p^2-(\bar n\cdot p)\ell,m^2)
\Bigl(\Scme^{(n_\ell)}(\ell)\Bigr)^{-1},
\end{equation}
where the collinear-soft ME is calculated in the $n_\ell$ flavor theory, i.e.\ containing the $n_\ell$ massless quarks but not accounting for the massive quark $Q$.
Like for the jet function $J_f^{\mathrm{uf},(n_\ell+1)}$, the primary jet-initiating quark $f$ can either be one of the $n_\ell$ massless quarks $q$ or the massive quark $Q$. The jet function $J_f^{\mathrm{mf},(n_\ell+1)}$ is IR divergent for $m\to 0$ and furthermore has rapidity singularities that are treated by rapidity renormalization. So upon renormalization, $J_f^{\mathrm{mf},(n_\ell+1)}$ depends on the (virtuality) renormalization scale $\mu$ and in addition on the (rapidity) renormalization scale $\nu$.
Following the philosophy of the SCET jet function being a matching function describing the fluctuations with virtuality $p^2-m^2 \sim p^2$, the jet function of \eq{jmmfdef} is the natural option for the case where $m^2\sim p^2\sim p^2-m^2$, when the mass is not an IR scale. It has also been advocated for this scale hierarchy in \rcite{Pietrulewicz:2017gxc} (originally developed for the massless primary quark and invariant mass dependent beam function \cite{Stewart:2009yx}). 
In the factorization approach of \rcite{Pietrulewicz:2017gxc} separate factorization theorems are formulated for each possible hierarchy of the mass scale $m$ with respect to the other kinematic scales. 
The approach is not designed to provide smooth dependence on $m$, but is more economical (in the sense of being minimalistic) concerning the appearance of collinear and soft mass mode contributions so that no cancellations arise between them in the context of a factorization theorem. Furthermore, the 
collinear and soft mass mode contributions adopt the status of genuine factorization functions.
The resulting structure of factorization theorems makes the structure of $\nu$-evolution to sum rapidity logarithms particularly transparent as it appears at the same level as the virtuality $\mu$-evolution. We call this the ``mass mode factorization'' approach and therefore give the jet function definition in \eq{jmmfdef} the superscript ``mf''.
In the following we call it the ``mass mode jet function''.

The difference between both $\overline{\rm MS}$-renormalized massive quark jet function definitions yields\footnote{The inverse jet function is defined in analogy to \eq{SSinv}.}
\begin{align}
S_{c}(\ell,m,\mu,\nu) ={}& \bar n\cdot p \int\!\dd p'^2\, J_f^{\mathrm{mf},(n_\ell+1)}((\bar n\cdot p)\ell-p'^2,m^2,\mu, \bar n\cdot p/\nu)\bigl(J_f^{\mathrm{uf},(n_\ell+1)}(p'^2,m^2,\mu)\bigr)^{-1}\nonumber\\
={}&\int\dd\ell\, \Scme^{(n_\ell+1)}(\ell-\ell',m,\mu,\nu)\Bigl(\Scme^{(n_\ell)}(\ell',\mu,\nu)\Bigr)^{-1},
\label{eq:Scdef}
\end{align}
and is called the ``collinear-soft function'' in \rcite{Pietrulewicz:2017gxc}.
It is IR-finite for finite $m$ and has a dependence on the virtuality and the rapidity renormalization scales $\mu$ and $\nu$, respectively.

In \sec{consistency} we illustrate the practical use of both types of jet functions and the two factorization approaches of \rcites{Pietrulewicz:2014qza,Gritschacher:2013pha} and \cite{Pietrulewicz:2017gxc} exemplarily for the double differential hemisphere mass distribution in $e^+e^-$ collisions for the hierarchies $m^2\sim p^2\sim p^2-m^2$ and $m^2\ll p^2$.
We also stress that the form of the collinear-soft ME of \eq{Sdef1} and
the two options of using either \eq{jmufdef} or \eq{jmmfdef} for IR finite jet functions in the presence of massive quarks can also be applied in a fully equivalent way for invariant mass dependent beam functions \cite{Stewart:2009yx} and the bHQET jet function \cite{Fleming:2007xt,Fleming:2007qr}.\footnote{In the context of bHQET \cite{Fleming:2007xt,Fleming:2007qr} we refer to the situation
that some of the $n_\ell$ light flavors have finite masses smaller than the mass $m$ of the primary HQET quark, but larger than the hadronization scale $\Lambda_\mathrm{QCD}$. The equivalence of the collinear-soft MEs (or the zero-bin subtractions) for SCET and bHQET is a consequence of a rescaling property of the bHQET and SCET soft gluon eikonal couplings that can e.g.\ be trivially seen from the structure of the zero-bin diagrams in both effective theories.}
The approach also applies in an analogous way for the treatment of massive gauge bosons which is conceptually related to the issues of secondary quark mass effects as was discussed in~\cite{Gritschacher:2013pha}.

\section{Analytic Results: Primary Massive Quark SCET Jet Functions}
\label{sec:results}

In this section we present the analytic results for the $\mathcal O(\alpha_s^2)$ primary quark SCET jet functions. Details on the computation of the $\mathcal O(\alpha_s^2)$ corrections, which are all new, are given in \sec{calculations}. We use the abbreviations
\begin{equation}
a_s^{(n_f)}\equiv \frac{\alpha_s^{(n_f)}(\mu)}{4\pi}\qquad\text{and}\qquad s\equiv p^2-m^2\,,
\end{equation}
for the strong coupling in the $n_f$-flavor scheme and the squared jet function invariant mass $p^2$, respectively,
\begin{equation}
L_m \equiv \log\frac{m^2}{\mu^2}\,, 
\end{equation}
as well as
\begin{equation}
\mathcal L_n(x) \equiv \frac{1}{\mu^\lambda}\left[\frac{\Theta(x)\log^n (x/\mu^\lambda)}{x/\mu^\lambda}\right]_+,
\end{equation}
for plus-distributions for a variable $x$ having mass dimension $\lambda$, where $\mu$ is the common $\overline{\mathrm{MS}}$ renormalization scale. They are defined such that
$\int_0^1 \dd x\, \mathcal L_n(x)=0$.
Furthermore we define
\begin{equation}
y\equiv \frac{m^2}{p^2}\,, \qquad
\bar y\equiv  \frac{s}{p^2}= 1-y\,, \qquad
L_y\equiv\log(y)\,, \qquad
L_{\bar y}\equiv\log(\bar y)
\,. 
\end{equation}
In the physical kinematic region we have $0\le y,\bar y\le 1$. Note that we also use the notation
\begin{equation}
Q\equiv (\bn\cdot p) = p^-\,,
\end{equation}
and that we carry out all computations in 
$d=4-2\varepsilon$ dimensions. 

\subsection{Mass Mode Jet Function}

The result for the renormalized primary massive mass mode jet function $J_Q^{\mathrm{mf},(n_\ell+1)}$, defined in \eq{jmmfdef}, in the pole mass scheme at $\mathcal O(\alpha_s^2)$ can be written in the form
\begin{align}
\label{eq:Jmfdef}
J_Q^{\mathrm{mf},(n_\ell+1)}(p^2,m^2,\mu,Q/\nu) &={} \delta(p^2-m^2) + a_s^{(n_\ell)} J_Q^{(1)}(p^2,m^2,\mu)\\
&+ \Bigl(a_s^{(n_\ell)}\Bigr)^2 \left(J_Q^{(2)}(p^2,m^2,\mu)+J_{Q,\mathrm{sec}}^{(2),\mathrm{mf}}(p^2,m^2,\mu,Q/\nu)\right) + \mathcal O\left(\alpha_s^3\right)\,, \notag
\end{align}
where we adopt the $n_\ell$ flavor scheme for the strong coupling. This can be considered the natural scheme choice for the mass mode jet function since it matches directly to the bHQET jet function, see \sec{ConsKine}. It is of course straightforward to switch to the $(n_\ell+1)$ flavor scheme using \eq{alphamatch}. At $\mathcal O(\alpha_s^2)$ we distinguish between the corrections from diagrams containing only one single quark $Q$ line ($J_Q^{(2)}$) and those from diagrams containing the $Q\bar Q$ vacuum polarization correction ($J_{Q,\mathrm{sec}}^{(2)}$). The mass mode jet function contains rapidity singularities from the $Q\bar Q$ vacuum polarization diagrams which are renormalized following the approach of \cite{Chiu:2011qc,Chiu:2012ir} and which lead to the dependence on the rapidity renormalization scale $\nu$. For the prototypical application of the mass mode jet function within factorization theorems where all logarithms are resummed by explicit renormalization group evolution factors the natural choice for the rapidity renormalization scale is $\nu\sim Q$. This is indicated by the argument $Q/\nu$. Throughout this paper we adopt this convention for the rapidity scale $\nu$ in an analogous way. In case the natural choice for $\nu$ is process (or factorization theorem) dependent, only $\nu$ appears as the argument.

The $\mathcal O(\alpha_s)$ contribution accounting for terms up to $\mathcal O(\varepsilon^2)$ in order to maintain all information mandatory for $\mathcal O(\alpha_s^2)$ calculations reads
\begin{align}
\label{eq:Jmf1}
&J_Q^{(1)}(p^2,m^2,\mu)=C_F \Bigg\{\left(2 L_m^2+L_m-\frac{\pi^2}{3}+8\right) \delta(s)+(-4 L_m-4) \mathcal L_{0}(s)+8 \mathcal L_{1}(s) \\
&\quad+G_1 \Theta(p^2-m^2)\nonumber\\\displaybreak[1]
&+\varepsilon\Bigg[\left(\frac{3}{2} L_m^2 - \frac{2 \pi^2}{3} L_m + 16 - \frac{5 \pi^2}{12} - \frac{16 \zeta_3}{3}\right)\delta(s)+\left(- 2 L_m^2 - 4 L_m - 8 + \pi^2\right)\mathcal L_0(s)\notag\\
&\quad+\left(8 + 8 L_m\right)\mathcal L_1(s) - 8\mathcal L_2(s)+G_1^{(\varepsilon)}\Theta(p^2-m^2)\Bigg]\nonumber\\\displaybreak[1]
&+\varepsilon^2\Bigg[\Bigg(\frac{1}{6} L_m^4 + \frac{1}{6}L_m^3 + \left(4 - \frac{\pi^2}{6}\right) L_m^2  + \left(-\frac{7 \pi^2}{12} - 4 \zeta_3\right) L_m+ 32 - \frac{2 \pi^2}{3} - \frac{\pi^4}{120}  - 5 \zeta_3\Bigg)\delta(s)\nonumber\\
&\quad+\left(- \frac{2}{3} L_m^3 - 2 L_m^2 + (-8 + \pi^2)L_m - 16 + \pi^2 + \frac{28 \zeta_3}{3}\right)\mathcal L_0(s)\notag\\
&\;+\left(4 L_m^2 + 8 L_m + 16 - 2 \pi^2\right)\mathcal L_1(s)+(-8 - 8 L_m) \mathcal L_2(s) + \frac{16}{3} \mathcal L_3(s)+G_1^{(\varepsilon^2)}\Theta(p^2-m^2)\Bigg]\Bigg\}.
\notag
\end{align}
Here, all distributional contributions are displayed explicitly and the non-distributional contributions are contained in the functions
\begin{align}
\label{eq:G1def}
G_1={}&\frac{1}{s}\left[\bar y^2 + 4 L_y\right] = -\frac{4}{m^2}+\mathcal O\leri{\frac{s}{m^4}},\displaybreak[1]\\
G^{(\varepsilon)}_1={}&\frac{1}{s}\left[ \left(-\bar{y}^2-4 L_y\right)L_m+L_y \left(-8 L_{\bar{y}}+y^2-2 y+5\right)-2 \bar{y}^2 L_{\bar{y}}+\bar{y}^2+6 L_y^2\right]\\
={}& -\frac{4}{m^2}\left(-2 \Lyby-L_m+1\right)+\mathcal O\leri{\frac{s}{m^4}},\notag\displaybreak[1]\\
G^{(\varepsilon^2)}_1={}&\frac{1}{s}\bigg[ \Big(L_y \left(8 L_{\bar{y}}-y^2+2 y-5\right)+2 \bar{y}^2 L_{\bar{y}}-\bar{y}^2-6L_y^2\Big)L_m+ \left(\frac{\bar{y}^2}{2}+2 L_y\right)L_m^2\\
&+L_y^2 \left(\frac{1}{2} \left(y^2-2y+13\right)-12 L_{\bar{y}}\right)+2 \bar{y}^2 L_{\bar{y}}^2-2 \bar{y}^2 L_{\bar{y}}-\frac{1}{4} \left(\pi ^2-8\right) \bar{y}^2+\frac{14}{3}L_y^3\nn\\
&+L_y \Big(-2 \left(y^2-2 y+5\right) L_{\bar{y}}+8L_{\bar{y}}^2+y^2-2 y-\pi ^2+9\Big)\bigg]\notag\\
={}&\frac{1}{m^2}\left[ \left(4-8 \Lyby\right)L_m-8 \LLyby+8\Lyby-2 L_m^2+\pi ^2-8\right]+\mathcal O\leri{\frac{s}{m^4}}.\notag
\end{align}
These functions do not contribute to the singular behavior in the bHQET limit $s=p^2-m^2\ll m^2$  and thus represent power corrections in this kinematic region. They constitute the so-called $\mathcal O(\alpha_s)$ bHQET non-singular corrections that are important for physical factorization predictions in the region $s\lesssim m^2$ to achieve a smooth transition between bHQET and SCET factorization theorems \cite{Fleming:2007qr,Dehnadi:2016snl}. For completeness we also quoted the $s\to 0$ limit of the $G_1$-functions. The $\varepsilon\to 0$ limit of \eq{Jmf1} was already given in \rcite{Fleming:2007xt}, and we refer to this reference for pictures of the Feynman diagrams and details of the calculation. The determination of the 
	${\cal O}(\varepsilon)$ and ${\cal O}(\varepsilon^2)$ contributions is straightforward.

The $\mathcal O(\alpha_s^2)$ contributions arising from Feynman diagrams containing only one single quark $Q$ line (see all diagrams in \fig{diagrams1} with the diagrams (o)-(r) containing only massless quark bubbles) read
\begin{align}
\label{eq:Jmf2}
&J_Q^{(2)}(p^2,m^2,\mu) = C_F^2 \Bigg\{\Bigg[2 L_m^4+2 L_m^3+\left(\frac{33}{2}-2 \pi^2\right) L_m^2+\left(-56 \zeta_3+\frac{13}{2}-\pi^2\right)L_m\\
&\qquad-38 \zeta_3-\frac{\pi^4}{2}+\frac{433}{8}+\pi^2 \left(-\frac{1}{3}-8 \log 2\right)\Bigg] \delta(s)\notag\\
&\quad+\left[-8 L_m^3-12 L_m^2+\left(\frac{20 \pi^2}{3}-36\right) L_m+64 \zeta_3+\frac{20 \pi^2}{3}-32\right] \mathcal L_{0}(s)\nonumber\\
&\quad+\left[32 L_m^2+40 L_m-\frac{40}{3} \left(\pi^2-6\right)\right] \mathcal L_{1}(s)+[-48 L_m-48] \mathcal L_{2}(s)+32 \mathcal L_{3}(s)\notag\\
&\quad+G_{C_F}^{(1Q)}\Theta(p^2-m^2)+G^{(3Q)}\Theta(p^2-(3m)^2)\Bigg\}\nonumber\displaybreak[1]\\
&+C_F C_A \Bigg\{\Bigg[\left(\frac{367}{18}-\frac{2 \pi^2}{3}\right) L_m^2+ \left(20 \zeta_3+\frac{341}{18}-\frac{56 \pi^2}{9}\right) L_m-\frac{188 \zeta_3}{9}-\frac{2 \pi^4}{15}+\frac{60221}{648}\notag\\
&\qquad+\pi^2 \left(4 \log 2-\frac{169}{54}\right)\Bigg] \delta(s)\nonumber\\
&\quad\!+\left[-\frac{22}{3}L_m^2+\frac{4}{9} \left(3 \pi^2-100\right) L_m+\frac{4}{27} \left(135 \zeta_3-547+42 \pi^2\right)\right] \mathcal L_{0}(s)\nonumber\\
&\quad\!+\left[\frac{88}{3}L_m-\frac{8 \pi^2}{3}+\frac{800}{9}\right]\! \mathcal L_{1}(s)-\frac{88}{3} \mathcal L_{2}(s)+G_{C_A}^{(1Q)}\Theta(p^2-m^2)-\frac{1}{2}G^{(3Q)}\Theta(p^2-(3m)^2)\Bigg\}\nonumber\displaybreak[1]\\
&+C_F T_F n_\ell \Bigg\{\left[-\frac{58 }{9}L_m^2+\frac{2}{9} \left(8 \pi^2-37\right) L_m+\frac{1}{162} \left(288 \zeta_3-6037+132 \pi^2\right)\right] \delta(s)\nonumber\\
&\quad+\left[\frac{8 }{3}L_m^2+\frac{128 }{9}L_m-\frac{16}{27} \left(3 \pi^2-47\right)\right] \mathcal L_{0}(s)+\left[-\frac{32 }{3}L_m-\frac{256}{9}\right] \mathcal L_{1}(s)+\frac{32}{3} \mathcal L_{2}(s)\notag\\
&\quad+G_{T_F}\Theta(p^2-m^2)\Bigg\}\,.\notag
\end{align}
The distributional contributions are again displayed explicitly. There are two sets of non-distributional contributions. One set ($\propto \Theta(p^2-m^2)$) is related to terms that are non-vanishing above the one-particle cut and the other ($\propto \Theta(p^2-(3m)^2)$) is related to terms that are non-vanishing above the three particle $QQ\bar Q$ cut. 
Even though the term $J_Q^{(2)}$ exclusively arises from diagrams with one single quark $Q$ line, see \fig{diagrams1}, three particle $QQ\bar Q$ cuts can arise due to the antiparticle component of the quark $Q$ propagator. The non-distributional contributions $\propto \Theta(p^2-m^2)$ read
\begin{align}
&G_{C_F}^{(1Q)} = \frac{1}{s}\Bigg\{L_m^2 \left(2 \bar{y}^2+8 L_y\right)+L_m \Big[4 \bar{y}^2 L_{\bar{y}}+L_y \left(16 L_{\bar{y}}-4 (2 (y-2) y+5)\right)-7 \bar{y}^2-24 L_y^2\Big]\notag\\
&\quad+L_y\bigg[\frac{4}{\bar{y}}\Bigl(3 y \left(y^2+y+5\right)-13\Bigr) L_{\bar{y}}+24 L_{\bar{y}}^2-8\,\text{Li}_2(-\bar{y}/y)-\frac{2}{\bar{y}}\nn\\
&\qquad-2 y \Big(y (3 y-17)+35\Big)-\frac{16 \pi ^2}{3}+47\bigg]+L_y^2 \left[-36 L_{\bar{y}}-\frac{32}{\bar{y}}+8 y (y+5)+76\right]\notag\\
&\quad+2 L_{\bar{y}} \Big[8\,\text{Li}_2(-\bar{y}/y)+y (y (3 y-17)+37)-7\Big]-8 (y-2) \bar{y}L_{\bar{y}}^2\notag\\
&\quad+\frac{4}{\bar{y}}\Big((2-21 y) y+7\Big)\,\text{Li}_2(-\bar{y}/y)+32\,\text{Li}_3(\bar{y})-16\,\text{Li}_3(-\bar{y}/y)+\frac{97}{90} \pi^4 y \bar{y}\label{eq:G1QCF}\\
&\quad+8 (3 y-13) \zeta_3 \bar{y}+\frac{1}{3} \pi^2 \bar{y} \Big(y (72 \log 2-49)+15\Big)+\frac{68}{3}L_y^3-\frac{3}{2}y^2-65 y+\frac{33}{2}+G_\mathrm{fit}(y)\Bigg\} \nn\displaybreak[1]\\
&\,\,\,=\frac{1}{m^2}\Big[L_m \Big(12-16 \Lyby\Big)-15.0 \LLyby+22.0992\Lyby-8 L_m^2-13.8568\Big]\nn\\
&\quad + \mathcal O\!\left(\frac{s}{m^4}\right),\notag\displaybreak[1]\\
&G_{C_A}^{(1Q)} = \frac{1}{s} \Bigg\{ L_m \frac{11}{3} \left(-\bar{y}^2-4 L_y\right)+2 L_y \Bigg[-2 L_{\bar{y}}^2+\left(-\frac{8}{\bar{y}}+12 y-\frac{5}{3}\right) L_{\bar{y}}\notag\\
&\qquad-\frac{1}{9 \bar{y}}\bigg(54 \bar{y} \,\text{Li}_2(-\bar{y}/y)-3 \pi ^2 \bar{y}+y \Big(3 y (5 y+18)+107\Big)-212\bigg)\Bigg]\notag\displaybreak[1]\\
&\quad+L_y^2 \left(-2L_{\bar{y}}+\frac{16}{\bar{y}}-28 y+\frac{25}{3}\right)+2 (y-3) \bar{y} L_{\bar{y}}^2-\frac{2}{3} \Big(y (5 y+17)+2\Big) L_{\bar{y}}\notag\displaybreak[1]\\
&\quad-\frac{2}{3 \bar{y}} \bigg[y \Big(3y (y-24)+40\Big)-7\bigg] \,\text{Li}_2(-\bar{y}/y)-32 \,\text{Li}_3(\bar{y})-24\,\text{Li}_3(-\bar{y}/y)\notag\displaybreak[1]\\
&\quad+\frac{4}{9} \Big(9 (13-3 y) \zeta_3 \bar{y}+46 y^2-17 y+43\Big)-\frac{97}{180} \pi ^4 y\bar{y}\notag\displaybreak[1]\\
&\quad-\frac{1}{3} \pi^2 \bar{y} \Big(y (36 \log 2-26)+9\Big)-2 L_y^3-\frac{G_\mathrm{fit}(y)}{2}\Bigg\}\label{eq:G1QCA}\displaybreak[1]\\
&=\frac{1}{m^2}\left[-0.5 \LLyby+26.2837 \Lyby+\frac{44}{3}L_m-37.1102\right] + \mathcal O\!\left(\frac{s}{m^4}\right)\notag,
\end{align}
\begin{align}
G_{T_F} ={}& \frac{1}{s}\Bigg\{L_m \left[\frac{4}{3}\bar{y}^2+\frac{16}{3}L_y\right]+\frac{8}{3} \bar{y}^2 L_{\bar{y}}+L_y \left[\frac{32}{3}L_{\bar{y}}+\frac{4}{9 \bar{y}}\Big(y (6 (y-3)y+29)-35\Big)\right]\notag\\
&\quad-\frac{2}{9} \Big[24 \,\text{Li}_2(-\bar{y}/y)+y (19 y-26)+43\Big]-\frac{32}{3} L_y^2\Bigg\}\label{eq:G1QTF} \displaybreak[1]\\
={}& -\frac{8}{9 m^2}\Big(6 L_m+12 \log \left(\bar y/y\right)-19\Big) + \mathcal O\!\left(\frac{s}{m^4}\right)
\,.\notag
\end{align}
The non-distributional contributions  $\propto \Theta(p^2-(3m)^2)$ read
\begin{align}
&G^{(3Q)} ={} \frac{1}{s}\Bigg[\frac{\sqrt{\left(1-\sqrt{y}\right) \left(1 + 3 \sqrt{y}\right)} \left(3 y^2+44 y-7\right)}{\left(\sqrt{y}+1\right)}\,\mathrm{E}\left(\frac{\left(1-3 \sqrt{y}\right)\left(1+\sqrt{y}\right)^3}{\left(1+3 \sqrt{y}\right)\left(1-\sqrt{y}\right)^3}\right)\nn\\
&\quad-\frac{8 \sqrt{1 + 3 \sqrt{y}} \left(3 y^{5/2}-18 y^{3/2}+3 y^3+43 y^2-y+3\sqrt{y}-1\right)}{3\left(1-\sqrt{y}\right)^{5/2} \left(1 + \sqrt{y}\right)}\,\mathrm{K}\left(\frac{\left(1-3 \sqrt{y}\right)\left(1+\sqrt{y}\right)^3}{\left(1+3 \sqrt{y}\right)\left(1-\sqrt{y}\right)^3}\right)\nn\\
&\quad+I_3\left(y\right)\frac{8 y \left(1+y\right)^2}{3 \bar y}\Bigg]\label{eq:G3QCF}\\
&\,\,\,={}\frac{\pi (p^2-(3 m)^2)^2}{3456 \sqrt{3} m^6}+\mathcal O\!\left(\frac{(p^2-(3 m)^2)^3}{m^8}\right).\notag
\end{align}
The non-distributional contributions $\propto \Theta(p^2-m^2)$ coming from diagrams (b) and (c) in \fig{diagrams1} could only be computed numerically for general values of $p^2$ and $m^2$, and are parametrized by the fit function $G_\mathrm{fit}$ which has the convenient form\footnote{We note that the function $G_\mathrm{fit}$ contains contributions arising both from $Q$ and $QQ\bar Q$ final states.}
\begin{align}
G_\mathrm{fit}&(y)= \bar y \Big[c_{11} \left(\frac{1}{6} \pi^2 y + L_{\bar y}\right)+c_{12} \left(L_{\bar y}^2-2 y \zeta_3\right)+c_{21} \left(\frac{1}{6} \left(\pi^2-6\right) y+\bar y L_{\bar y}\right)\notag\\
&+c_{22} \left(\bar y L_{\bar y}^2-2 y (\zeta_3-1)\right)+c_{23} \left(\frac{1}{15} \left(\pi^4-90\right) y+\bar y L_{\bar y}^3\right)\Big]\,,
\label{eq:Gfit}
\end{align}
with
\begin{equation}
c_{11} =-41.9008,\; c_{12} =1.0,\; c_{21} =-20.2744,\; c_{22} =-10.5870,\; c_{23} =-7.1277
\,.
\label{eq:cij}
\end{equation}
With the values of the fitting parameters $c_{ij}$ quoted in \eq{cij} the fit function $G_\mathrm{fit}$ approximates the corresponding numerical results obtained in \sec{topob} with a relative precision of better than $0.5\%$. 
Also accounting for the intrinsic uncertainty of the numerical results we reach an accuracy of at least $1\%$ for $G_\mathrm{fit}$.
Besides that we note that the contributions parametrized in the fit function account for less then $0.5\%$ of the $\mathcal O(\alpha_s^2)$ corrections for all physical values of $y=m^2/p^2$. So for all practical purposes the uncertainties associated to $G_\mathrm{fit}$ are negligible, and we did not quote uncertainties in \eq{cij}.

The $\mathcal O(\alpha_s^2)$ contributions arising from Feynman diagrams containing the $Q\bar Q$ vacuum polarization subdiagrams  (see diagrams  (o)-(r) in \fig{diagrams1}  with only massive quark bubbles) read
\begin{align}
J_{Q,\mathrm{sec}}^{(2),\mathrm{mf}}&(p^2,m^2,\mu,Q/\nu) ={} C_F T_F\Bigg\{\Bigg[2 L_m^2+\bigg ( \frac{2}{3}+ \frac{8 \pi^2}{9}  \bigg) L_m +\frac{3139}{162} -\frac{4 \pi^2 }{3} + \frac{8 \zeta_3}{3}\notag\\
&- \bigg (  \frac{8}{3}  L_m^2+ \frac{80}{9}  L_m+\frac{224}{27} \bigg) \log \bigg ( \frac{Q}{\nu}\bigg)\Bigg]\delta(s)+G_{\mathrm{sec}}^{(3Q)}\Theta\big(p^2-(3m)^2\big)\Bigg\}\,,
\label{eq:Jmf2sec}
\end{align}
where again the distributional contributions are displayed explicitly. The dependence on the rapidity renormalization scale $\nu$ is displayed explicitly in terms of the logarithm $\log(Q/\nu)$. In the $(n_\ell+1)$ flavor scheme for the strong coupling the coefficient $J_{Q,\mathrm{sec}}^{(2),\mathrm{mf}}$ is the only one that is modified: $J_{Q,\mathrm{sec}}^{(2),\mathrm{mf}}\to J_{Q,\mathrm{sec}}^{(2),\mathrm{mf}}+4/3 L_m J_Q^{(1)}$.

The non-distributional contributions arise entirely from the three particle $QQ\bar Q$ cut and its result reads
\begin{align}
&G_{\mathrm{sec}}^{(3Q)} = \frac{1}{s}\Bigg[-\frac{8 \left(252 y^{9/2}+570 y^{7/2}+822y^{5/2}+118 y^{3/2}+45 y^5+357 y^4+720 y^3+524 y^2\right)}{27 \sqrt{1 + 3 \sqrt{y}} \left(1 - \sqrt{y}\right)^{5/2}\left(\sqrt{y}+1\right)^3}\nn\\
&\quad\times\frac{8 \left(y+10 \sqrt{y}+5\right)}{9 \sqrt{1 + 3 \sqrt{y}} \left(1 - \sqrt{y}\right)^{5/2}\left(\sqrt{y}+1\right)^3}\,\mathrm{K}\left(\frac{\left(1-3 \sqrt{y}\right)\left(1+\sqrt{y}\right)^3}{\left(1+3 \sqrt{y}\right)\left(1-\sqrt{y}\right)^3}\right)\nonumber\\
&+\frac{2 \sqrt{\left(1-\sqrt{y}\right)^3 \left(1 + 3 \sqrt{y}\right)} \left(81 y^4+54 y^3-24 y^2-54 y+71\right)}{27 \bar y^3}\,\mathrm{E}\left(\frac{\left(1-3 \sqrt{y}\right)\left(1+\sqrt{y}\right)^3}{\left(1+3 \sqrt{y}\right)\left(1-\sqrt{y}\right)^3}\right)\nonumber\\
&+\frac{32 y^{1/4} \left(15 y^3+27 y^2+25 y+5\right)}{9 \sqrt{\left(1-\sqrt{y}\right)^3 \left(3 \sqrt{y}+1\right)} \bar y}\,\Pi \left(\frac{\left(1-3 \sqrt{y}\right)\left(1+\sqrt{y}\right)}{\left(1-\sqrt{y}\right)^2};\frac{\left(1-3 \sqrt{y}\right)\left(1+\sqrt{y}\right)^3}{\left(1+3 \sqrt{y}\right)\left(1-\sqrt{y}\right)^3}\right)\notag\\
&+\frac{16}{3}\hat I(y)\Bigg]
= \frac{11\pi (p^2-(3m)^2)^2}{5184 \sqrt{3}m^6}+\mathcal O\!\left(\frac{(p^2-(3m)^2)^3}{m^8}\right).
\label{eq:G3QTF}
\end{align}
The non-distributional three particle $QQ\bar Q$ cut contributions in $J_Q^{(2)}$ and $J_{Q,\mathrm{sec}}^{(2)}$ involve the elliptic functions
\begin{align}
\mathrm K(m)\equiv{}&\int_0^\frac{\pi}{2}\dd t\,\frac{1}{\sqrt{1-m \sin^2 t}}\,,\\
\mathrm E(m)\equiv{}&\int_0^\frac{\pi}{2}\dd t\,\sqrt{1-m \sin^2 t}\,,\\
\Pi(n;m)\equiv{}&\int_0^\frac{\pi}{2}\dd t\,\frac{1}{(1-n \sin^2 t)\sqrt{1-m \sin^2 t}}\,,
\end{align}
and the integral functions
\begin{align}
I_3(y)={}&\int_9^{1/y}\dd t\,\frac{2\,(t-9)}{(t-1) \sqrt{(\sqrt{t} + 3)(\sqrt{t} - 1)^3}}\,\mathrm K\left(\frac{\left(\sqrt{t}-3\right) \left(\sqrt{t}+1\right)^3}{(\sqrt{t} + 3)(\sqrt{t} - 1)^3}\right),\displaybreak[1]\\
\tilde I(y)={}&\int_{y}^{\frac{1}{4}(1-\sqrt{y})^2}\mathrm dz\,\frac{\sqrt{z-y} \, \log\biggl(\frac{y-4 z-1-\sqrt{(y+4 z-1)^2-16 y z}}{y-4z-1+\sqrt{(y+4 z-1)^2-16 y z}}\biggr)}{z^{3/2}}\,,
\label{eq:Itilde}
\end{align}
which can be readily evaluated numerically. For related discussions see e.g.\ \rcite{Ablinger:2017bjx}. In practical applications it is, however, useful to have an approximate representation of $G^{(3Q)}$ and $G_{\mathrm{sec}}^{(3Q)}$ involving only elementary functions. 
Such approximate representations are provided by 
\begin{align}
G^{(3Q)}\approx{}&\frac{1}{m^2}\Bigg\{c_1^{C_F(3Q)} \left[81 y^2 \left(9 y+\frac{1}{4}\right) \log (9 y)-\frac{81}{8} y^2 (9 y-1) (27 y+7)\right]\notag\\
&+c_2^{C_F(3Q)} \left[81 y^2 \left(81y^2-\frac{1}{3}\right) \log (9 y)-27 y^2 (9 y-1) (45 y-3)\right]\notag\\
&-\frac{1}{256} y (9 y-1) \bigg[-243 y^2 \Big(-64 \left(176 \zeta_3-209+8 \log^2 3-8 \log 3\right)+9 \sqrt{3} \pi\notag\\
&\qquad +128 \pi^2 (16 \log 2-11)\Big)+9 y \Big(-64 \left(1584 \zeta_3-1869+72 \log^2 3-88 \log 3\right)\notag\\
&\qquad+45\sqrt{3} \pi +1152 \pi^2 (16 \log 2-11)\Big)+256 (4\log 3-7)\bigg]\notag\\
&+9 y \log (9 y) \Bigg[9 y \bigg(\frac{1}{18} \left(-528 \zeta_3+611-24 \log^2 3+40 \log 3\right)+\frac{\sqrt{3} \pi }{128}\notag\\
&\quad+\frac{1}{3} \pi^2 (16 \log 2-11)\bigg)-\frac{2}{9}\Bigg]\Bigg\}\label{eq:G3QCFapprox} \\
={}& \frac{9y}{m^2}\big[-1.4175 (1-9y) + 1.6758 (1-9y)^2 - 0.5478 (1-9y)^3 - 1.4175 \log(9y)\notag\\
& + 2.3420 (1-9y) \log(9y) - 1.3552 (1-9y)^2 \log(9y) + 0.2085 (1-9y)^3 \log(9y)\big]\,, \nn
\end{align}
\begin{align}
G_{\mathrm{sec}}^{(3Q)}\approx{}&\frac{1}{m^2}\Bigg\{c^{T_F(3Q)} \left[81 y^2 \left(9 y+\frac{1}{4}\right) \log (9 y)-\frac{81}{8} y^2 (9 y-1) (27 y+7)\right]\notag\\
&+\frac{y (9 y-1)}{3456}\bigg[-64 \Big(81 y^2 \left(432 \zeta_3-2053+576 \log^3 3+72 \log^2 3+816 \log 3\right)\notag\\
&\quad+9 y \left(-1296 \zeta_3+6491-1728 \log^3 3+360 \log^2 3-2688 \log 3\right)\notag\\
&\qquad+4\left(359+144 \log^2 3-348 \log 3\right)\Big)+8019 \sqrt{3} \pi  y (27y-5)\notag\\
&\quad+384 \pi^2 \Big(243 y^2 (8 \log 3-9)+9 y (89-72 \log 3)+8\Big)\bigg]\notag\\
&+9 y \log (9 y) \Bigg[9 y \Big(\frac{1}{81} \left(-144 \zeta_3+863-192 \log^3 3+168 \log^2 3-448 \log 3\right)\notag\\
&\quad+\frac{11 \pi}{64 \sqrt{3}}+\frac{2}{81} \pi^2 (24 \log 3-35)\Big)-\frac{4}{81} (24 \log 3-29)\Bigg]+\frac{8}{3} y \log^2(9 y)\Bigg\}
\label{eq:G3QTFapprox}\displaybreak[1] \\
={}&\frac{9y}{m^2}\big[-0.2586 (1-9y) + 0.5931 (1-9y)^2 - 0.0707 (1-9y)^3 - 0.2586 \log(9y)\notag\\
& + 0.7070 (1-9y) \log(9y) - 0.3183 (1-9y)^2 \log(9y) + 0.2963 \log^2(9y)\big]\,, \nn
\end{align}
with $c_1^{C_F(3Q)}=-0.7298$, $c_2^{C_F(3Q)}=-0.2085$, $c^{T_F(3Q)}=-0.3183$.
They both approximate the exact expressions to better than $0.5\%$ over the physical kinematic range and provide exact analytic results in the limit $p^2\to 9m^2$ and $m\to 0$. We also quoted a fully numerical version of the expressions useful for practical applications.

In analogy to the non-distributional terms at $\mathcal O(\alpha_s)$ the non-distributional $G$-functions at $\mathcal O(\alpha_s^2)$ do not contribute to the singular contributions in the bHQET limit $s=p^2-m^2\ll m^2$. They represent power corrections in this kinematic region and constitute the $\mathcal O(\alpha_s^2)$ bHQET non-singular corrections. For completeness we also quoted the respective results in the limit $p^2\to m^2$ (\eqss{G1QCF}{G1QTF}) and 
in the limit $p^2\to 9m^2$ (\eqs{G3QCF}{G3QTF}).

For completeness we also provide the result for the renormalized mass mode jet function $J_q^{\mathrm{mf},(n_\ell+1)}(p^2,m^2,\mu,Q/\nu)$ for massless primary quarks in \eq{Jmfmassless}, which can be extracted from the calculations in \rcite{Pietrulewicz:2014qza}, but was not given in the literature before. The $\mu$ (virtuality) and $\nu$ (rapidity) anomalous dimensions 
of the mass mode jet functions for massive and massless primary quarks 
agree identically.

\subsection{Universal Jet Function}

The result for the renormalized primary massive universal jet function $J_Q^{\mathrm{uf},(n_\ell+1)}$, defined in \eq{jmufdef}, in the pole mass scheme at $\mathcal O(\alpha_s^2)$ can be written in the form
\begin{align}
J_Q^{\mathrm{uf},(n_\ell+1)}(p^2,m^2,\mu) ={}& \delta(p^2-m^2)+a_s^{(n_\ell+1)} J_Q^{(1)}(p^2,m^2,\mu)\label{eq:Jufdef}\\
& + \left(a_s^{(n_\ell+1)}\right)^2 \left(J_Q^{(2)}(p^2,m^2,\mu)+J_{Q,\mathrm{sec}}^{(2),\mathrm{uf}}(p^2,m^2,\mu)\right) + \mathcal O(\alpha_s^3)\,, \nn
\end{align}
where we adopt the $(n_\ell+1)$ flavor scheme for the strong coupling, which is the natural scheme choice for the universal jet function. The universal jet function does not contain rapidity divergences and thus only depends on the UV renormalization scale $\mu$ coming from dimensional regularization. The $\mathcal O(\alpha_s)$ coefficient $J_Q^{(1)}$ and the $\mathcal O(\alpha_s^2)$ contribution from diagrams containing only one single quark $Q$ line $J_Q^{(2)}$  (see all diagrams in \fig{diagrams1} with the diagrams (o)-(r) containing only massless quark bubbles)  are identical to the mass mode jet function and given in \eqs{Jmf1}{Jmf2}, respectively.

The $\mathcal O(\alpha_s^2)$ contributions arising from Feynman diagrams containing the $Q\bar Q$ vacuum polarization subdiagrams (see diagrams  (o)-(r) in \fig{diagrams1} containing only massive quark bubbles) read
\begin{align}
J_{Q,\mathrm{sec}}^{(2),\mathrm{uf}}(p^2,m^2,\mu) ={}& C_F T_F\Bigg[\left(\frac{32}{9}L_m^3+\frac{70}{9} L_m^2+\frac{754}{27} L_m-\frac{32 \zeta_3}{9}-\frac{46\pi^2}{27}+\frac{7075}{162}\right) \delta(s)\notag\\
&+\left(-8 L_m^2-\frac{128}{9}L_m-\frac{224}{27}\right) \mathcal L_0(s)+\frac{32}{3} L_m \mathcal L_1(s)\notag\\
& + \frac{4}{3} L_m G_1 \Theta(p^2-m^2) + G_{\mathrm{sec}}^{(3Q)}\Theta(p^2-(3m)^2)\Bigg]\,,
\label{eq:Juf2sec}
\end{align}
where again the distributional contributions are displayed explicitly. The non-distributional contributions come from the single heavy quark $Q$ cut and from the three particle $QQ\bar Q$ cut. The $Q$ cut term is given in \eq{G1def} and arises from using the $(n_\ell+1)$ flavor scheme for the strong coupling. The $QQ\bar Q$ cut term $G_{\mathrm{sec}}^{(3Q)}$ is given in  \eq{G3QTF}.

The non-distributional $G$-functions represent power corrections in the bHQET limit $s=p^2-m^2\ll m^2$ and represent $\mathcal O(\alpha_s^2)$ bHQET non-singular corrections. 
For completeness we also quoted the respective results in the limit $p^2\to m^2$ (\eqss{G1QCF}{G1QTF}) and 
in the limit $p^2\to 9m^2$ (\eqs{G3QCF}{G3QTF}).

For completeness we also quote the result for the renormalized universal jet function $J_q^{\mathrm{uf},(n_\ell+1)}(p^2,m^2,\mu)$ for primary massless quarks in \eq{Jufmassless}, which was computed in \rcite{Pietrulewicz:2014qza}.
The renormalization $Z$-factor as well as the $\mu$ (virtuality) anomalous dimension of 
of the universal jet functions (to all orders as well as for primary massive and massless quarks) agree with those of the well known massless quark jet functions~\cite{Becher:2006qw}.

\subsection{Consistency Checks and Kinematic Limits}
\label{sec:ConsKine}

There are a number of essential consistency properties the $\mathcal O(\alpha_s^2)$ primary massive quark jet functions have to satisfy and which provide important consistency checks. These concern the relation between the mass mode and universal jet functions, the limit of massless quarks and the bHQET limit of a supermassive quark.  We discuss them in the following. 

As explained in \sec{notation}, the difference between the mass mode and the universal jet functions is related to the collinear-soft function $S_c$ \cite{Pietrulewicz:2017gxc} via \eq{Scdef}. Using the results of \eqs{Jmfdef}{Jufdef}
and accounting for the result for the renormalized collinear-soft function determined in \rcite{Pietrulewicz:2017gxc}, which in our notation has the form
\begin{align}
S_c(\ell,m,\mu,\nu) ={}& \delta(\ell) + \left(a_s^{(n_\ell)}\right)^2 C_F T_F \Bigg\{\Bigg[\left(\frac{8}{3}L_m^2+\frac{80}{9}L_m+\frac{224}{27}\right) \log\left(\frac{\nu}{\mu}\right)-\frac{8}{9}L_m^3-\frac{40}{9}L_m^2\notag\\
&\qquad+\frac{4}{27} \left(3 \pi^2-112\right) L_m+\frac{2}{27} \left(84 \zeta_3-328+5 \pi^2\right)\Bigg]\delta(\ell)\notag\\
&\quad + \left[\frac{8}{3}L_m^2+\frac{80}{9}L_m+\frac{224}{27}\right] \mathcal L_0(\ell)\Bigg\}+\mathcal O(\alpha_s^3)\,,
\label{eq:Sc}
\end{align}
it is straightforward to check the validity of \eq{Scdef}.
Here one has to account for the decoupling relation
\begin{align}
\alpha_s^{(n_\ell+1)}(\mu) ={}& \alpha_s^{(n_\ell)}(\mu)\Bigg\{1 - \frac{\alpha_s^{(n_\ell)}(\mu)}{4\pi} T_F\Bigg[\frac{4}{3}L_m+\varepsilon\left(-\frac{2}{3}L_m^2-\frac{\pi^2}{9}\right)\notag\\
&\quad + \varepsilon^2 \left(\frac{2}{9}L_m^3+\frac{\pi^2}{9}L_m+\frac{4 \zeta_3}{9}\right) + \ord{\varepsilon^3}\Bigg]\Bigg\} + \mathcal O(\alpha_s^3)\,,
\label{eq:alphamatch}
\end{align}
for $\alpha_s$ in the $n_\ell$ and $(n_\ell+1)$ flavor schemes
(where we also displayed the $\mathcal O(\varepsilon)$ and $\mathcal O(\varepsilon^2)$ terms useful for divergent expressions). 
The sole dependence of the collinear-soft function on the rapidity scale $\nu$ indicates that the natural choice for $\nu$ in general depends on the process and the structure of the factorization theorem.

The definition of the universal jet function $J_f^{\mathrm{uf},(n_\ell+1)}$ given in \eq{jmufdef} entails that for $m\to 0$ it converges to the jet function $J_f^{(n_\ell+1)}$ for $(n_\ell+1)$ massless quarks, see \eq{jm0def}, which at $\mathcal O(\alpha_s^2)$ was computed in \rcite{Becher:2006qw}. 
In the limit $m\to 0$, the non-distributional $G$-functions also yield distributions and the corresponding limiting expressions are given in \app{Gmto0}. Accounting for these results our expression for universal jet function correctly approaches the massless quark two-loop jet function for $m\to 0$.

Finally, we discuss the bHQET limit $s=p^2-m^2\ll m^2$, the jet mass threshold region. Taking the double differential hemisphere jet mass distribution in $e^+ e^-$ annihilation (w.r.t.\ the thrust axis) as an example, one can consider the kinematic region where both jet masses are close to threshold. The corresponding factorization theorem for boosted top quarks ($Q\gg m$) has the form \cite{Fleming:2007qr,Hoang:2015vua}\footnote{Note that throughout this paper we do not account for renormalization group functions to sum (virtuality or rapidity) logarithms in our discussions of factorization theorems in order to keep the notations brief. 
}
\begin{align}
\frac{1}{\sigma_0}\frac{\dd \sigma}{\dd M^2_Q\,\dd M^2_{\bar Q}} ={}& H_Q^{(n_\ell+1)}(Q,\mu) H_m\left(m,\frac{Q}{m},\mu\right)\int\dd\ell^+\,\dd\ell^-\, S^{(n_\ell)}(\ell^+,\ell^-,\mu)\notag\\
&\times J_B^{(n_\ell)}\left(\frac{M^2_Q-Q\ell^+}{m}-m,\mu\right) J_B^{(n_\ell)}\left(\frac{M^2_{\bar Q}-Q\ell^-}{m}-m,\mu\right)\notag\\
&\times\left[1+\mathcal O\!\left(\frac{m\alpha_s}{Q}\right)+\mathcal O\!\left(\frac{m^2}{Q^2}\right)+\mathcal O\!\left(\frac{s_{Q,\bar Q}^2}{m^4}\right)\right],
\label{eq:facttheorembHQET}
\end{align}
where $M_{Q,\bar Q}$ stand for the hemisphere masses, $\sigma_0$  denotes the tree level cross section for $e^+ e^- \to Q \bar Q$. The term $H_Q^{(n_\ell+1)}$ is the hard function related to the matching of the QCD $Q\bar Q$ current to the SCET massive quark dijet current at the scale $Q$ and the term $H_m$ is the mass threshold hard function related to the matching of the SCET dijet current to the bHQET current at the scale $m$. The terms $J_B^{(n_\ell)}$ and $S^{(n_\ell)}$ denote the bHQET jet function~\cite{Fleming:2007qr,Jain:2008gb} and dijet soft function, respectively. The soft function $S^{(n_\ell)}$ describes ultrasoft cross talk between the two jets at the scale $s_Q/Q$. The bHQET jet  function $J_B^{(n_\ell)}$~\cite{Fleming:2007qr,Jain:2008gb} describes the effects of the ultra-collinear radiation inside the jets (i.e.\ soft radiation in the rest frame of the massive quarks) at the scales $s_{Q,\bar Q}/m$ and contain all the dynamical (ultra) collinear effects for $s_{Q,\bar Q}\ll m^2$. The mass threshold hard function $H_m(m,Q/m,\mu)$ consists of $n$-collinear, $\bar n$-collinear and soft mass mode contributions which can be written in factorized form as~\cite{Hoang:2015vua}
\begin{equation}
\label{eq:Hmfactors}
H_m\left(m,\frac{Q}{m},\mu\right) = H_{m,n}\left(m,\mu,\frac{Q}{\nu}\right) H_{m,\bar n}\left(m,\mu,\frac{Q}{\nu}\right) H_{m,s}\left(m,\mu,\frac{m}{\nu}\right).
\end{equation}
Each of the collinear ($n,\bar n$) and soft ($s$) mass mode factors depends on the rapidity renormalization scale $\nu$. The natural scaling for the collinear factors is $\nu\sim Q$ and for the soft factor it is $\nu\sim m$. They were calculated at $\mathcal O(\alpha_s^2)$ in \rcite{Hoang:2015vua} (see eqs.~(5.7) and (5.8) in \rcite{Hoang:2015vua} and note that $H_{m,n}=H_{m,\bar n}$ for the symmetric regulator of \rcite{Chiu:2012ir}, see \eq{XVdef}).

Up to power-suppressed (for $s_{Q,\bar Q}\ll m^2$) and non-distributional contributions the factorization theorem in \eq{facttheorembHQET} is also valid in the kinematic region $s_{Q,\bar Q}=p_{Q,\bar Q}^2-m^2\sim m^2\lesssim  p_{Q,\bar Q}^2$. Interestingly, in \rcite{Pietrulewicz:2017gxc} it was pointed out in the context of invariant mass dependent beam functions for exclusive Drell-Yan gauge boson production, that for this kinematic region the collinear-soft function $S_c$ describes dynamical fluctuations located in the same phase space region as those contained in the jet function. The most economical formulation (w.r.t.\ the number of factorization functions containing the dynamical effects of the collinear and soft mass mode contributions) for $s_{Q,\bar Q}=p_{Q,\bar Q}^2-m^2\sim m^2\sim p^2$ is therefore the one where the contributions of the collinear-soft function are \textit{included} in the definition of the (invariant mass dependent) beam function, which corresponds to the mass mode definition of \eq{jmmfdef}.
For the double differential hemisphere jet function this implies that for $s_{Q,\bar Q}=p_{Q,\bar Q}^2-m^2\sim m^2\sim p_{Q,\bar Q}^2$ one can formulate a factorization theorem of the form
\begin{align}
\frac{1}{\sigma_0}\frac{\dd^2 \sigma}{\dd M^2_Q\,\dd M^2_{\bar Q}} ={}& H_Q^{(n_\ell+1)}(Q,\mu) H_{m,s}\left(m,\mu,\frac{m}{\nu}\right)\int\dd\ell^+\,\dd\ell^-\, S^{(n_\ell)}(\ell^+,\ell^-,\mu)\notag\\
&\times J_Q^{\mathrm{mf},(n_\ell+1)}\left(M^2_Q-Q\ell^+,m^2,\mu,\frac{Q}{\nu}\right) J_Q^{\mathrm{mf},(n_\ell+1)}\left(M^2_{\bar Q}-Q\ell^-,m^2,\mu,\frac{Q}{\nu}\right)\notag\\
&\times\left[1+\mathcal O\left(\frac{m\alpha_s}{Q}\right)+\mathcal O\left(\frac{m^2}{Q^2}\right)
\right].
\end{align}
This in turn implies the relation
\begin{align}
J_Q^{\mathrm{mf},(n_\ell+1)}\left(p^2\to m^2,m^2,\mu,\frac{Q}{\nu}\right) = H_{m,n}\left(m,\mu,\frac{Q}{\nu}\right)J_B^{(n_\ell)}\left(\frac{p^2}{m}-m,\mu\right)
 +  \mbox{non-sing.\ terms},
\label{eq:JBJmf}
\end{align}
between the primary massive quark mass mode jet function and the bHQET jet function up to non-singular terms integrable in $p^2$ at the threshold $m^2$. Using the $\mathcal O(\alpha_s^2)$ results for the bHQET jet function (naturally given in the $n_\ell$-flavor scheme for $\alpha_s$) in \rcite{Jain:2008gb} (see eq.~(39)), the collinear mass mode hard functions $H_{m,n}$ in \rcite{Hoang:2015vua} (see eq.~(5.7)) and our result for the mass mode jet function in \eq{jmmfdef}, accounting for the fact that \textit{all} $G$-functions are power-suppressed in the bHQET limit, it is straightforward to see that the relation in \eq{JBJmf} is indeed satisfied. 
As discussed in more detail in \sec{topob}, our calculations of the ${\cal O}(\alpha_s^2)$ corrections for the mass mode jet function provide a high precision semi-analytic check of \eq{JBJmf} as some loop integrals could only be determined by numerical methods. The relation~(\ref{eq:JBJmf}) was then used as an input to analytically parametrize the jet functions result of \eqs{jmmfdef}{jmufdef}.

\section{Alternative Factorization Approaches and Practical Use}
\label{sec:consistency}

The non-trivial conceptual aspect of the $\mathcal O(\alpha_s^2)$ primary massive quark SCET jet function concerns the mass mode contribution due to secondary radiation. Since these are tied to corrections arising from collinear as well as soft phase space sectors with invariant masses of order the quark mass $m$, their treatment entails rapidity singularities (and the resummation of associated logarithms) within a $\mathrm{SCET_{II}}$ type approach regardless of whether the observable is $\mathrm{SCET_{I}}$ or $\mathrm{SCET_{II}}$. As already pointed out in \sec{notation}, in the literature two alternative factorization formulations to account for mass mode corrections have been advocated which are (or can be rendered) conceptually and phenomenologically equivalent due to consistency relations. We emphasize that none of these approaches may be considered superior to the respective other and, by emphasizing different and complementary aspects of the mass mode effects, together provide a thorough and comprehensive treatment and understanding from the conceptual as well as practical point of view.

In this section we discuss the relation between the approaches employing the mass mode and the universal jet functions for primary massive quarks and using the double differential hemisphere mass distribution already introduced in \sec{ConsKine} as an example. To keep the discussion brief we also restrict the discussion to the kinematic regions $s^2/Q^2\ll m^2\ll s\sim p^2\ll Q^2$  and $s=p^2-m^2\sim p^2\sim m^2$ where the quark mass effects calculated in this paper are most relevant. We emphasize, however, that the discussion can be extended in a straightforward way to all kinematic situations and also to other types of factorization theorems where mass mode effects shall be treated (and where factorization functions other than jet functions arise).

The universal factorization approach of \rcites{Gritschacher:2013pha,Gritschacher:2013tza,Pietrulewicz:2014qza,Hoang:2015iva} was devised with the motivation to obtain universal hard, jet and soft functions defined such that they can smoothly cover all possible choices for the quark mass $m$ in the context of the strict hierarchy $s^2/Q^2\ll s\ll Q^2$. The resulting factorization theorems therefore contain power corrections w.r.t.\ mass mode contributions in at least some of the factors in any kinematic region, which, in case they are known, may be expanded away accordingly if a smooth description of mass effects is not wanted. The approach was also constructed to allow for the formulation of universal and simple rules to implement the summation of logarithms through flavor number dependent renormalization evolution factors of the hard, jet and soft function for arbitrary choices of the global renormalization scale $\mu$. The approach also entails that only the hard, jet and soft functions and, depending on the choice of $\mu$, their respective threshold matching factors appear in the factorization theorems (in case they are evolved through the massive quark flavor threshold). By construction in the universal factorization approach, all rapidity logarithms (and their summations) {\it are fully contained within} these individual quark mass threshold matching factors at the scale $\mu_m\sim m$.

Let us consider the situation $s^2/Q^2\ll m^2\lesssim s\sim p^2\ll Q^2$. The corresponding leading power factorization theorem in the universal factorization function approach for $s^2/Q^2<\mu^2<m^2$ (when the SCET current and the jet functions are evolved below $m$) has the form
\begin{align}
\label{eq:uffacth1}
\frac{1}{\sigma_0}\frac{\dd \sigma}{\dd M^2_Q\,\dd M^2_{\bar Q}} ={}& H_Q^{(n_\ell+1)}(Q,\mu) H_m\left(m,\frac{Q}{m},\mu\right) \int\dd s_Q\, \dd s_{\bar Q}\, \int\dd\ell^+\,\dd\ell^-\notag\\
&\times J_Q^{\mathrm{uf},(n_\ell+1)}(M^2_Q-s_Q-Q\ell^+,m^2,\mu)\, J_Q^{\mathrm{uf},(n_\ell+1)}(M^2_{\bar Q}-s_{\bar Q}-Q\ell^-,m^2,\mu)\notag\\
&\times \mathcal M_{J_Q}(s_Q,m,\mu) \mathcal M_{J_Q}(s_{\bar Q},m,\mu) S^{(n_\ell)}(\ell^+,\ell^-,\mu)\,,
\end{align}
where $H_m$ and $\mathcal M_{J_Q}$ are the quark mass threshold matching factors for the SCET current and the universal jet function $J_Q^\mathrm{uf}$ respectively. For $m^2<\mu^2<s$ (where the soft function is evolved above $m$) the \textit{same} factorization theorem has the form
\begin{align}
\label{eq:uffacth2}
\frac{1}{\sigma_0}\frac{\dd \sigma}{\dd M^2_Q\,\dd M^2_{\bar Q}} ={}& H_Q^{(n_\ell+1)}(Q,\mu) \int\dd\ell^+\,\dd\ell^-\dd\ell'^+\,\dd\ell'^-\notag\\
&\times J_Q^{\mathrm{uf},(n_\ell+1)}(M^2_Q-Q\ell^+,m^2,\mu) J_Q^{\mathrm{uf},(n_\ell+1)}(M^2_{\bar Q}-Q\ell^-,m^2,\mu)\notag\\
&\times \mathcal M_{S}(\ell'^+,\ell'^-,m,\mu) S^{(n_\ell)}(\ell^+-\ell'^+,\ell^--\ell'^-,\mu)\,,
\end{align}
where $\mathcal M_S$ is the mass threshold matching factor for the soft function $S$.
It is straightforward to consider other choices of $\mu$ as well. 

For a strong hierarchy $m^2\ll s$ one may replace the universal primary massive quark jet function $J_Q^{\mathrm{uf},(n_\ell+1)}$ by the jet function $J_q^{(n_\ell+1)}$ for $(n_\ell+1)$ massless quark flavors (calculated at $\mathcal O(\alpha_s^2)$ in \cite{Becher:2006qw}). But using $J_Q^{\mathrm{uf},(n_\ell+1)}$ provides validity of \eqs{uffacth1}{uffacth2} smoothly covering \textit{all} cases between $m^2\ll s$ and $m^2\sim s$.
The resummation of virtuality logarithms (with respect to $\mu$) is achieved by the flavor number dependent $\mu$-anomalous dimensions of the hard currents and jet functions known from SCET (with $(n_\ell+1)$ flavors) and bHQET (with $n_\ell$ flavors) and of the soft function (with $(n_\ell+1)$ flavors for scales above $m$ and $n_\ell$ flavors for scales below $m$). Furthermore one has to account for the natural scales of the hard, jet and soft function being $\mu_Q\sim Q$, $\mu_J\sim s^{1/2}$ and $\mu_S\sim s/Q$, respectively. In \eq{uffacth1} the mass threshold matching functions $H_m$ and $\mathcal M_J$ account for the change between $(n_\ell+1)$-flavor SCET evolution (above the mass threshold) and $n_\ell$-flavor bHQET evolution (below the mass threshold) for the hard current (see \rcite{Hoang:2015vua}) and universal jet function, respectively, at the mass threshold scale $\mu_m\sim m$. For the universal jet function for a massive primary quark this matching relation reads
\begin{align}
J_B^{(n_\ell)}\left(\frac{p^2}{m}-m,\mu\right)  = &\int\dd s\, \mathcal M_{J_Q}(p^2-s,m,\mu)\, J_Q^{\mathrm{uf},(n_\ell+1)}\left(s,m^2,\mu\right)\\\nonumber &
+  \mbox{non-singular\ terms for $p^2\to m^2$}\,.
\end{align}
(The analogous relation for the massless primary quark case has been given in eq.~(119) of \rcite{Pietrulewicz:2014qza}.)
In \eq{uffacth2} 
the mass threshold matching function $\mathcal M_S$ accounts for the corresponding change between the corresponding $n_\ell$- and $(n_\ell+1)$-flavor evolution of the soft function (see eqs.~(136) and (139) of \rcite{Pietrulewicz:2014qza}). Rapidity logarithms and their summation take place only inside each of the mass threshold matching functions $H_m$, $\mathcal M_{J_Q}$ and $\mathcal M_S$ (see for example \eq{Hmfactors}), and the jet and soft functions are free of any rapidity logarithms.
The $\nu$-anomalous dimensions are the same for primary massive and primary massless quarks and were provided in \rcite{Hoang:2015iva,Hoang:2015vua}.
Finally, consistency between \eqs{uffacth1}{uffacth2} implies the consistency identity
\begin{equation}
\label{eq:consist1}
H_m\left(m,\frac{Q}{m},\mu\right) \mathcal M_J(Q \ell^+,m,\mu) \mathcal M_J(Q\ell^-,m,\mu) = \mathcal M_S(\ell^+,\ell^-,m,\mu)\,.
\end{equation}

The mass mode factorization approach of \rcite{Pietrulewicz:2017gxc} was devised with the motivation to formulate distinct and unique factorization theorems for the different possible scale hierarchies of the mass $m$ with respect to $Q$, $s^{1/2}$ and $s/Q$, where the physically relevant $n$- and $\bar n$-collinear and soft mass mode effects appear explicitly within three distinct factors to achieve a transparent and economic representation of their contributions. In contrast to the universal factorization approach, the mass mode factorization approach does not result in 
two options to formulate the same factorization theorem, such as \eqs{uffacth1}{uffacth2},
which are equivalent due to consistency relations such as \eq{consist1}. The factorization theorems in the mass mode approach contain functions that have explicit $\mu$ and $\nu$ dependence, and the summation of virtuality and rapidity logarithms is carried out on equal footing. The summation of the associated logarithms, on the other hand (and in contrast to the universal factorization approach), does explicitly involve consistency relations among the anomalous dimensions to allow for a transparent treatment of renormalization group evolution with either $n_\ell$ or $(n_\ell+1)$ dynamical flavors with respect to the mass threshold scale $m$. 
In addition, in the formulation of the mass mode factorization approach of \rcite{Pietrulewicz:2017gxc} the resulting factorization theorems have been free of any power corrections in the quark mass $m$
and did not provide a smooth dependence on the quark mass $m$. The latter is mandatory for smooth predictions of differential cross sections when the quark mass threshold is crossed. 

Let us consider the situation $s^2/Q^2\ll m^2\ll s\sim p^2\ll Q^2$. The corresponding factorization theorem in the mass mode factorization approach has the form
\begin{align}
\label{eq:mffacth1}
\frac{1}{\sigma_0}\frac{\dd \sigma}{\dd M^2_Q\,\dd M^2_{\bar Q}} ={}& H_Q^{(n_\ell+1)}(Q,\mu) H_{m,s}\left(m,\mu,\frac{m}{\nu}\right) \int\dd\ell^+\,\dd\ell^-\dd\ell'^+\,\dd\ell'^-\notag\\
&\times J_q^{(n_\ell+1)}(M^2_Q-Q\ell^+-Q\ell'^+,\mu)\, J_q^{(n_\ell+1)}(M^2_{\bar Q}-Q\ell^--Q\ell'^-,\mu)\notag\\
&\times S_c(\ell'^+,m,\mu,\nu) S_c(\ell'^-,m,\mu,\nu) S^{(n_\ell)}(\ell^+,\ell^-,\mu)\,,
\end{align}
where $H_{m,s}$ is the soft mass mode contribution of the mass threshold hard function $H_m$ quoted in \eq{Hmfactors}, $S_c$ is the collinear-soft function of \eqs{Scdef}{Sc}, and $J_q^{(n_\ell+1)}$ is the jet function for $(n_\ell+1)$ {\it massless} quarks which coincides with the universal jet function for vanishing quark mass, i.e.\ $J_Q^{\mathrm{uf},(n_\ell+1)}(p^2,0,\mu)$. Since the factorization theorems \eqref{eq:mffacth1} and \eqref{eq:uffacth1} as well as \eqref{eq:uffacth2} provide identical descriptions for the situations $s^2/Q^2\ll m^2\ll s\sim p^2\ll Q^2$ (up to suppressed quark mass power corrections), they imply the following consistency relations for the mass threshold matching factors $\mathcal M_{J_Q}$ for the universal function $J_Q^{\mathrm{uf},(n_\ell+1)}$ and $\mathcal M_S$ for the soft function $S^{(n_\ell)}$:
\begin{align}
\mathcal M_{J_Q}(s,m,\mu) ={}& \frac{1}{Q}H_{m,n}^{-1}\left(m,\mu,\frac{Q}{\nu}\right)S_c\left(\frac{s}{Q},m,\mu,\nu\right),\label{eq:MjHmSc}\\
\mathcal M_S(\ell^+,\ell^-,m,\mu) ={}&  H_{m,s}\left(m,\mu,\frac{m}{\nu}\right) S_c(\ell'^+,m,\mu,\nu) S_c(\ell'^-,m,\mu,\nu)\label{eq:MsHsSc}\,,
\end{align}
where $H_{m,n}$ and $H_{m,s}$ are the collinear and soft mass mode contributions of the mass threshold hard function $H_m$, see \eq{Hmfactors}, which appears in the universal factorization theorem \eq{uffacth1}.

In the situation $s^2/Q^2\ll m^2\sim s\sim p^2\ll Q^2$ the mass mode factorization approach states the factorization theorem
\begin{align}
\label{eq:mffacth2}
\frac{1}{\sigma_0}\frac{\dd \sigma}{\dd M^2_Q\,\dd M^2_{\bar Q}} &={} H_Q^{(n_\ell+1)}(Q,\mu) H_{m,s}\left(m,\mu,\frac{m}{\nu}\right) \int\dd\ell^+\,\dd\ell^-\nn\\
&\times J_Q^{\mathrm{mf},(n_\ell+1)}\left(M^2_Q-Q\ell^+,m^2,\mu, \frac{Q}{\nu}\right) J_Q^{\mathrm{mf},(n_\ell+1)}\left(M^2_{\bar Q}-Q\ell^-,m^2,\mu, \frac{Q}{\nu}\right)\nn\\
&\times S^{(n_\ell)}(\ell^+,\ell^-,\mu)\,.
\end{align}
Using the consistency relations \eqref{eq:MjHmSc} and \eqref{eq:MsHsSc} as well as the relation between the mass mode jet function $J_Q^\mathrm{mf}$, the universal jet function $J_Q^\mathrm{uf}$ and the collinear-soft function of \eq{Scdef}, it is easy to see that \eq{mffacth2} and the universal factorization theorems \eqref{eq:uffacth1} and \eqref{eq:uffacth2} provide identical descriptions for $s^2/Q^2\ll m^2\sim s\sim p^2\ll Q^2$.
The relations and the analogous equivalence for \eq{mffacth1} show that the factorization theorem \eqref{eq:mffacth1} -- upon replacing the massless quark jet function $J_q^{(n_\ell+1)}$ by the universal jet function $J_Q^{\mathrm{uf},(n_\ell+1)}$ -- fully accounts for the content of the factorization theorem \eqref{eq:mffacth2} for $s^2/Q^2\ll m^2\ll s\sim p^2\ll Q^2$. Thus in this modified form, \eqref{eq:mffacth1} is applicable for the entire region $s^2/Q^2\ll m^2\lesssim s\sim p^2\ll Q^2$ and exactly equivalent to the universal factorization theorems \eqref{eq:uffacth1} and \eqref{eq:uffacth2}.

Thus from a practical application point of view the mass mode factorization theorems \eqref{eq:mffacth1} and \eqref{eq:mffacth2} can be considered as special cases of the universal factorization theorems \eqref{eq:uffacth1} and \eqref{eq:uffacth2}. On the other hand, the mass mode factorization approach for the case $s^2/Q^2\ll m^2\ll s\sim p^2\ll Q^2$ makes explicit use of the collinear-soft function $S_c$ and provides a representation of the quark mass threshold matching factors for the jet and soft functions that disentangle the relevant process-independent collinear and soft mass mode contributions. So, \eqs{MjHmSc}{MsHsSc} in connection with \eq{Hmfactors} clarify the interplay of the different elementary (collinear and soft) mass mode functions $(H_{m,n},H_{m,\bar n},H_{m,s}, S_c)$. Interestingly, by simply replacing the collinear mass mode factors $H_{m,n}$ and $H_{m,\bar n}$ valid for primary massive quarks by the corresponding mass mode factors for primary massless quarks (see eqs.~(4.10) and (4.12) in \rcite{Pietrulewicz:2017gxc} for analytic expressions) the analogous versions of eqs.~(\ref{eq:Hmfactors}), \eqref{eq:MjHmSc} and \eqref{eq:MsHsSc} are also valid for the corresponding factorization theorems for primary massless quarks. In fact, upon making these modifications and replacing the primary massive quark jet functions $J_Q^{\mathrm{mf},(n_\ell+1)}$ and $J_Q^{\mathrm{uf},(n_\ell+1)}$ by the corresponding jet functions for primary massless quarks, \eq{uffacth1}, \eqref{eq:uffacth2} (and \eqref{eq:mffacth1}, \eqref{eq:mffacth2}) agree identically with the results given in \rcite{Pietrulewicz:2014qza}. This demonstrates the universal structure of the mass mode corrections and illustrates how to obtain them in an economic way in factorization theorems for other applications.

Our final comment concerns the summation of logarithms. Combining the information from the universal factorization and the mass mode factorization approaches provides a modular and transparent method to resum virtuality as well as rapidity logarithms (where we acknowledge, however, that complete treatments of correctly summing both types of logarithms is available in both approaches, see \cite{Pietrulewicz:2017gxc} and \cite{Hoang:2015iva}).

From the point of view of summing virtuality logarithms, the universal factorization theorems \eqref{eq:uffacth1} and \eqref{eq:uffacth2} may be seen as more transparent and practical because all anomalous dimensions coincide with
those known from factorization theorems where secondary massive quark effects (which entail rapidity singularities) are absent: one only has to keep track of the dependence on the number of dynamical flavors $n_f$  being either equal to $n_\ell$, when the evolution is below the quark mass $m$ or $(n_\ell+1)$ when it is above.
If we write the virtuality $\mu$-anomalous dimensions of the SCET and bHQET quark-antiquark production currents ($\mathcal J_\mathrm{SCET}$ and $\mathcal J_\mathrm{bHQET}$, respectively~\cite{Fleming:2007qr,Fleming:2007xt}), the hard factor, the universal and bHQET jet function, and the soft function in the form ($\hat s=s/m$)
\begin{align}
\label{eq:MuAnomDims}
\frac{\dd}{\dd \log\mu} \mathcal J_\mathrm{SCET}^{(n_f)}(\mu) ={}& \gamma_c^{(n_f)}(Q,\mu) \mathcal J_\mathrm{SCET}^{(n_f)}(\mu)\,,\notag\\
\frac{\dd}{\dd \log\mu} \mathcal J_\mathrm{bHQET}^{(n_f)}(\mu) ={}& \gamma_{c_B}^{(n_f)}(Q,m,\mu) \mathcal J_\mathrm{bHQET}^{(n_f)}(\mu)\,,\notag\displaybreak[1]\\
\frac{\dd}{\dd \log\mu} H_Q^{(n_f)}(Q,\mu) ={}& \left(\gamma_Q^{(n_f)}(Q,\mu)+\gamma_Q^{(n_f)*}(Q,\mu)\right) H_Q^{(n_f)}(Q,\mu)\,,\displaybreak[1]\\
\frac{\dd}{\dd \log\mu} J_Q^{\mathrm{uf},(n_f)}(p^2,m^2,\mu) ={}& \int\dd p'^2\, \gamma_J^{(n_f)}(p^2-p'^2,\mu) J_Q^{\mathrm{uf},(n_f)}(p'^2,m^2,\mu)\,,\notag\displaybreak[1]\\
\frac{\dd}{\dd \log\mu} J_B^{(n_f)}(\hat s,\mu) ={}& \int\dd \hat s'\, \gamma_{J_B}^{(n_f)}(\hat s-\hat s',\mu) J_B^{(n_f)}(\hat s',\mu)\,,\notag\displaybreak[1]\\
\frac{\dd}{\dd \log\mu} S^{(n_f)}(\ell^+,\ell^-,\mu) ={}& \int\dd {\ell^+}'\,\dd{\ell^-}'\, \gamma_S^{(n_f)}(\ell^+-{\ell^+}',\ell^--{\ell^-}',\mu) S^{(n_f)}({\ell^+}',{\ell^-}',\mu)\,,\notag
\end{align}
the $\mu$-anomalous dimensions of the corresponding mass threshold matching factors appearing in the factorization theorems \eqref{eq:mffacth1} and \eqref{eq:mffacth2} read
\begin{align}
\label{eq:MuAnomDims2}
\frac{\dd}{\dd\log\mu} H_m\left(m,\frac{Q}{m},\mu\right) ={}& \bigg[\gamma_{c_B}^{(n_\ell)}(Q,m,\mu) + \gamma_{c_B}^{(n_\ell)*}(Q,m,\mu)\\
&\quad - \left(\gamma_{Q}^{(n_\ell+1)}(Q,\mu) + \gamma_{Q}^{(n_\ell+1)*}(Q,\mu)\right)\bigg] H_m(Q,m,\mu)\,,\notag\displaybreak[1]\\
\frac{\dd}{\dd\log\mu} \mathcal M_{J_Q}\left(s,m,\mu\right) ={}& \int \dd s'\, \left[\gamma_{J_B}^{(n_\ell)}\left(\frac{s}{m}-\frac{s'}{m},\mu\right) - \gamma_{J}^{(n_\ell+1)}(s-s',\mu)\right] \mathcal M_{J_Q}(s',m,\mu)\,,\notag\displaybreak[1]\\
\frac{\dd}{\dd\log\mu} \mathcal M_{S}\left(\ell^+,\ell^-,\mu\right) ={}& \int \dd {\ell^+}'\,\dd{\ell^-}' \Big[\gamma_{S}^{(n_\ell+1)}\left(\ell^+-{\ell^+}',\ell^--{\ell^-}',\mu\right)\notag\\
&\quad - \gamma_{S}^{(n_\ell)}(\ell^+-{\ell^+}',\ell^--{\ell^-}',\mu)\Big] \mathcal M_{S}({\ell^+}',{\ell^-}',\mu)\,.\notag
\end{align}
Their form expresses the fact that the mass threshold matching functions $H_m$, $\mathcal M_J$ and $\mathcal M_S$ simply account for the appropriate change between $(n_\ell+1)$ and $n_\ell$ flavor $\mu$-evolution. In this context one just has to take into account that for primary massive quarks the  $n_\ell$-flavor evolution for the jet and the hard functions is carried out in bHQET. For sake of completeness we have displayed all $\mu$-anomalous dimensions up to $\mathcal O(\alpha_s^2)$ in \app{AnoDim}.

From the point of view of summing rapidity logarithms the structure provided by the mass mode factorization theorems \eqref{eq:mffacth1} and \eqref{eq:mffacth2} may be seen as most transparent because the universal structure of the $\nu$-anomalous dimensions of the collinear and soft mass mode factors is fully determined by quoting the results
\begin{align}
\label{eq:NuAnomDims}
\frac{\dd}{\dd\log\nu} H_{m,n}\left(m,\mu,\frac{Q}{\nu}\right) ={}& \gamma_{\nu}(m,\mu) \,H_{m,n}\left(m,\mu,\frac{Q}{\nu}\right),\\
\frac{\dd}{\dd\log\nu} H_{m,s}\left(m,\mu,\frac{m}{\nu}\right) ={}& -2\gamma_{\nu}(m,\mu)\, H_{m,s}\left(m,\mu,\frac{m}{\nu}\right),\notag\\
\frac{\dd}{\dd\log\nu} S_{c}\left(\ell,m,\mu,\nu\right) ={}& \gamma_{\nu}\left(m,\mu\right)\, S_{c}\left(\ell,m,\mu,\nu\right),\notag\\
\frac{\dd}{\dd\log\nu} J_Q^{\mathrm{mf},(n_\ell+1)}\left(p^2,m^2,\mu,\frac{Q}{\nu}\right) ={}& \gamma_{\nu}(m,\mu)\, J_Q^{\mathrm{mf},(n_\ell+1)}\left(p^2,m^2,\mu,\frac{Q}{\nu}\right)\,.\notag
\end{align}
Interestingly, owing to the consistency relations (\ref{eq:Hmfactors}), (\ref{eq:MjHmSc}) and (\ref{eq:MsHsSc}), the $\nu$-anomalous dimensions of all collinear and soft mass mode factors are identical to all orders in $\alpha_s$ up to trivial factors and signs which are determined from the fact that the rapidity divergences within each of the threshold matching functions $H_m$, $\mathcal M_{J_Q}$ and $\mathcal M_{S}$ cancel.
The consistency relations also entail that the $\nu$-anomalous dimensions of the collinear-soft function $S_{c}$ and the mass mode jet function $J_Q^{\mathrm{mf},(n_\ell+1)}$ are proportional to a $\delta$-function. The convolution in their anomalous dimension therefore reduces to a simple multiplication which we have already accounted for in the formulae above. The universal $\nu$-anomalous dimension up to $\mathcal O(\alpha_s^2)$ reads
\begin{align}
\label{eq:NuAnomDim}
\gamma_{\nu}(m,\mu) ={}& a_s^2 C_F T_F\left(\frac{8}{3}L_m^2+\frac{80}{9}L_m+\frac{224}{27}\right)+\mathcal O(\alpha_s^3)\,.
\end{align}

To resum virtuality as well as rapidity logarithms in the universal factorization approach, one evaluates the  factorization theorems of  eqs.~(\ref{eq:uffacth1}) or (\ref{eq:uffacth2}) and uses the $\mu$-anomalous dimensions in \eq{MuAnomDims} to sum all virtuality logarithms by evolving the hard ($H_Q^{(n_\ell+1)}$), jet ($J_Q^{\mathrm{uf},(n_\ell+1)}$) and soft  ($S^{(n_\ell)}$) functions from their natural scale to the global renormalization scale. Whenever one of their $\mu$-evolution crosses the mass threshold $m$, so that the anomalous dimension is modified, the corresponding matching factor $H_m$, $ \mathcal M_{J_Q}$ or $ \mathcal M_S$ has to be included. 
The rapidity logarithms are fully contained within each of the mass threshold matching factors $H_m$, $ \mathcal M_{J_Q}$ and $ \mathcal M_S$ evaluated at $\mu\sim m$ and can be resummed by using (\ref{eq:Hmfactors}), (\ref{eq:MjHmSc}) and (\ref{eq:MsHsSc}) and the $\nu$-anomalous dimensions of \eq{NuAnomDims}.

\section{Calculation of Massive Quark Jet Functions at ${\cal O}(\alpha_s^2)$}
\label{sec:calculations}

\subsection{Comment on the Calculations and Summary of Matrix Element Results} 
\label{sec:commentssummary}

In this section we provide details on the calculation of the SCET primary massive quark jet functions at ${\cal O}(\alpha_s^2)$. The universal and mass modes SCET jet functions differ concerning the 
${\cal O}(\alpha_s^2 C_F T_F)$ corrections coming from diagrams with secondary massive quarks (which for the jet functions refers to diagrams with a closed massive quark-antiquark loop subdiagram). These corrections involve a number of conceptual and technical subtleties, and we briefly address these in the following before presenting details of the computations in the subsequent sections. We also summarize  the individual results for the jet-function ME (calculated in the subsequent sections) and the two collinear-soft MEs (extracted from~\cite{Pietrulewicz:2017gxc}) which are needed to determine the universal and mass mode SCET jet functions. We believe that this makes the conceptual subtleties of the calculations explicit. 

The conceptual starting point is the calculation of the jet-function ME $\Jme_f(p^2, m^2)$, defined in~\eq{JmeSCET}, which is {\it infrared divergent} and defined in the $(n_\ell+1)$ flavor theory. The {\it infrared finite} universal jet function $J_f^{\mathrm{uf},(n_\ell+1)}(p^2,m^2)$ and the mass mode jet function $J_f^{\mathrm{mf},(n_\ell+1)}(p^2,m^2)$ are then given via subtractions related to the likewise {\it infrared divergent} collinear-soft MEs $\Scme^{(n_\ell+1)}(\ell,m)$ and  $\Scme^{(n_\ell)}(\ell,m)$, respectively, defined in
the $(n_\ell+1)$ flavor theory (i.e.\ containing $n_\ell$ massless quarks and one quark $Q$ with mass $m$, see \eq{Sdef1}) and in the $n_\ell$ flavor theory (i.e.\ containing only $n_\ell$ massless quarks), respectively. The corresponding definitions are given in \eq{jmufdef} and \eq{jmmfdef}, respectively. We stress that the concept of zero-bin subtractions~\cite{Manohar:2006nz} {\it does not play any role} in our calculation, and we also believe that our approach can be generalized to other calculations in the context of SCET factorization theorems. As already explained in \sec{intro}, the universal jet function $J_f^{\mathrm{uf},(n_\ell+1)}(p^2,m^2)$ by construction approaches the well-known massless quark jet function~\cite{Becher:2006qw} in the limit $m\to 0$. It can therefore be thought of as the mandatory jet function definition for $m^2\ll q^2$ that involves  subtractions related to the mass singularity coming from the secondary quark mass effects in this limit. On the other hand, the mass mode jet function $J_f^{\mathrm{mf},(n_\ell+1)}(p^2,m^2)$ does not involve subtractions related to the secondary quark mass effects and can be thought of as the jet function definition suitable for $m^2\sim q^2$, where the quark mass $m$ does not represent an infrared scale. It is useful for the computations to keep these conceptual aspects of the two jet functions in mind, even though we emphasize that
both jet functions can be applied in a broader kinematic context and are related through the collinear-soft function $S_{c}(\ell,m)$ via \eq{Scdef} (see the discussion in \sec{consistency}).\footnote{In previous literature on the factorization concerning secondary massive quark corrections these subtleties were hidden.} 

On the technical level an additional subtlety arises because it turns out that - at least at this time - the only feasible way to determine the full set of ${\cal O}(\alpha_s^2)$ corrections is to use two completely separate calculational schemes: one for the purely gluonic and the massless quark corrections at ${\cal O}(\alpha_s^2 C_F^2, \alpha_s^2 C_F C_A, \alpha_s^2 C_F T_F n_\ell)$ and another one for the secondary massive quark corrections at ${\cal O}(\alpha_s^2 C_F T_F)$. This separation arises because the ${\cal O}(\alpha_s^2 C_F T_F)$ secondary massive quark corrections involve rapidity singularities. To quantify them analytically it is mandatory to introduce an explicit dimensionful infrared regulator (for which we adopt a gluon mass $\Lambda$), such that UV and rapidity divergences can be disentangled and identified unambiguously. The treatment of subdivergences at  ${\cal O}(\alpha_s^2 C_F T_F)$ then also requires a calculation of the ${\cal O}(\alpha_s)$ corrections with a finite gluon mass for consistency. These calculations have to be carried out for the jet-function ME $\Jme_f(p^2, m^2)$ and the collinear-soft MEs $\Scme^{(n_\ell+1)}(\ell,m)$ and  $\Scme^{(n_\ell)}(\ell,m)$ which also entails infrared gluon mass dependent $\overline{\rm MS}$ renormalization $Z$-factors for each of them.
For the renormalized universal and mass mode jet functions as well as for their virtuality and rapidity renormalization group equations contributing at ${\cal O}(\alpha_s^2 C_F T_F)$ it can then be explicitly checked that the dependence on the infrared gluon mass cancels. 
In contrast, the purely gluonic as well as the massless quark corrections at  ${\cal O}(\alpha_s^2 C_F^2, \alpha_s^2 C_F C_A, \alpha_s^2 C_F T_F n_\ell)$ (as well as the ${\cal O}(\alpha_s)$ corrections) do not involve rapidity singularities and just using dimensional regularization without an additional infrared regulator is in principle sufficient. In fact using this simpler regularization is the only feasible way to compute the gluonic ${\cal O}(\alpha_s^2)$ corrections due to their complexity. For these corrections 
then all diagrams contributing to the collinear-soft MEs $\Scme^{(n_\ell+1)}(\ell,m)$ and $\Scme^{(n_\ell)}(\ell,m)$  lead to vanishing scaleless integrals which just turn all $1/\varepsilon^n$ IR singularities in the infrared divergent jet-function ME $\Jme_f(p^2, m^2)$ into $1/\varepsilon^n$ UV singularities of the infrared finite jet functions $J_f^{\mathrm{uf},(n_\ell+1)}(p^2, m^2)$ and $J_f^{\mathrm{mf},(n_\ell+1)}(p^2, m^2)$. Both jet functions thus agree concerning their ${\cal O}(\alpha_s)$ and ${\cal O}(\alpha_s^2 C_F^2, \alpha_s^2 C_F C_A, \alpha_s^2 C_F T_F n_\ell)$ corrections.
This justifies that for the ${\cal O}(\alpha_s^2)$ corrections involving only diagrams with massless partons (in addition to the primary massive quark line), at least at the computational level, the subtractions associated to collinear-soft MEs or zero-bins can be ignored. The treatment of subdivergences in this calculation also requires the calculations of the ${\cal O}(\alpha_s)$ corrections, which differ from the corresponding results with a gluon mass regulator. 
The lack of having an infrared regularization scheme that allows the computation of all ${\cal O}(\alpha_s^2)$ corrections in a uniform and manageable way entails that - at least at the present time - the computation of the ${\cal O}(\alpha_s^3)$ corrections to the 
SCET jet functions in the presence of massive quarks would represent an extremely challenging task.

We present the calculations of the purely gluonic and massless quark corrections at ${\cal O}(\alpha_s^2 C_F^2, \alpha_s^2 C_F C_A, \alpha_s^2 C_F T_F n_\ell)$ for the jet-function ME $\Jme_f(p^2, m^2)$ in \sec{massless}, and those of the secondary massive quark corrections at ${\cal O}(\alpha_s^2 C_F T_F)$ in \sec{secondary}. The reader not interested in the computational technical details may skip these two subsections. In the following we summarize the analytic results for jet-function ME and the two collinear-soft MEs within the two calculational schemes.
Results needed for the calculation of the ${\cal O}(\alpha_s^2 C_F^2, \alpha_s^2 C_F C_A, \alpha_s^2 C_F T_F n_\ell)$ corrections to the jet function are computed in the massless gluon computational scheme (where the ${\cal O}(\alpha_s^2 C_F T_F)$ corrections cannot be given) and are quoted with the subscript ``$\cancel{C_F T_F}$''.
In contrast all results needed for the calculation of the ${\cal O}(\alpha_s^2 C_F T_F)$ terms from secondary mass effects are computed in the massive gluon computational scheme (where the ${\cal O}(\alpha_s^2 C_F^2, \alpha_s^2 C_F C_A, \alpha_s^2 C_F T_F n_\ell)$ corrections cannot be given) and are quoted with the subscript ``$C_F T_F$''.
Note that we use this labeling also for the corresponding one-loop results that contribute (through the treatment of subdivergences) to the determination of the respective renormalized two-loop jet functions and their anomalous dimensions.\footnote{We note that all statements made before in \sec{commentssummary} apply for massless ($f=q$) and massive primary quarks ($f=Q$).}

In the computational scheme with a massless gluon the renormalized jet-function ME can be written in the form 
\begin{align}
\left.\Jme_Q^{(n_\ell+1)}(p^2,m^2,\Lambda^2=0,\mu,Q/\nu)\right|_{\cancel{C_F T_F}} ={} &\delta(p^2-m^2) + a_s^{(n_\ell+1)} J_Q^{(1)}(p^2,m^2,\mu)\notag\\
&
+ \left(a_s^{(n_\ell+1)}\right)^2 J_Q^{(2)}(p^2,m^2,\mu) + \mathcal O(\alpha_s^3)\,,
\label{eq:JdefLam0}
\end{align}
where the ${\cal O}(\alpha_s)$ coefficient $J_Q^{(1)}(p^2,m^2,\mu)$ agrees with the well-known one-loop SCET primary massive quark jet function result from \rcite{Fleming:2007xt} and is given up to ${\cal O}(\varepsilon^2)$ in \eq{Jmf1}. 
The two-loop coefficient $J_Q^{(2)}$ containing the ${\cal O}(\alpha_s^2 C_F^2, \alpha_s^2 C_F C_A, \alpha_s^2 C_F T_F n_\ell)$ corrections is given in \eq{Jmf2}. 
Both are free of rapidity divergences and therefore do not carry an argument with respect to the rapidity renormalization scale $\nu$.
For convenience, we also provide the corresponding renormalized result for the massless primary quark jet-function ME $\left.\Jme_q^{(n_\ell+1)}(p^2,m^2,\Lambda^2=0,\mu,Q/\nu)\right|_{\cancel{C_F T_F}}$  in \app{OtherJet}.

The divergent $Z$-factor of the jet-function ME reads
\begin{align}
&\left. Z_{\Jme_Q}^{(n_\ell+1)}(p^2,m^2,\Lambda^2=0,\mu,Q/\nu)\right|_{\cancel{C_F T_F}} = \delta(p^2)\notag\\
&\quad + a_s^{(n_\ell+1)}C_F\left\{\left[\frac{4}{\varepsilon^2}+\frac{3}{\varepsilon}\right] \delta(p^2) - \frac{4}{\varepsilon} \mathcal L_0(p^2)\right\}\notag\\
&\quad + \left(a_s^{(n_\ell+1)}\right)^2 \Bigg\{\!C_F T_F n_\ell\! \left[\left(\frac{4}{\varepsilon^3}-\frac{2}{9 \varepsilon ^2}+\left(-\frac{121}{27}-\frac{2 \pi ^2}{9}\right)\!\frac{1}{\varepsilon}\right) \delta (p^2) + \left(-\frac{8}{3 \varepsilon ^2} + \frac{40}{9 \varepsilon }\right)\! \mathcal L_0(p^2)\right]\nn\\
&\qquad+C_F C_A \Bigg[\left(-\frac{11}{\varepsilon ^3}+\left(\frac{35}{18}-\frac{\pi ^2}{3}\right)\frac{1}{\varepsilon^2}+\left(-20 \zeta_3+\frac{1769}{108}+\frac{11 \pi ^2}{18}\right)\frac{1}{\varepsilon }\right)\delta (p^2)\nn\\
&\qquad\quad + \left(\frac{22}{3 \varepsilon ^2}+\left(\frac{2 \pi ^2}{3}-\frac{134}{9}\right)\frac{1}{\varepsilon }\right) \mathcal L_0(p^2)\Bigg]\nn\\
&\qquad+C_F^2 \Bigg[\left(\frac{8}{\varepsilon ^4}+\frac{12}{\varepsilon ^3}+\left(\frac{9}{2}-\frac{4 \pi ^2}{3}\right)\frac{1}{\varepsilon ^2}+\left(12 \zeta_3+\frac{3}{4}-\pi^2\right)\frac{1}{\varepsilon}\right)\delta (p^2)\nn\\
&\qquad\quad+\left(-\frac{16}{\varepsilon^3}-\frac{12}{\varepsilon ^2}\right) \mathcal L_0(p^2) + \frac{16}{\varepsilon ^2} \mathcal L_1(p^2)\Bigg]\Bigg\} + \ord{\alpha_s^3}\,.
\label{eq:ZJmassless}
\end{align}
By definition it contains {\it all} $1/\varepsilon$-divergent terms and quantifies the difference between the renormalized and bare jet-function MEs,
\begin{align}
\label{eq:Jren}
\Jme_Q^{(n_\ell+1),\mathrm{bare}}(p^2,m^2,\mu) = \int\dd p'^2\, Z_{\Jme,Q}^{(n_\ell+1)}(p^2-p'^2,\mu)\Jme_Q^{(n_\ell+1)}(p'^2,Z_m^2 m^2)\,.
\end{align}
In the context of our jet-function ME computation it includes UV and IR divergent $1/\varepsilon$ terms. We stress 
that, upon (literally) replacing the color factor $C_F T_F n_\ell$ by $C_F T_F(n_\ell+1)$ in \eq{ZJmassless}, one obtains the {\it full}  
${\cal O}(\alpha_s^2)$ $\overline{\rm MS}$ $Z$-factor for the universal SCET jet functions $J_f^{\mathrm{uf},(n_\ell+1)}$ (for massless or massive primary quarks) as well as for the SCET jet function for $(n_\ell+1)$ massless quark flavors, where {\it all} $1/\varepsilon$ divergent terms are UV divergences.\footnote{Note that that the $Z$-factor for the mass mode jet function $J_f^{\mathrm{mf},(n_\ell+1)}$ can be obtained from the results for the bare and renormalized jet-function ME $\Jme_Q^{(n_\ell+1)}$ and collinear-soft ME $\Scme^{(n_\ell)}$. Because both are defined in theories with different number of quark flavors, the $Z$-factor for the mass mode jet function cannot be quoted in the $\overline{\rm MS}$ scheme and in general has finite contributions.} This relation is required by consistency and can be used as a cross check for calculations.

In the massless gluon computational scheme the loop corrections to the bare collinear-soft ME $\Scme^{(n_\ell)}$ are given by scaleless integrals 
and thus vanish to all orders in perturbation theory
\begin{align}
&\Scme^{(n_\ell),{\rm bare}}(\ell,m,\Lambda=0) = \delta(\ell)\,.
\end{align}
The corresponding bare collinear-soft ME $\Scme^{(n_\ell+1)}$, on the other hand, corresponds to scaleless integrals only
at ${\cal O}(\alpha_s)$ and ${\cal O}(\alpha_s^2 C_F^2, \alpha_s^2 C_F C_A,\alpha_s^2 C_F T_F n_\ell)$: 
\begin{align}
&\left.\Scme^{(n_\ell+1),{\rm bare}}(\ell,m,\Lambda=0)\right|_{\cancel{C_F T_F}} =
 \delta(\ell)
+ \ord{\alpha_s^3}
\,.
\end{align}

In the computational scheme with an infinitesimal gluon mass $\Lambda$ the terms of the renormalized jet-function ME relevant for the calculation of the ${\cal O}(\alpha_s^2 C_F T_F)$ secondary quark mass effects read
\begin{align}
&\left. \Jme_Q^{(n_\ell+1)}(p^2,m^2,\Lambda^2,\mu,Q/\nu)\right|_{C_F T_F} = \delta(s) + a_s^{(n_\ell+1)}C_F\Bigg\{\Bigg[2 L_m^2 +L_m+2 \log ^2\left(\frac{\Lambda^2}{\mu ^2}\right)\notag\\
&\quad\quad+4 \log\left(\frac{\Lambda^2}{\mu ^2}\right) \log \left(\frac{Q}{\nu }\right)+8\Bigg]\delta (s)+\left[-4 L_m-4 \log \left(\frac{\Lambda^2}{\mu^2}\right)-4\right] \mathcal L_0(s)+8 \mathcal L_1(s)\notag\\
&\quad+\Theta (p^2-m^2)\left[\frac{s}{\left(m^2+s\right)^2}-\frac{4 \log\left(\frac{s}{m^2}+1\right)}{s}\right]\Bigg\}\notag\\
&+ \left(a_s^{(n_\ell+1)}\right)^2 C_F T_F\Bigg\{\Bigg[\left(\frac{8}{3} \log ^2\left(\frac{\Lambda ^2}{\mu^2}\right)+\frac{8 \pi ^2}{9}+\frac{34}{3}\right) L_m+\frac{8}{3} L_m^3+\frac{10}{3} L_m^2\notag\\
&\quad\quad+\log \bigg(\frac{Q}{\nu }\bigg) \left(\frac{16}{3}\log \left(\frac{\Lambda ^2}{\mu ^2}\right) L_m-\frac{80}{9} L_m-\frac{8}{3} L_m^2-\frac{224}{27}\right)+\frac{8\zeta_3}{3}-\frac{4 \pi ^2}{3}+\frac{3139}{162}\Bigg]\delta (p^2)\notag\\
&\quad + \left[\left(-\frac{16}{3} \log \left(\frac{\Lambda ^2}{\mu^2}\right)-\frac{16}{3}\right) L_m-\frac{16}{3} L_m^2\right]\mathcal L_0(p^2)+\frac{32}{3} L_m \mathcal L_1(p^2)\notag\\
&\quad+L_m \left[\frac{4 s}{3 \left(m^2+s\right)^2}-\frac{16 \log\left(\frac{s}{m^2}+1\right)}{3 s}\right]\Theta (p^2-m^2) + G_{\mathrm{sec}}^{(3Q)}\Theta(p^2-(3m)^2)\Bigg\} + \ord{\alpha_s^3}\,,
\label{eq:Jme1}
\end{align}
where $G_{\mathrm{sec}}^{(3Q)}$ is given in \eq{G3QTF}. For convenience, we also provide the analogous results for the renormalized massless primary quark jet-function ME $\left.\Jme_q^{(n_\ell+1)}(p^2,m^2,\Lambda^2,\mu,Q/\nu)\right|_{C_F T_F}$ in \app{OtherJet}. The ${\overline{\rm MS}}$ renormalization $Z$-factor associated with the jet-function ME contributions in \eq{Jme1} (which is identical for massive and massless primary quarks) accounts for UV and rapidity divergences and reads
\begin{align}
&\left. Z_{\Jme}(p^2,m^2,\Lambda^2,\mu,Q/\nu)\right|_{C_F T_F} = \delta(p^2)\notag\\
&\quad + a_s^{(n_\ell+1)}C_F\left\{\left[\frac{4}{\varepsilon }-4 \log \left(\frac{\Lambda ^2}{\mu^2}\right)+\ord{\varepsilon}\right]\frac{1}{\eta}+\left[3-4 \log \left(\frac{Q}{\nu }\right)\right]\frac{1}{\varepsilon}\right\}\delta (p^2)\notag\\
&\quad +\left(a_s^{(n_\ell+1)}\right)^2 C_F T_F \Biggl\{\left[\frac{8}{3 \varepsilon^2}-\frac{40}{9 \varepsilon}+\frac{8}{3}L_m^2+\left(\frac{80}{9}-\frac{16}{3}\log\left(\frac{\Lambda^2}{\mu^2}\right)\right)L_m+\frac{224}{27} + \ord{\varepsilon}\right]\frac{1}{\eta}\notag\\
&\qquad+\left[2-\frac{8}{3}\log\left(\frac{Q}{\nu}\right)\right]\frac{1}{\varepsilon^2}+\left[\frac{40}{9}\log\left(\frac{Q}{\nu}\right)-\frac{4 \pi^2}{9}-\frac{1}{3}\right]\frac{1}{\varepsilon}\Biggr\}\delta(p^2) + \ord{\alpha_s^3}\,.
\label{eq:ZJme1}
\end{align}
The corresponding results for the two different renormalized collinear-soft MEs are
\begin{align}
&\left.\Scme^{(n_\ell)}(\ell,m,\Lambda,\mu,\nu)\right|_{C_F T_F} = \delta(\ell)\notag\\
&\quad + a_s^{(n_\ell)} C_F \left\{\left[-4 \log \left(\frac{\nu }{\mu }\right)\log \left(\frac{\Lambda^2}{\mu ^2}\right)+2 \log ^2\left(\frac{\Lambda^2}{\mu ^2}\right)+\frac{\pi^2}{3}\right]\delta(\ell)-4 \mathcal L_0(\ell) \log \left(\frac{\Lambda^2}{\mu ^2}\right)\right\}\notag\\
&\quad+ \ord{\alpha_s^3}\,,
\end{align}
and
\begin{align}
&\left.\Scme^{(n_\ell+1)}(\ell,m,\Lambda,\mu,\nu)\right|_{C_F T_F} = \delta(\ell)\notag\\
&\quad + a_s^{(n_\ell+1)} C_F \left\{\left[-4 \log \left(\frac{\nu }{\mu }\right)\log \left(\frac{\Lambda^2}{\mu ^2}\right)+2 \log ^2\left(\frac{\Lambda^2}{\mu ^2}\right)+\frac{\pi^2}{3}\right]\delta(\ell)-4 \mathcal L_0(\ell) \log \left(\frac{\Lambda^2}{\mu ^2}\right)\right\}\notag\\
&\quad+ \left(a_s^{(n_\ell+1)}\right)^2 C_F T_F \Bigg\{\bigg[-\frac{8}{9} L_m^3+\left(\frac{8}{3} \log \left(\frac{\nu}{\mu }\right)-\frac{40}{9}\right) L_m^2\notag\\
&\qquad\quad+L_m \left(-\frac{16}{3} \log \left(\frac{\Lambda^2}{\mu ^2}\right) \log \left(\frac{\nu }{\mu }\right)+\frac{8}{3} \log^2\left(\frac{\Lambda ^2}{\mu ^2}\right)+\frac{80}{9} \log \left(\frac{\nu }{\mu}\right)+\frac{8 \pi ^2}{9}-\frac{448}{27}\right)\notag\\
&\qquad\quad+\frac{56 \zeta_3}{9}+\frac{10 \pi^2}{27}-\frac{656}{27}+\frac{224}{27} \log \left(\frac{\nu }{\mu}\right)\bigg]\delta(\ell)\notag\\
&\qquad+ \left[\left(\frac{80}{9}-\frac{16}{3} \log \left(\frac{\Lambda ^2}{\mu ^2}\right)\right)L_m+\frac{8}{3} L_m^2+\frac{224}{27}\right] \mathcal L_0(\ell)\Bigg\}+\ord{\alpha_s^3}\,.
\end{align}
Their respective ${\overline{\rm MS}}$ renormalization $Z$-factors accounting for UV and rapidity divergences read
\begin{align}
&\left. Z_{\Scme}^{(n_\ell)}(\ell,m,\Lambda,\mu,\nu) \right|_{C_F T_F} ={} \delta(\ell) + 4\, a_s^{(n_\ell)}C_F\Bigg\{\left[\frac{1}{\varepsilon} - \log\left(\frac{\Lambda^2}{\mu^2}\right)+\ord{\varepsilon}\right]\frac{1}{\eta}\delta(\ell)\nn\\
&\qquad+\left[-\frac{1}{\varepsilon^2}+ \log \left(\frac{\nu}{\mu}\right) \right]\delta(\ell)+\mathcal L_0(\ell)\frac{1}{\varepsilon}\Bigg\} +\ord{\alpha_s^3}\,,  
\end{align}
and
\begin{align}
&\left. Z_{\Scme}^{(n_\ell+1)}(\ell,m,\Lambda,\mu,\nu) \right|_{C_F T_F} ={} \delta(\ell) + 4\, a_s^{(n_\ell)}C_F \Bigg\{\left[\frac{1}{\varepsilon} - \log\left(\frac{\Lambda^2}{\mu^2}\right)+\ord{\varepsilon}\right]\frac{1}{\eta}\delta(\ell)\nn\\
&\qquad+\left[-\frac{1}{\varepsilon^2}+ \log \left(\frac{\nu}{\mu}\right) \right]\delta(\ell)+\mathcal L_0(\ell)\frac{1}{\varepsilon}\Bigg\}\notag\\
&\quad + \left(a_s^{(n_\ell+1)}\right)^2 C_F T_F\Bigg\{\left[\frac{8}{3 \varepsilon ^2}-\frac{40}{9 \varepsilon }-\frac{16}{3}\log\left(\frac{\Lambda^2}{\mu^2}\right)L_m+\frac{8 L_m^2}{3}+\frac{80 L_m}{9}+\frac{224}{27}\right]\frac{1}{\eta}\delta(p^2)\nn\\
&\qquad + \left[-\frac{4}{\varepsilon^3}+\left(\frac{20}{9}+\frac{8}{3}\log\left(\frac{\nu}{\mu}\right)\right)\frac{1}{\varepsilon ^2}+\left(-\frac{40}{9}\log\left(\frac{\nu}{\mu}\right)-\frac{2 \pi^2}{9}+\frac{112}{27}\right)\frac{1}{\varepsilon}\right] \delta \left(p^2\right)\nn\\
&\qquad + \left[\frac{8}{3 \varepsilon ^2}-\frac{40}{9 \varepsilon }\right] \mathcal L_0(p^2)\Bigg\} + \ord{\alpha_s^3}
\,.
\end{align}
These results can be extracted from \rcite{Pietrulewicz:2017gxc}, and we have explicitly cross checked their results.\footnote{Note that eq.~(B.55) of \rcite{Pietrulewicz:2017gxc} has a typo due to a missing global factor of $4$.}

From the results quoted above it is straightforward to obtain the results for the universal (see \eqs{jmufdef}{Jufdef}) and the mass mode SCET jet functions (see \eqs{jmmfdef}{Jmfdef}). The virtuality and rapidity anomalous dimensions for the universal and mass mode SCET jet functions (and the collinear-soft function) 
(see \eqs{MuAnomDims}{NuAnomDims} and \app{AnoDim}) are related to the corresponding anomalous dimensions for the jet-function and the collinear-soft MEs (see \app{AnoDim} as well).

\subsection{Gluonic and Massless Quark Corrections}
\label{sec:massless}

\begin{figure}
\centering
\includegraphics[width=0.22\textwidth]{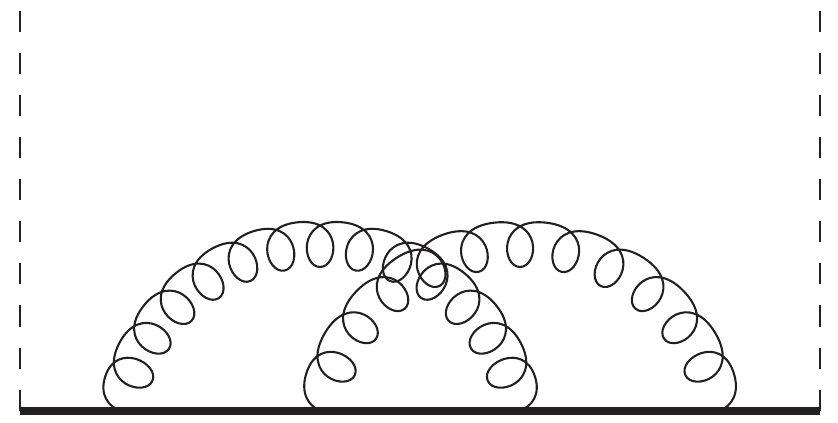}
\put(-54,38){(a)}\hspace{7pt}
\includegraphics[width=0.22\textwidth]{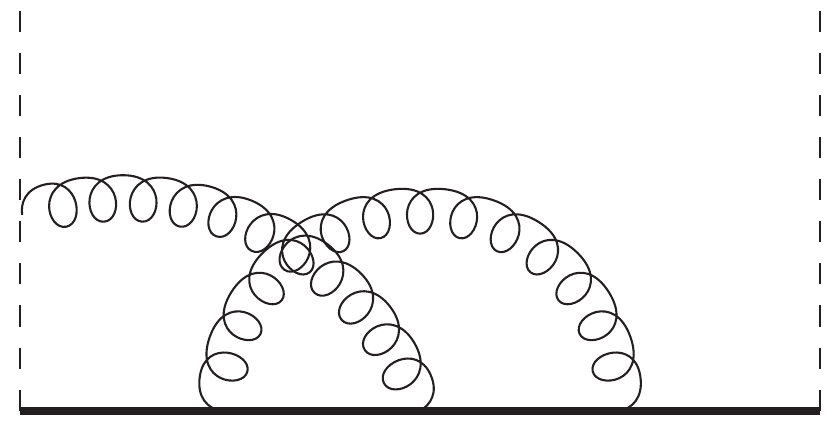}
\put(-54,38){(b)}\hspace{7pt}
\includegraphics[width=0.22\textwidth]{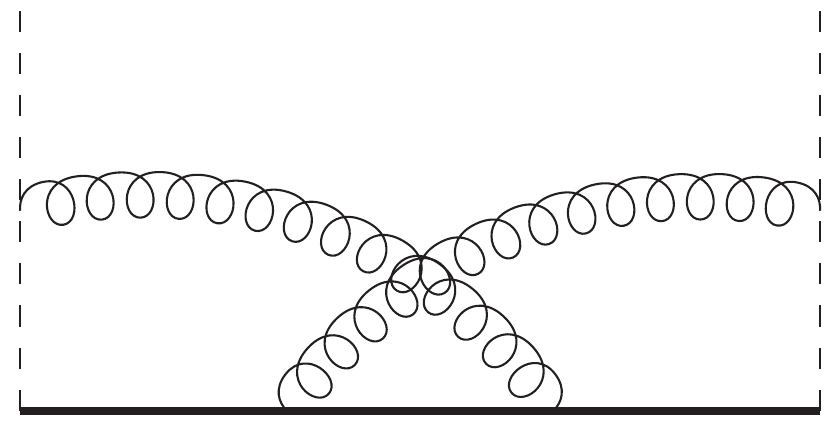}
\put(-54,38){(c)}\hspace{7pt}
\includegraphics[width=0.22\textwidth]{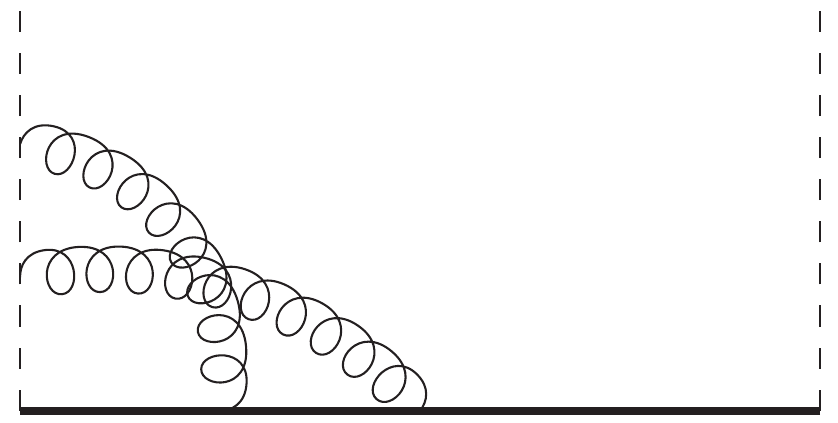}
\put(-54,38){(d)}\hspace{7pt}\\[7pt]
\includegraphics[width=0.22\textwidth]{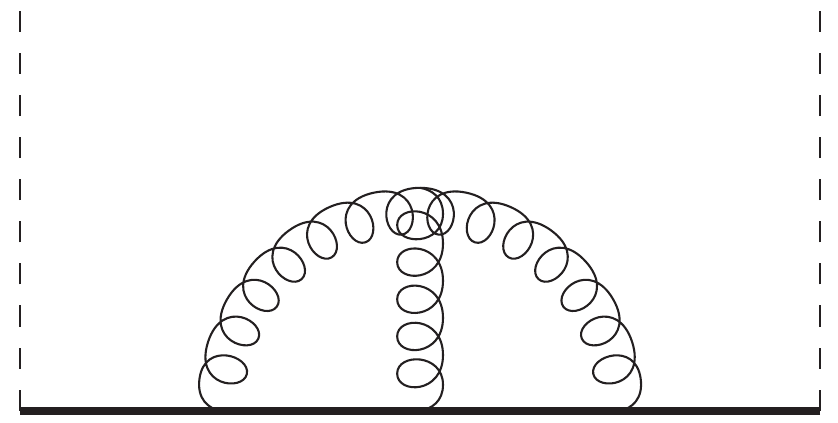}
\put(-54,38){(e)}\hspace{7pt}
\includegraphics[width=0.22\textwidth]{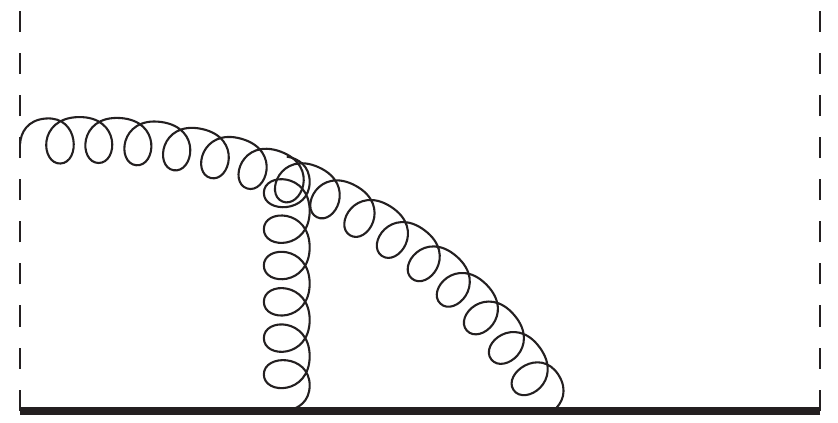}
\put(-54,38){(f)}\hspace{7pt}
\includegraphics[width=0.22\textwidth]{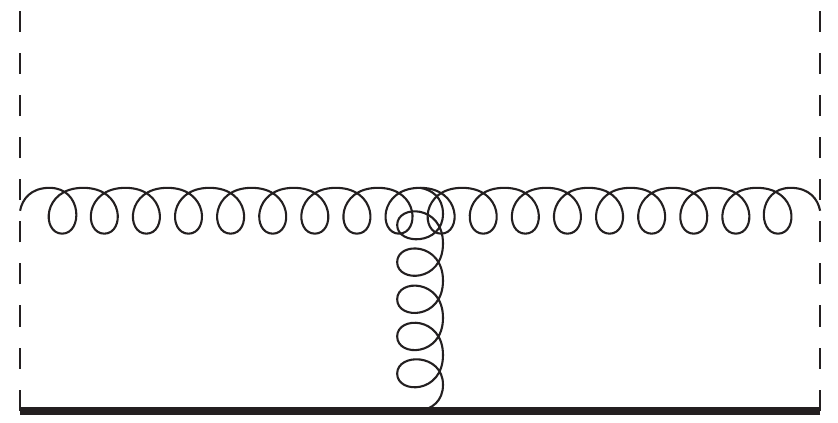}
\put(-54,38){(g)}\hspace{7pt}
\includegraphics[width=0.22\textwidth]{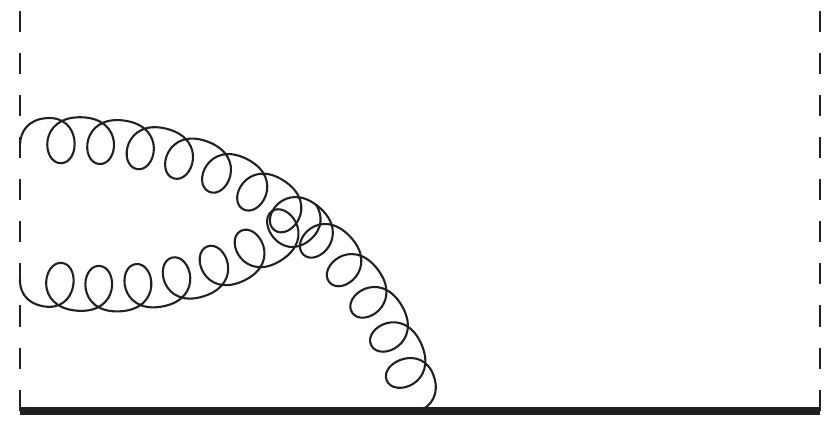}
\put(-54,38){(h)}\hspace{7pt}\\[7pt]
\includegraphics[width=0.22\textwidth]{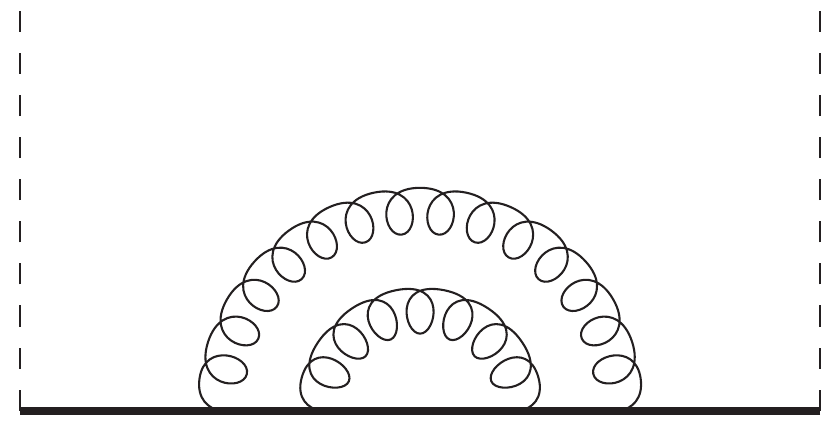}
\put(-54,38){(i)}\hspace{7pt}
\includegraphics[width=0.22\textwidth]{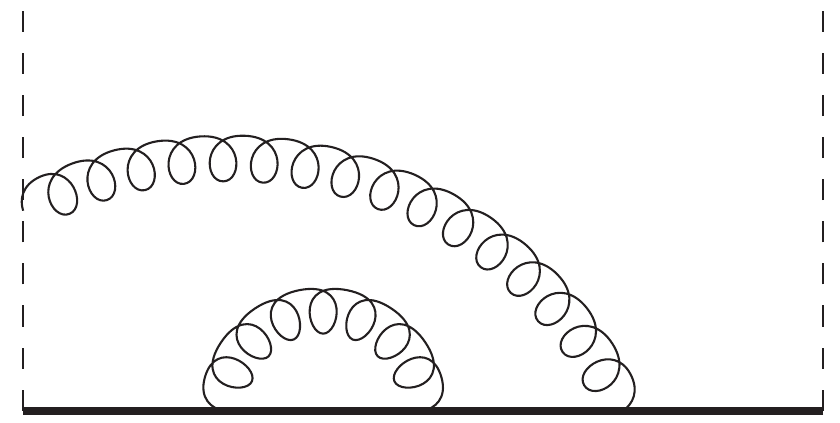}
\put(-54,38){(j)}\hspace{7pt}
\includegraphics[width=0.22\textwidth]{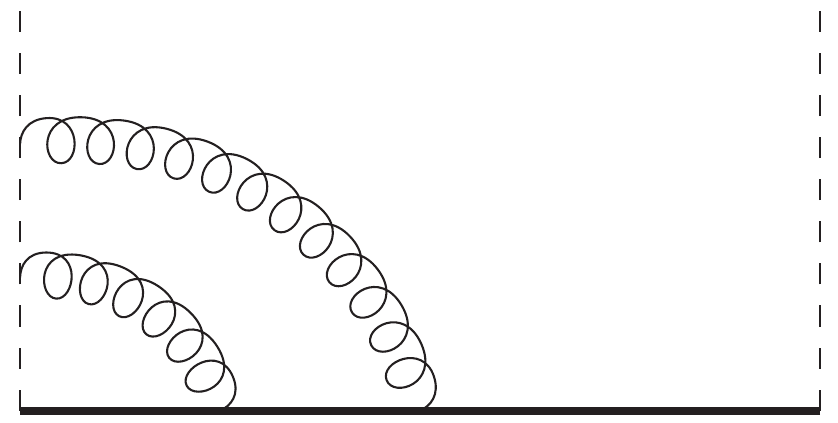}
\put(-54,38){(k)}\hspace{7pt}
\hphantom{\includegraphics[width=0.22\textwidth]{figs/i1}}\\[7pt]
\includegraphics[width=0.22\textwidth]{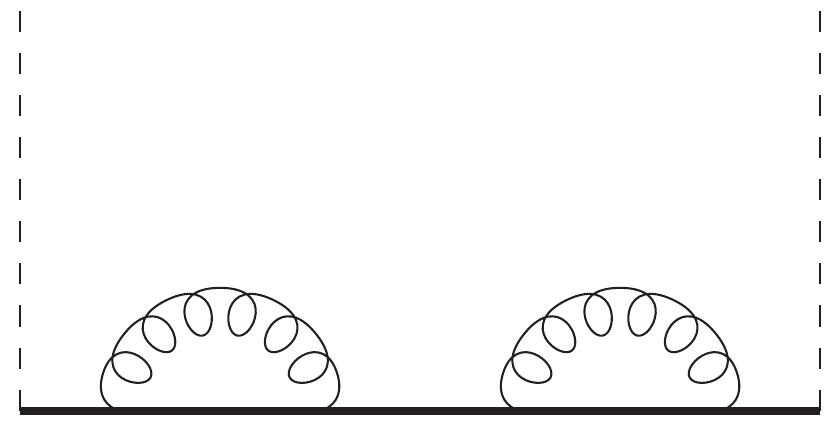}
\put(-54,38){(l)}\hspace{7pt}
\includegraphics[width=0.22\textwidth]{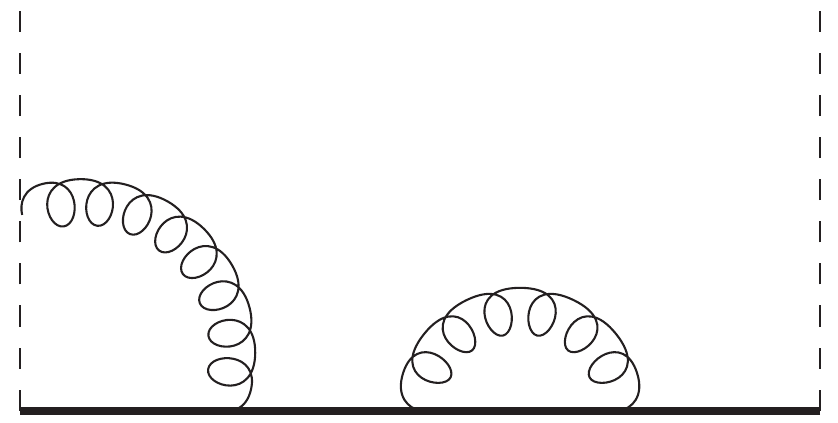}
\put(-54,38){(m)}\hspace{7pt}
\includegraphics[width=0.22\textwidth]{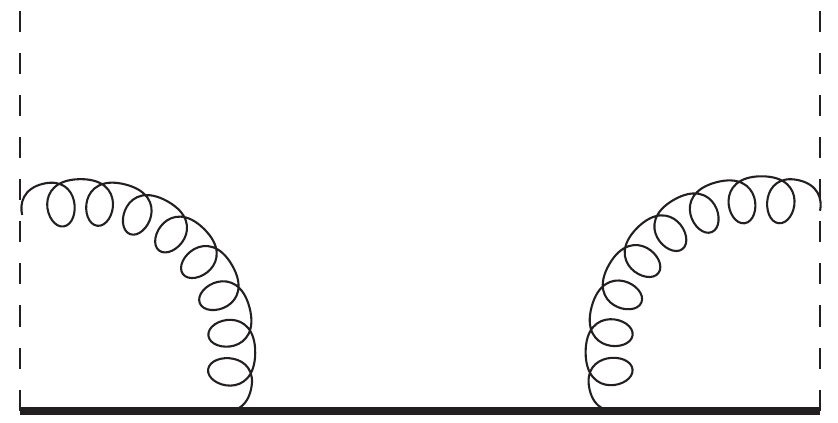}
\put(-54,38){(n)}\hspace{7pt}
\hphantom{\includegraphics[width=0.22\textwidth]{figs/i1}}\\[7pt]
\includegraphics[width=0.22\textwidth]{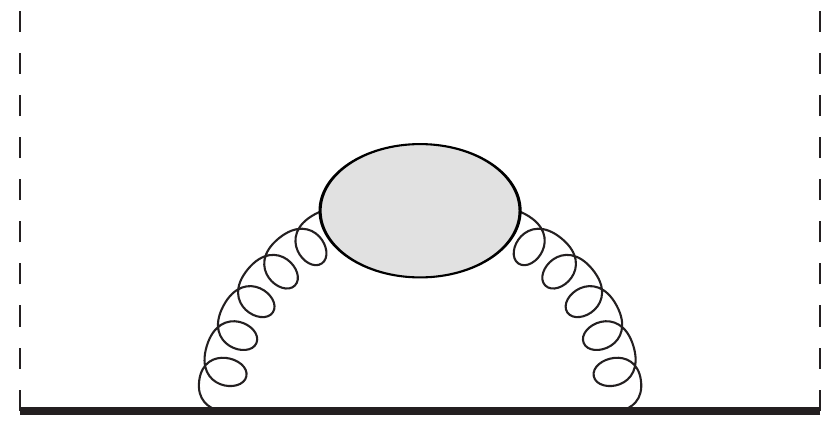}
\put(-54,38){(o)}\hspace{7pt}
\includegraphics[width=0.22\textwidth]{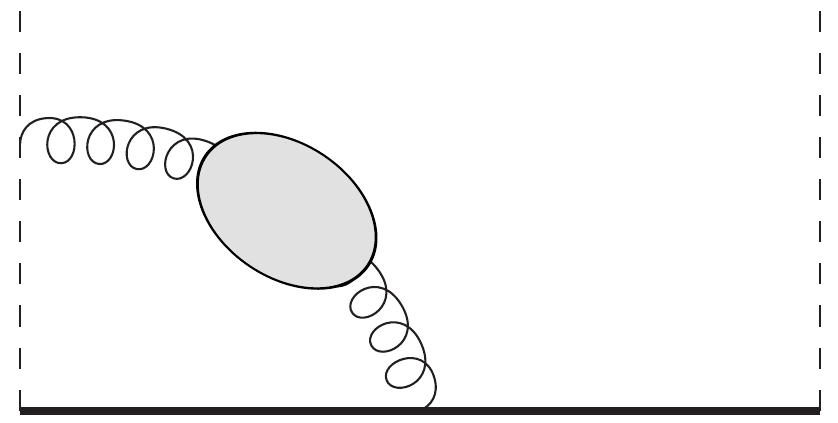}
\put(-54,38){(p)}\hspace{7pt}
\includegraphics[width=0.22\textwidth]{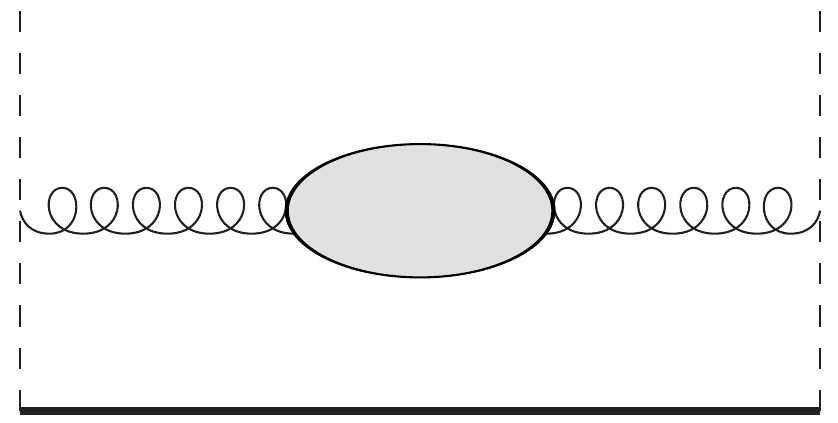}
\put(-54,38){(q)}\hspace{7pt}
\includegraphics[width=0.22\textwidth]{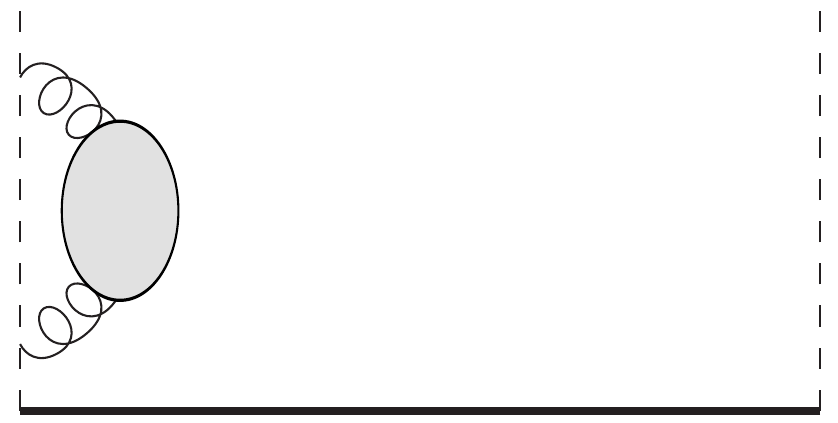}
\put(-54,38){(r)}\hspace{7pt}\\[7pt] 
\caption{QCD Feynman diagrams contributing to the jet field correlator in \eq{JmeQCD} at two-loop order in Feynman gauge. 
The dashed lines represent $n$-collinear (lightlike) Wilson lines. 
The gray blob symbolizes the set of all 1-loop gluon self-energy subdiagrams. 
Left-right mirror graphs are understood.
Diagrams that are trivially zero (as $\bn^2=0$) are not shown.}
\label{fig:diagrams1}
\end{figure}

In \fig{diagrams1} all two-loop Feynman diagrams relevant for the calculation of the two-loop jet-function ME in \eq{JmeQCD} using QCD Feynman rules and Feynman gauge are displayed.
In this section we discuss the computation of the  ${\cal O}(\alpha_s^2 C_F^2, \alpha_s^2 C_F C_A, \alpha_s^2 C_F T_F n_\ell)$ corrections using the computational scheme with a massless gluon, as explained in the previous subsection.

After evaluating the Dirac trace in \eq{JmeQCD} the contribution of each diagram for the jet function forward scattering ME can be expressed as a linear combination of dimensionally regularized ($d=4-2\varepsilon$) scalar integrals of the generic form
\begin{align}
\int&\!\frac{\dd^d k\,\dd^d\ell\;\; [-\bar n\cdot k]^{-c_1}[-\bar n\cdot \ell]^{-c_2}}{[-k^2]^{a_1}[-\ell^2]^{a_2}[-(k-\ell)^2]^{a_3}[-(p+k)^2+m^2]^{b_1}[-(p+\ell)^2+m^2]^{b_2}[-(p+k+\ell)^2+m^2]^{b_3}} \nn\\
&\!\!\!\!\!\!\!\!= -\pi^d(-s)^{d-a_1-a_2-a_3-b_1-b_2-b_3}(\bar n\cdot p)^{-c_1-c_2}\mathcal I(-m^2/p^2;a_1,a_2,a_3,b_1,b_2,b_3,c_1,c_2)
 \,, 
 \label{eq:integral1}
\end{align}
where all propagator denominators have the usual $-i0$ term
and the prescription $s\to s+i0$ for the jet virtuality $s= p^2-m^2 \ge 0$ is understood.
In the second line we pulled out the conventional factor of $i\pi^{d/2}$ per loop integral.
We also pulled out the factor $(\bar n\cdot p)^{-c_1-c_2}$ as this dependence is fixed by the behavior of the integral under the rescaling of $\bn$ or Lorentz covariance.
The power of $(-s)$ is then determined by choosing $\mathcal I$ to be dimensionless.
This factor encodes the full $s$ dependence in the massless limit ($s\to p^2$), where $\mathcal I$ becomes a function of $\eps$ only.

The set of propagator denominators in \eq{integral1} does not represent a linearly independent basis. 
Either $a_3$ or $b_3$ can be rendered zero by partial fraction decomposition.
We therefore can arrange all integrals into two integral families: family A, where $b_3=0$, and family B, where $a_3=0$.
Integrals where $a_3=b_3=0$ are for convenience assigned to family A.
Some family B integrals with only two massive propagators can be mapped by loop momentum shifts to family A.
Diagrams (g) and (c) contain the top sector integrals of family A and B, respectively  (i.e.\ where all propagator powers are positive).
We distinguish two types of diagrams:
the ``planar'' diagrams (e)-(q), which contain the leading color contributions, and the ``nonplanar'' diagrams (a)-(d) which are proportional to $C_F^2-C_F C_A/2$ and therefore color-suppressed. The planar diagrams only involve family A integrals, while
the ``nonplanar'' diagrams contain integrals of family A and B. Diagrams (r) are scaleless and vanish.

The classification in family A and B integrals is useful for two reasons.
The first is that they are closed under integration by parts (IBP) reductions and require different sets of IBP identities.
The second is related to the different mathematical properties of family A and B integrals.
Family A integrals have at most two massive propagators and involve at most harmonic polylogarithms in the $\varepsilon$ expanded results. In contrast, family B integrals can contain elliptic functions in the $\varepsilon$ expanded expressions, which makes their evaluation more complicated.
This is connected to the fact that they admit a triple massive-particle cut.
Moreover, we observed that some family B integrals (including cases with only two massive propagators) are ill-defined in pure dimensional regularization due to rapidity singularities even though the Feynman diagrams are individually rapidity finite. Further, using IBP reductions on rapidity finite family B integrals can lead to rapidity singular integrals.
Therefore, the computations of family B integrals in general require a rapidity regulator.

Let us briefly illustrate this issue for the following rapidity-finite family B integral
\begin{align}
&\int\!\!\frac{\dd^d k\,\dd^d\ell}{[-\ell^2][-(p+k)^2+m^2][-(p+\ell)^2+m^2][-(p+k+\ell)^2+m^2][-\bar n\cdot k]^{1+\eta}} \nn\\
&\quad =-\pi^d(-s)^{d-5}(\bar n\cdot p)^{-1-\eta}\,\mathcal I(-y;0,1,0,1,1,1,1+\eta,0)\,,
\end{align}
where we introduced an analytic rapidity regulator $\eta$ for the $k$-integration, see e.g.\ \rcites{Becher:2011dz,Chiu:2011qc,Chiu:2012ir}.
Applying the IBP relation
\begin{align}
 0={}& 
 \Big( a_1 - b_1 + a_1 \mathbf{a_1^+} - \frac{1+y}{1-y} b_1 \mathbf{b_1^+} - 2 \frac{y}{1-y} b_3 \mathbf{b_3^+} + c_1 \mathbf{c_1^+} + b_1 \mathbf{b_1^+} \mathbf{a_1^-} + b_3 \mathbf{b_3^+} \mathbf{a_1^-}\nonumber\\
 & + b_3 \mathbf{b_3^+} \mathbf{a_2^-}- a_1 \mathbf{a_1^+} \mathbf{b_1^-} - b_3 \mathbf{b_3^+} \mathbf{b_1^-} - b_3 \mathbf{b_3^+} \mathbf{b_2^-} \Big)\, \mathcal{I}
 \,,
 \label{eq:IBPrap}
\end{align}
where the bold letters represent operators that increase ($+$) or decrease ($-$) the corresponding propagator powers in $\mathcal{I}$,
we obtain
\begin{align}
 &\mathcal I(-y;0,1,0,1,1,1,1+\eta,0)= (1+\eta)\, \mathcal I(-y;0,1,0,1,1,1,2+\eta,0)\nonumber\\
 &\quad - \mathcal I(-y;0,1,0,0,1,2,1+\eta,0) + \mathcal I(-y;0,0,0,1,1,2,1+\eta,0)\nonumber\\
 &\quad - \frac{1+y}{1-y}\, \mathcal I(-y;0,1,0,2,1,1,1+\eta,0) - 2 \frac{y}{1-y} \,\mathcal I(-y;0,1,0,1,1,2,1+\eta,0) \nn\\ 
 &\quad+ \mathcal I(-y;-1,1,0,2,1,1,1+\eta,0) 
 + \mathcal I(-y;-1,1,0,1,1,2,1+\eta,0)\nonumber\\
 &\quad- \mathcal I(-y;0,1,0,1,0,2,1+\eta,0)
 \,.
 \label{eq:IBPexample}
\end{align}
The first three integrals on the right hand side turn out to be individually rapidity divergent, i.e.\ they have $1/\eta$ poles as can e.g.\ be checked numerically with the sector decomposition program \texttt{pySecDec}~\cite{Borowka:2017esm}.
These $1/\eta$ poles cancel in the sum.
Without a rapidity regulator, i.e.\ naively setting $\eta=0$ in \eq{IBPexample} literally, however, gives an incorrect relation because the term $\eta\,\mathcal I(-y;0,1,0,1,1,1,2+\eta,0)$ leads to finite contributions which are missed when the rapidity regularization is introduced after using the IBP relation. An analogous issue also arises when using a dimensionful rapidity regulator such as the
$\Delta$ regulator proposed in \rcite{Chiu:2009yx}.
We stress that the IBP reduction with a rapidity regulator (no matter of what kind) in general leads to a substantially increased number of master integrals. 
Moreover, these are typically more complicated to compute than comparable integrals where this kind of problem does not arise and an additional rapidity regularization is not necessary.

Interestingly, the IBP reduction of family A integrals does not give rise to spurious rapidity divergences and works consistently without any rapidity regulator in the standard way.
This is (currently) an empirical observation as we are lacking a simple systematic criterion that would allow us to identify when the problem of spurious rapidity divergences arises and to possibly avoid or treat
them in an automatized way.
We therefore explicitly verified all family A integral reductions used in our computations numerically in order to dispel any doubt about their validity.
In this context the known (or easily computed) massless primary quark limit of each individual diagram provides a valuable cross check of our calculations.  

The calculation of the planar diagrams  (e)-(q) can thus be performed with standard modern multi-loop technology as explained in \sec{topoa}.
For the  nonplanar diagrams (a)-(d), the issue of the spurious rapidity divergences and the elliptic nature of certain family B integrals forced us to use different and less uniform methods as explained in \sec{topob}.  
As a general strategy, we analytically compute the two-loop master integrals for the jet field correlator to the required order in $\varepsilon$ first, insert them into the different diagrams, 
and finally take the imaginary part of the total contribution to the jet field correlator according to \eq{JmeQCD} for each color factor.
The only exceptions are diagrams (b) and (c), for which we take a different semi-numerical approach, as explained in \sec{topob}.
We renormalize the resulting jet-function MEs $\Jme_Q^{(n_\ell+1)}\big|_{\cancel{C_F T_F}}$ according to \eq{Jren}, i.e.\ we absorb {\it all} divergent terms into the $Z$-factor, and we use the pole mass scheme for the quark mass.
The necessary counterterms for the pole mass $m$ (and the $\msb$ coupling $\alpha^{(n_\ell+1)}_s$) can e.g.\ be found in \rcite{Melnikov:2000zc}. We note that in the pole mass scheme derivatives of the distributions $\delta(s)$ and $\mathcal L_{n}(s)$ do not arise in the final result.

\subsubsection{Planar Diagrams}
\label{sec:topoa}

\begin{figure}[t]
 \centering
\includegraphics[width=0.22\textwidth]{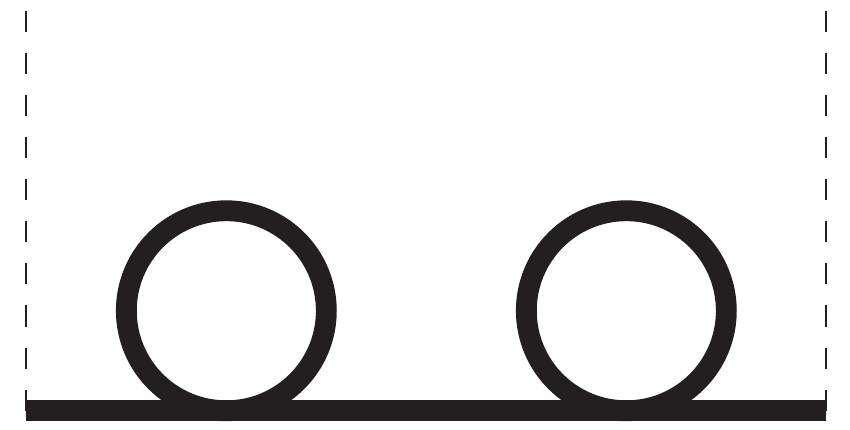}
\put(-54,38){$M_1$}\hspace{7pt}
\includegraphics[width=0.22\textwidth]{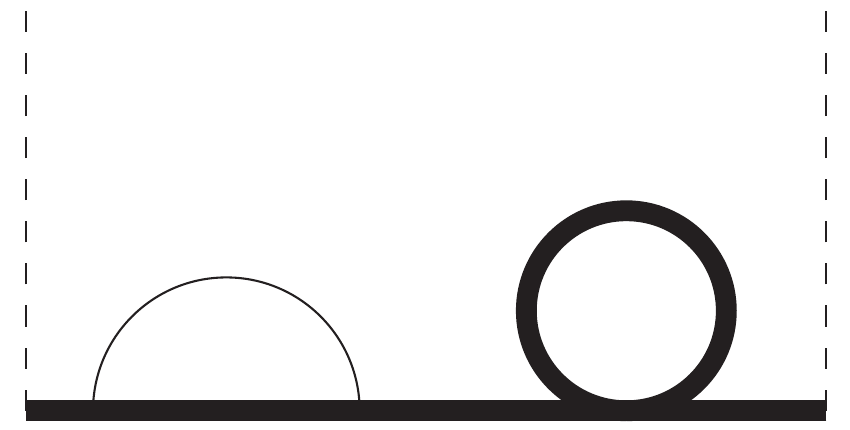}
\put(-54,38){$M_2$}\hspace{7pt}
\includegraphics[width=0.22\textwidth]{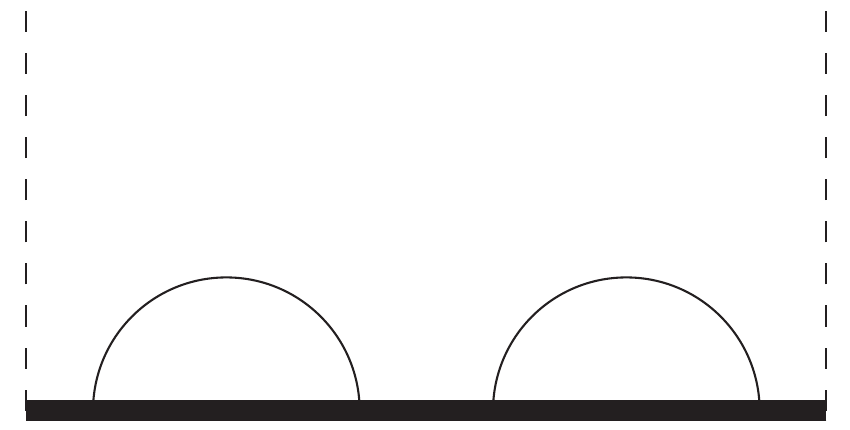}
\put(-54,38){$M_3$}\hspace{7pt}
\includegraphics[width=0.22\textwidth]{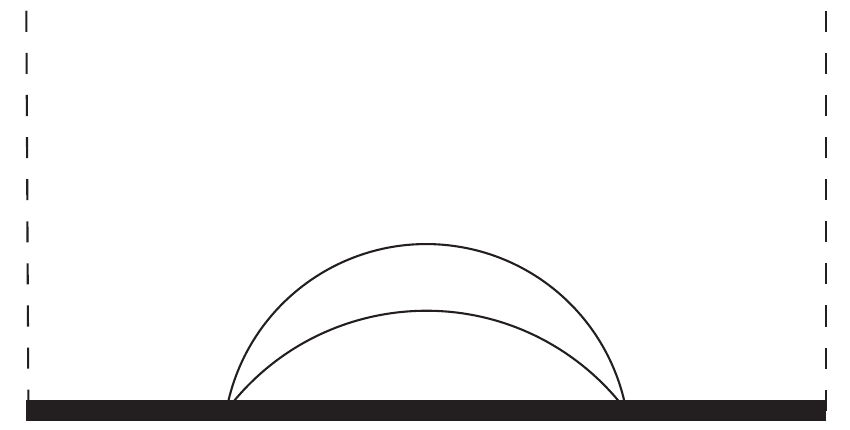}
\put(-54,38){$M_4$}\\[7pt]
\includegraphics[width=0.22\textwidth]{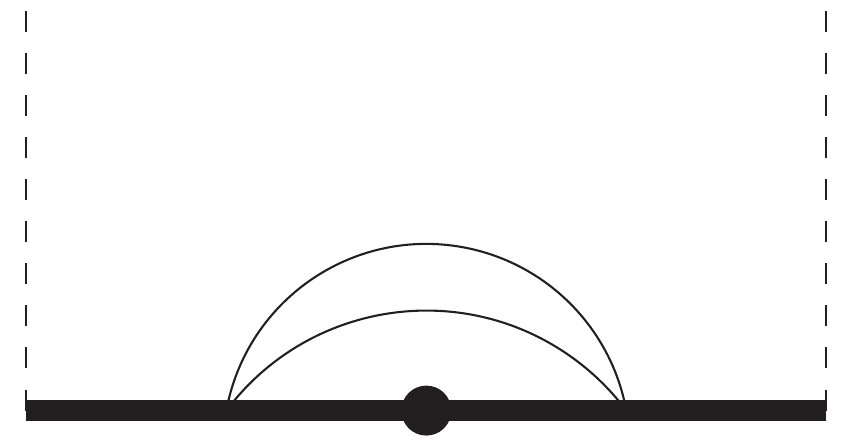}
\put(-54,38){$M_5$}\hspace{7pt}
\includegraphics[width=0.22\textwidth]{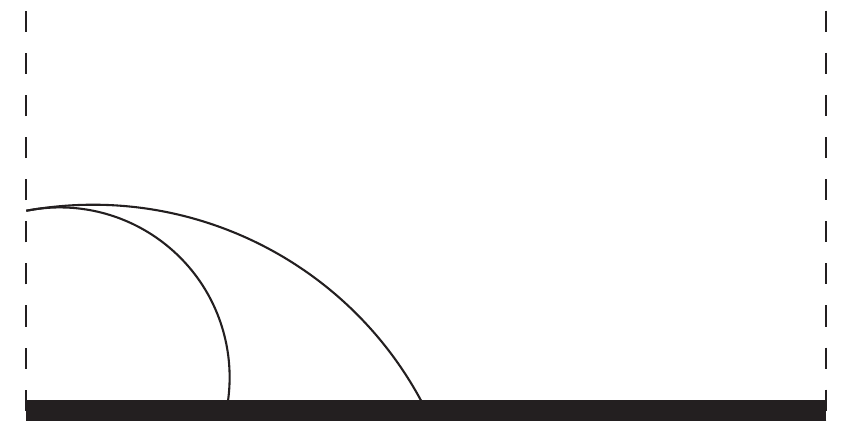}
\put(-54,38){$M_6$}\hspace{7pt}
\includegraphics[width=0.22\textwidth]{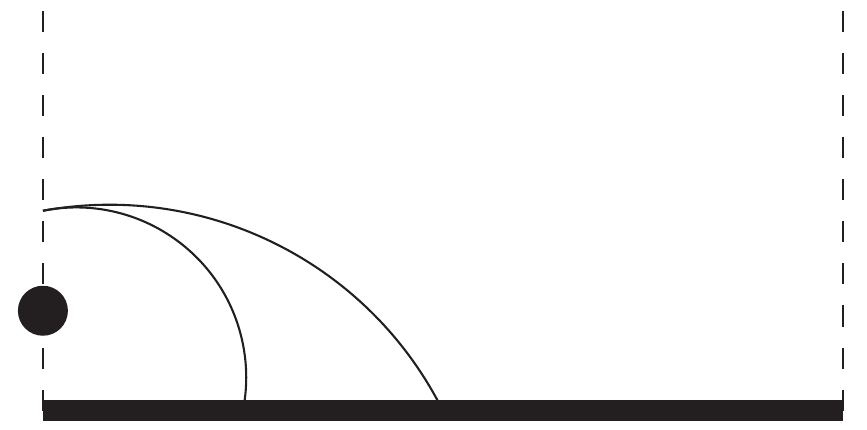}
\put(-54,38){$M_7$}\hspace{7pt}
\caption{The seven master integrals of family A. 
 Bold solid lines represent massive propagators, thin solid lines represent massless propagators and dashed lines represent (lightlike) Wilson line propagators. A dot on a line means that the associated propagator is squared.
 Note that the offshell quark propagators $\propto 1/s$ are not included in the definition of the master integrals $M_i$ and are only drawn here for illustration.
 The corresponding solutions for the $\mathcal I_i$ according to \eq{integral1} are given in \app{MI}.
 \label{fig:mis1}}
 \end{figure}

The calculation of the  planar diagrams  (e)-(q) (including only massless partonic loops in the gluon self-energy) is rather straightforward.
They involve $\ord{100}$ family A integrals.
We reduced this set of integrals to the seven master integrals depicted in \fig{mis1} using the IBP program \texttt{FIRE5}~\cite{Smirnov:2014hma}.

The master integrals $M_1$-$M_5$ can be computed using Feynman parameters for arbitrary $\varepsilon$ in terms of hypergeometric functions.
They can be expanded in terms of harmonic polylogarithms (HPLs)~\cite{Gehrmann:2000zt,Remiddi:1999ew} with the help of the \texttt{Mathematica} package \texttt{HypExp}~\cite{Huber:2005yg}.
We also used the \texttt{Mathematica} package \texttt{HPL}~\cite{Maitre:2005uu} to exploit relations among the HPLs for the simplification of expressions or to make their singular behavior for $s \to 0$ explicit, see below.

For $M_6$, $M_7$ we used the method of differential equations~\cite{Kotikov:1990kg,Remiddi:1997ny,Gehrmann:1999as}.
To this end we take their derivatives with respect to $y = m^2/p^2$. The result can be expressed via IBP reduction in terms of $M_1$-$M_7$.
The coupled system of seven homogeneous first-order differential equations can also be rewritten in terms of a coupled system of two first-order differential equations for $M_6$, $M_7$ with linear combinations of $M_1$-$M_5$ as (known) inhomogeneous terms:
\begin{align}
\frac{\partial M_6}{\partial y} ={}& - \frac{(1 - \varepsilon)^2 (1-y)}{2\varepsilon  y^2} M_1 + \frac{(1 - 2 \varepsilon) (1 - \varepsilon)}{2 \varepsilon y^2} M_2 - \frac{\left(2 - 7\varepsilon + 6 \varepsilon ^2\right) (1 - \varepsilon (5-y))}{4 \varepsilon ^2 (1-y) y} M_4\nonumber\\
&- \frac{(1 - 2 \varepsilon) \left((2 - y^2) \varepsilon - (2-7 \varepsilon )y\right)}{4 \varepsilon ^2 (1-y)^2 y} M_5 - \frac{\varepsilon (3 - y)}{y(1-y)} M_6 + \frac{3}{2 y} M_7\,,\\
\frac{\partial M_7}{\partial y} ={}& \frac{1 - 3 \varepsilon + 2 \varepsilon^2}{y^2} M_2 + \frac{2 - 7\varepsilon + 6 \varepsilon^2}{y(1-y)} M_4 - \frac{(1 - 2 \varepsilon) (1+y)}{y(1-y)^2} M_5\nonumber\\
& - \frac{4 \varepsilon^2}{y(1-y)} M_6 + \frac{2 \varepsilon}{y(1-y)} M_7\, .
\end{align}
These can be decoupled into two separate second-order inhomogeneous differential equations for $M_6$ and $M_7$. Upon expansion in $\varepsilon$ these can be iteratively rewritten in second-order inhomogeneous differential equations where only derivatives of expansion coefficients of $M_6$ or $M_7$ arise. These differential equations happen to have a particularly simple form because these derivatives appear in a combination, such that upon integration the resulting differential equations only involve first derivatives of the expansion coefficients.
Another integration w.r.t.\ $y$ then yields the result depending on two integration constants (boundary values) that can be fixed by taking the (massless) limit $y \to 0$, where the integrals are known~\cite{Becher:2006qw}.
We give the explicit solutions for $M_1$-$M_7$ in \app{MI}.
We checked all of them numerically to the required order in $\varepsilon$ using the Sector Decomposition codes \texttt{FIESTA4} \cite{Smirnov:2015mct} and \texttt{pySecDec} \cite{Borowka:2017esm}.
 
In order to take the imaginary part of the planar contributions we proceed as follows.
Defining $f(s)$ as the contribution of the planar diagrams at the correlator level 
we want to determine
\begin{align}
\mathrm{Im}[f(s+i0)] = \frac1{2i} \mathrm{Disc}[f(s)] = \frac1{2i} \lim_{\beta\to0} \big[f(s+i \beta)-f(s-i \beta)\big]
\,.
\label{eq:ImDisc}
\end{align}
The result involves the distributions $\delta(s)$ and $\mathcal L_n(s)$ and their derivatives prior to pole mass renormalization.
(These derivatives only arise from diagrams that contain quark self-energy subdiagrams.)
In practice we compute \eq{ImDisc} for $s>0$ first.
We write the HPLs from the solutions of the master integrals with the help of the \texttt{HPL} package in a form, where the branch cuts  are made explicit as $\log^n(-s/\mu^2)$ terms.
We can now easily take the imaginary part of these terms according to \eq{ImDisc}. The remaining terms are regular in $s$ and thus do not contribute.
In the next step we promote the terms $\log^n(s/\mu^2)/s$ and $[n \log^{n-1}(s/\mu^2)-\log^n(s/\mu^2)]/s^2$ for $s>0$ to the corresponding plus distributions $\mathcal L_{n}(s)$ and their derivatives, respectively.
We fix the coefficient of the $\delta(s)$ term by evaluating the imaginary part of the integral of $f(s+i0)$ over a small region around $s=0$:
\begin{align}
\label{eq:lambdazero}
 \mathrm{Im} \bigg[ \int_{-\Lambda_1}^{\Lambda_2} \!\! \dd s\; f(s+i0) \bigg]\quad \text{with}\quad
\Lambda_1, \Lambda_2 >0 \,.
\end{align}
The result is independent of $\Lambda_1$, as the imaginary part of $f(s+i0)$ does not have support for $s<0$.
For $\Lambda_2\to 0$ the result involves a constant and terms $\log^k(\Lambda_2)/\Lambda_2^m$ which match the structure of the distributions $\mathcal L_{n}(s)$ and their derivatives.
The constant term determines the coefficient of $\delta(s)$ in $\mathrm{Im}[f(s+i0)]$.
The coefficients of first and second derivatives of $\delta(s)$ are determined in an analogous way considering 
\eq{lambdazero} with additional weight functions $s$ and $s^2$, respectively.
Upon imposing pole mass renormalization all derivatives of $\delta(s)$ and  $\mathcal L_{n}(s)$ consistently cancel, which provides an important cross check.

\subsubsection{Nonplanar Diagrams}
\label{sec:topob}

For the calculation of the nonplanar diagrams (a) and (d) we proceed in the same way as described for the planar diagrams in the previous subsection.
Regarding the IBP reductions, no rapidity regulator is required for diagram (a), because it does not involve any Wilson line propagators.  
In the reduction of diagram (d) issues related to spurious rapidity divergences may in principle arise.
However, we were able to exclude by hand IBP relations such as \eq{IBPrap}, which generate (possibly ill-defined) rapidity divergent integrals at intermediate steps.
Interestingly, diagram (d) then happens to reduce to already known family A master integrals.
We explicitly checked the IBP reduction numerically using \texttt{pySecDec}~\cite{Borowka:2017esm} together with our analytical results for the master integrals.

\begin{figure}[t]
 \centering
 \parbox{3.5cm}{\includegraphics[width=0.22\textwidth]{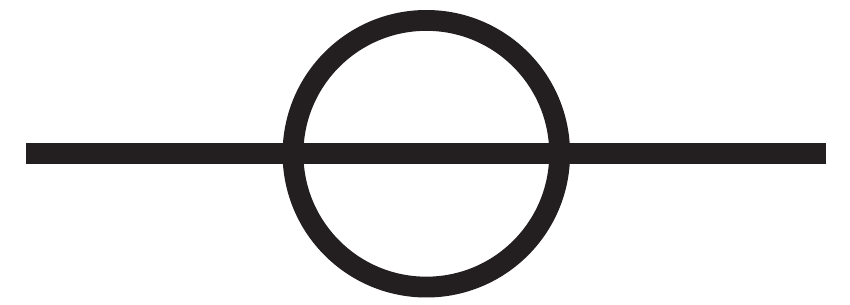}\\ \centering $M_8$}\hspace{7pt}
 \parbox{3.5cm}{\raisebox{1pt}{\includegraphics[width=0.22\textwidth]{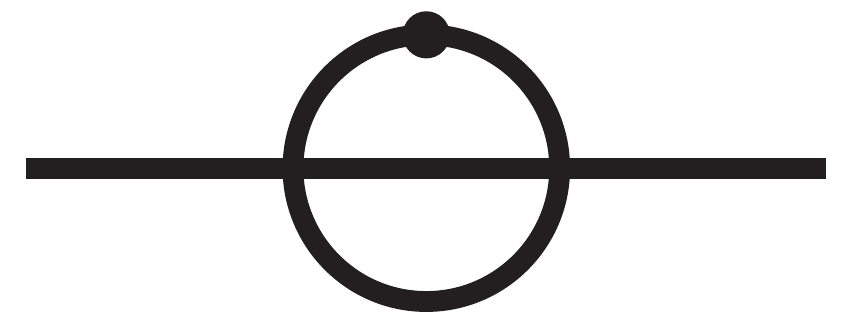}}\\ \centering $M_9$}\hspace{7pt}
 \parbox{3.5cm}{\raisebox{-0.5pt}{\includegraphics[width=0.22\textwidth]{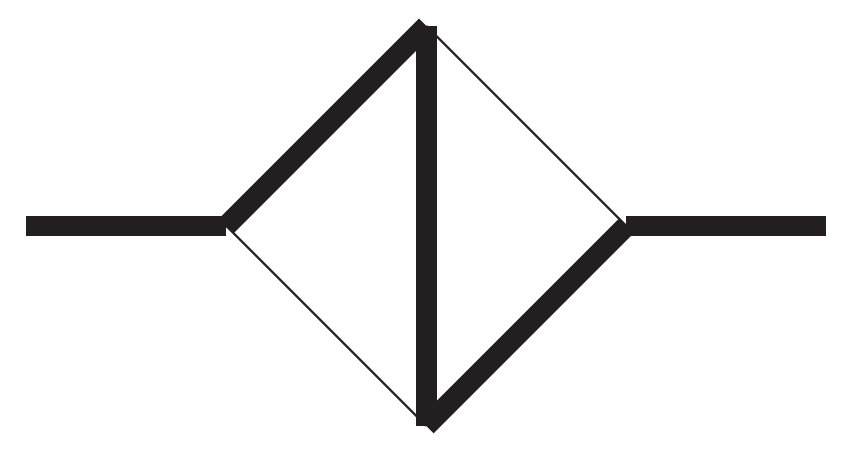}}\\ \centering $M_{10}$}
 \caption{Master integrals of family B contributing to diagram (a).
 Bold solid lines represent massive propagators, thin solid lines represent massless propagators. A dot on a line means that the associated propagator is squared.
 The solutions for these offshell sunrise and kite type integrals can be found in the literature.
 We give the results for their imaginary parts and small $s$ expansion necessary for our calculation in \app{MI}.
 \label{fig:topoBmis}}
\end{figure}

For diagram (a) family B master integrals cannot be avoided and 
a suitable basis for them is depicted in \fig{topoBmis}.
It contains so-called sunrise and kite integrals for which generic results can be found in the literature.  All three master integrals admit a triple massive particle cut, which is reflected by the $\Theta(p^2-(3m)^2)=\Theta(1-9y)$ terms in their imaginary part, and evaluate to (integrals of) elliptic functions. The master integrals contribute to the jet field correlator with prefactors proportional to $(-s-i0)^{-2\varepsilon}$ and upon $\varepsilon$-expansion in practice we only need the imaginary part of the combination $(-s-i0)^{-2\varepsilon}M_i$ when $s>0$. To determine the coefficients of the $\delta(s)$ and its derivatives we additionally need the first few terms in the expansion of the master integrals in $s = p^2 \bar{y} = p^2(1-y)$. All these results are given in \app{MI}. They are straightforwardly determined or directly taken from \rcites{Laporta:2004rb,Bauberger:1994by,Remiddi:2016gno}.

To compute diagrams (b) and (c) we are forced to take a different approach.
The reason is that diagram (b) and (c) already involve rapidity divergent family B integrals before any reduction.
These rapidity divergences cancel in the sum for each diagram. 
An IBP reduction method without additional regulator, as used for diagram (d) therefore seems impractical.
Using a rapidity regulator, on the other hand, results in new types of master integrals 
which involve three massive propagators, several massless propagators, as well as regularized Wilson line propagators.
Such integrals appear too difficult to be solved with the tools at hand.
We therefore decided to refrain from using IBP reductions and compute them directly using numerical methods and exploiting consistency conditions on their analytical structure as explained in the following.\\

\begin{samepage}
\noindent {\it Numerical calculation of the contributions from diagrams (b) and (c)}\\

In a first step we use the sector decomposition code \texttt{pySecDec} \cite{Borowka:2017esm} to obtain numerical results for the real and imaginary parts of the contributions of diagrams (b) and (c) to the jet field correlator expanded in $\varepsilon$ up to 
$\ord{\varepsilon^0}$.
To this end we set $\mu^2=s$, separately write the integrands of the two diagrams in Feynman parameter representation and solve as many integrals analytically as possible. For both diagrams four integrations over rational functions of the parameters are remaining and are carried out with \texttt{pySecDec}.
We furthermore multiply the integrands by $p^2(1-y)^2=s^2/p^2$ to render them dimensionless and to damp the singular behavior in the (bHQET) limit $y\to1$. 
We then obtained numerical results at $100$ points in the physical region $0\le y<1$.
From these numerical data points we verified that the coefficients of the $1/\eps$ poles together with the divergences of the already computed diagrams correctly reproduce within numerical uncertainties (which are typically at the sub-percent level) the divergence structure related to the renormalization $Z$-factor in \eq{ZJmassless}. This provides an important cross check for our approach, in particular confirms the correctness concerning the $\mu$ dependence.
\end{samepage}

Due to the multiplication of the factor  $p^2(1-y)^2=s^2/p^2$ we do not have direct numerical access to the coefficient of the
$\delta(s)$ term. We will come back to this point at the end of this subsection. However, analytic information on the contribution of diagrams (b) and (c) to the jet-function ME is known in various kinematic limits. Using IBP reduction\footnote{Note that in the massless case no rapidity divergences arise.} and the master integrals given in \cite{Becher:2006qw}, it is straightforward to compute the diagrams in the massless limit. Upon taking the imaginary part the result, including mirror diagrams, reads
\begin{align}
\Jme^{\mathrm{b}+\mathrm{c}}_\mathrm{fin}(p^2,0) ={}& a_s^2 \left(C_F^2 - \frac{C_F C_A}{2}\right)\Bigg[\left(-\frac{10 \zeta_3}{3}-68+2 \pi^2+\frac{\pi^4}{6}\right) \delta(p^2)\notag\\
&+\left(-40 \zeta_3+24+\frac{4 \pi^2}{3}\right) \mathcal L_0(p^2)+\frac{16}{3} \pi^2 \mathcal L_1(p^2)-8 \mathcal L_2(p^2)\Bigg]
\,, \label{eq:JBCml}
\end{align}
where the $1/\eps$ divergences have been subtracted.
We can also extract the corresponding expression in the bHQET limit ($y\to1$) from \eq{JBJmf}.
Combining the results for the two-loop bHQET jet function $J_B^{(n_\ell)}$~\cite{Jain:2008gb} and the collinear mass mode matching coefficient $H_{m,n}$~\cite{Hoang:2015vua} with our results for the other diagrams, we find
\begin{align}
\Jme^{\mathrm{b}+\mathrm{c}}_\mathrm{fin}(p^2,m^2) ={}& a_s^2\left(C_F^2 - \frac{C_F C_A}{2}\right)\Bigg\{ \delta(s) \Bigg[\left(-72 \zeta_3+24 +18 \pi^2\right) L_m\notag\\
&\quad+\left(\frac{8 \pi^2}{3}-16 \right) L_m^2-\frac{154 \zeta_3}{3}-\frac{\pi^4}{45}+\frac{50 \pi^2}{3}-68 -24 \pi^2 \log 2\Bigg]\notag\displaybreak[1]\\
&+\left[16 L_m^2+\left(32 -\frac{16 \pi^2}{3}\right) L_m+64 \zeta_3-18 \pi^2 \right] \mathcal L_0(s)\notag\\
&+\left[-64 L_m+\frac{32 \pi^2}{3}-32 \right] \mathcal L_1(s)+40 \mathcal L_2(s) + \mathcal O\!\left(m^{-2}\right)\Bigg\}
\,.
\label{eq:JBChq}
\end{align}
Note that this expression contains all distributional terms in $\Jme^{\mathrm{b}+\mathrm{c}}(p^2,m^2)$.

\begin{figure}[t]
\centering
\includegraphics[width=0.49\textwidth]{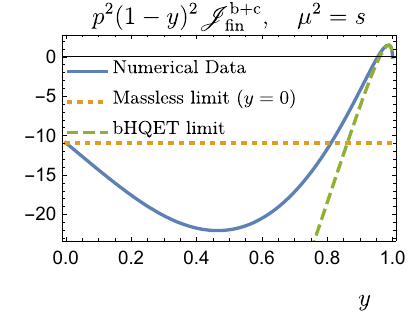}
\includegraphics[width=0.49\textwidth]{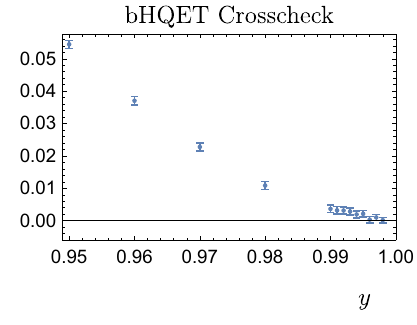}
\caption{\textit{Left:} Contribution of diagrams (b) and (c) to the jet-function ME finite part multiplied with the singularity-damping factor $p^2(1-y)^2$ as a function of $y \equiv m^2/p^2$ for $\mu^2=s$. 
The numerical data from \texttt{pySecDec} is shown as the solid blue line. 
The associated numerical uncertainty is too small to be visible in the plot.
The corresponding heavy quark (bHQET) limit, according to \eq{JBChq}, is shown as dashed green line. 
The dotted orange line represents the value of the jet-function ME times damping factor in the massless limit ($y=0$), i.e.\ to the $\mathcal L_0(p^2)$ term in \eq{JBCml}.
\textit{Right:} Difference between numerical data points and analytic bHQET limit, normalized to the massless limit, for $\mu^2=s$. The error bars indicate the uncertainty of the numerical integration using \texttt{pySecDec}.
\label{fig:numdata}}
\end{figure}
In the left panel of \fig{numdata} we plot our (interpolated) numerical result for $p^2(1-y)^2 \Jme^{\mathrm{b}+\mathrm{c}}_\mathrm{fin}$  (solid blue line) and the corresponding analytical expression for $s \ll m^2$ according to \eq{JBChq} (dashed green line) for $\mu^2=s$ in the range $0<y<1$.
Note that for $s>0$ and $\mu^2=s$ only the $\mathcal L_0(s)$ term in \eq{JBChq} survives.
With the factor $p^2(1-y)^2$ and for $\mu^2=s\to p^2>0$ the massless result in \eq{JBCml} collapses to a single value, namely the coefficient of the $\mathcal L_0(p^2)$ distribution.
This is indicated by the dotted orange line which agrees with the numerical data point at $y=0$ within numerical uncertainty.
We also observe that in the limit $y\to 1$ the numerical data and the bHQET curve correctly approach each other nicely.
To better illustrate this we show in the right panel of \fig{numdata} the difference between numerical data points and bHQET result normalized to the $y=0$ value (dotted orange line in the left panel) in the range $0.95<y<1$ including the numerical uncertainties as quoted by \texttt{pySecDec}.
We clearly see that for $y \gtrsim 0.996$ the difference is zero within the numerical uncertainties.
This represents a strong cross check of our combined numerical and analytical evaluation.

In order to obtain a practical fit function that parametrizes our numerical results
respecting at the same time the analytic constraints from the bHQET and massless limits we proceed as follows.
We start with an ansatz for $\Jme^{\mathrm{b}+\mathrm{c}}_\mathrm{fin}(p^2,m^2,\mu^2=s)$ valid for $s>0$ of the form 
\begin{align}
\label{eq:ansatz}
s\, \Jme^{\mathrm{b}+\mathrm{c}}_\mathrm{fin,ansatz}(p^2,m^2,\mu^2\!=\!s) = a_s^2\left(C_F^2 - \frac{C_F C_A}{2}\right) \sum_{m=0}^2 \sum_{n=0}^3 c_{mn} (1-y)^m \log^n(1-y)
\,.
\end{align}
Note that we multiplied $\Jme^{\mathrm{b}+\mathrm{c}}_\mathrm{fin}(p^2,m^2)$ on the LHS with the factor $s$.
We can now fix\footnote{We note that a numerical fit of the coefficients in \eq{JBChqcon} from the numerical data points provides results that agree with the values shown within numerical uncertainties.}
\begin{equation}
c_{00} = -18 \pi^2 + 64 \zeta_3,\qquad c_{01} = \frac{16}{3} (\pi^2 - 6),\qquad c_{02} = 16,\qquad c_{03} = 0,
\label{eq:JBChqcon}
\end{equation}
by comparing \eq{ansatz} in the (bHQET) limit $y \to 1$ to the coefficient of $\mathcal L_0(s)$ in \eq{JBChq}.
Similarly, we can take the massless limit ($y\to 0$) of \eq{ansatz} and compare it to the coefficient of $\mathcal L_0(p^2)$ in \eq{JBCml}, which fixes
\begin{equation}
c_{10} = -c_{00} - c_{20} + 24 + \frac{4}{3}\pi^2 - 40 \zeta_3
\,.
\label{eq:DBWmlcon}
\end{equation}

It is straightforward to generalize the ansatz function to arbitrary values of $\mu^2 \neq s$ using 
the renormalization $Z$-factor in \eq{ZJmassless}
and our results from the other diagrams.
The full distributional ansatz $\Jme^{\mathrm{b}+\mathrm{c}}_\mathrm{fin,ansatz}(p^2,m^2,\mu^2)$ for arbitrary $\mu$ and $s\ge 0$ is then obtained by promoting the $\log^n(s/\mu^2)/s$ terms to $\mathcal L_n(s)$ distributions and adding the $\delta(s)$ term of \eq{JBChq}.
To also include the nontrivial analytic information from the $\delta(p^2)$ term in the massless quark result of \eq{JBCml}, we integrate
the full distributional ansatz over a finite interval around $s=0$, take the limit $m \to 0$ and compare to the corresponding integral of \eq{JBCml}.
In this way we find the additional constraint
\begin{align}
\label{eq:delConstr}
c_{20} ={}& -\frac{\pi^2}{6} c_{01}+2 \zeta_3 c_{02}-\frac{\pi^4}{15} c_{03}-\frac{\pi^2}{6} c_{11}+2 \zeta_3 c_{12}-\frac{\pi^4}{15} c_{13}+\left(1-\frac{\pi^2}{6}\right) c_{21}+2 (\zeta_3-1) c_{22}\notag\\
&+\left(6-\frac{\pi^4}{15}\right) c_{23}-56 \zeta_3-\frac{17 \pi^4}{90}-8 \pi^2 (3\log 2-2)
\,.
\end{align}
From the same computation (using arbitrary integration limits) one obtains analogous constraints for  $c_{01}$, $c_{02}$, and $c_{03}$ 
which fully agree with  \eq{JBChqcon}. This represents a nontrivial check of our values for $c_{01}$, $c_{02}$, and $c_{03}$ and implies  consistency with the $L_m$ dependent terms in the $\delta(s)$ coefficient of \eq{JBChq}.

\begin{figure}[t]
	\centering
	\includegraphics[width=0.5\textwidth]{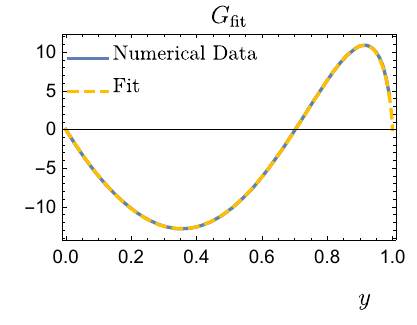}
	\caption{
		Fit function $G_\mathrm{fit}$ (dashed orange line) as given in \eq{Gfit} compared to the (interpolated) numerical data from \texttt{pySecDec} (solid blue line). The numerical uncertainties are too small to be visible in the plot.
		\label{fig:Gfit}}
\end{figure}

Furthermore, one can argue on physical grounds that
\begin{align}
c_{13} = 0\,,\label{eq:c130}
\end{align}
because we expect the first subleading bHQET power correction at two loops to have a similar logarithmic structure as \eq{JBChq}, 
i.e.\ is should not contain any cubic logarithmic terms.\footnote{Analogously, one could argue that there is no $c_{23}$ term. Unlike the $c_{13}$ term it is however regular for $y \to 1$ and we will use it to model other regular terms in our fit.}
A dedicated numerical analysis in the bHQET regime confirms \eq{c130} and moreover yields
\begin{equation}
c_{12} = 1.0\,.
\label{eq:c12}
\end{equation}
The values for the remaining coefficients in \eq{ansatz} are obtained by a weighted fit to the numerical data from \texttt{pySecDec}. We find
\begin{equation}
c_{11} ={} -41.9008,\qquad c_{21} ={} -20.2744,\qquad c_{22} ={} -10.5870,\qquad c_{23} ={} -7.1277
\,.
\label{eq:cnum}
\end{equation}
With these values the ansatz approximates our numerical data points to a precision better than $0.5\%$, while the uncertainty of the points themselves is at most $1\%$.

To check consistency of our numerical computation with the coefficient of $\delta(s)$ term given in \eq{JBChq}, we performed the same fit as described above, but without including the additional constraint of \eq{delConstr}.
We find that the fitted value for $c_{20}$ satisfies \eq{delConstr} within the numerical uncertainties. Since the massless limit for diagrams (b) and (c) was explicitly computed, this consistency check reliably verifies the analytic input for the $\delta(s)$ term in \eq{JBChq}.

Finally, we separate the analytical part in \eq{ansatz} from the purely numerical part given by the terms associated with the coefficients in \eqs{c12}{cnum}, which we call $G_\mathrm{fit}$, see \eq{Gfit}.
We plot the fit function $G_\mathrm{fit}(y)$ together with the associated numerical results in \fig{Gfit}.
Since the relative contribution of $G_\mathrm{fit}(y)$ to the entire $\ord{\alpha_s^2}$ corrections of the jet-function ME is smaller then $0.5\%$ in the whole physical range, we do not bother to quote uncertainties in \eqs{c12}{cnum}.

\subsection{Secondary Massive Quark Corrections}
\label{sec:secondary}

\begin{figure}[t]
 \centering
 \includegraphics[width=0.22\textwidth]{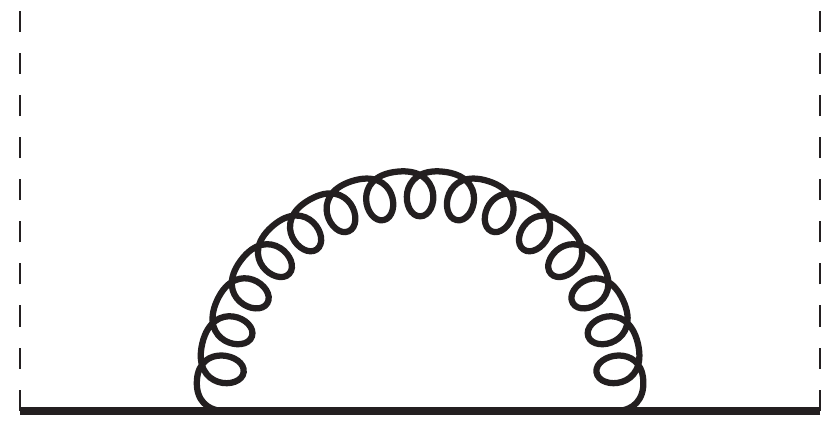}
 \put(-54,38){(o$'$)}\hspace{7pt}
 \includegraphics[width=0.22\textwidth]{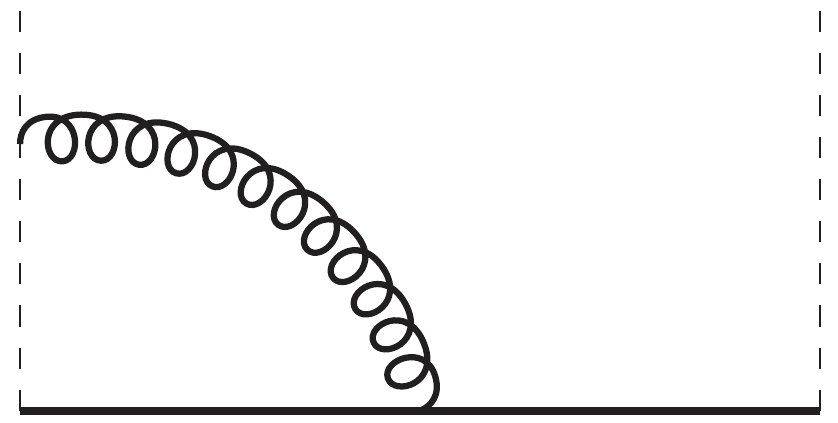}
 \put(-54,38){(p$'$)}\hspace{7pt}
 \includegraphics[width=0.22\textwidth]{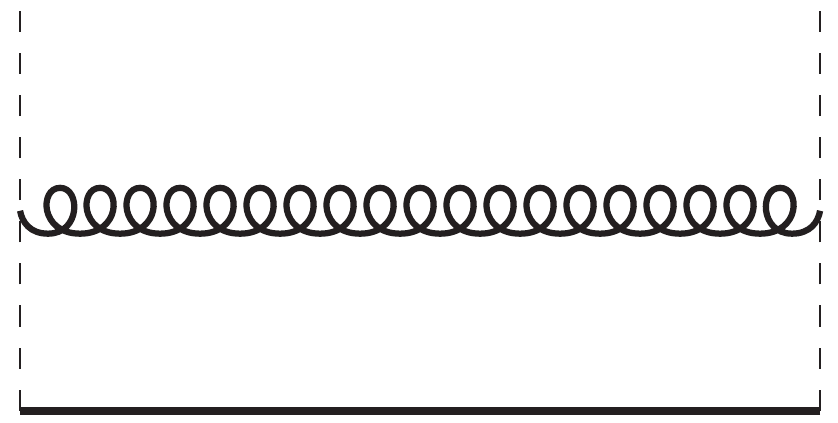}
 \put(-54,38){(q$'$)}\hspace{7pt}
 \includegraphics[width=0.22\textwidth]{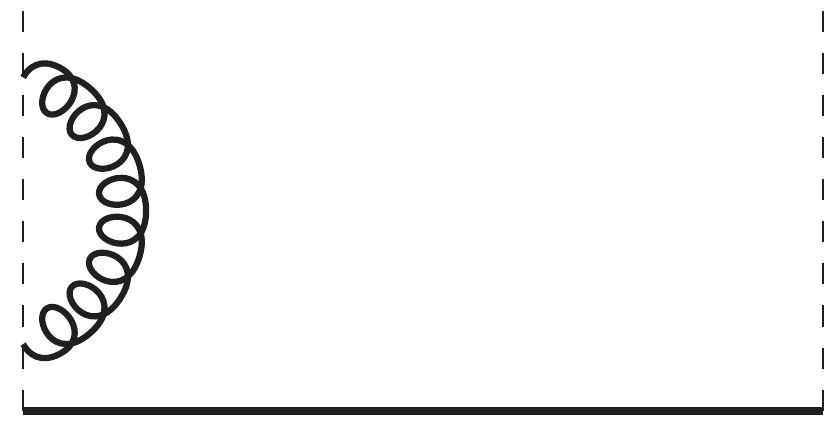}
 \put(-54,38){(r$'$)}\hspace{7pt}
 \caption{One-loop Feynman diagrams with massive gluons required for the calculation of the secondary mass effects, i.e.\ the $\mathcal{O}(\alpha_s^2 C_F T_F)$ terms, in the jet-function ME using the dispersion relation in \eq{dispersion1}. 
 They are associated with the diagrams (o), (p), (q) and (r) in \fig{diagrams1} with a massive quark $Q$ bubble. Left-right mirror graphs are understood.
 \label{fig:secdia}
 }
\end{figure}

In this section we compute the  $\mathcal{O}(\alpha_s^2 C_F T_F)$ secondary massive quark corrections to the jet-function ME $\Jme_f(p^2, m^2)$. 
The relevant Feynman graphs are diagrams (o), (p), (q) and (r) in \fig{diagrams1}, where the vacuum polarization subdiagram is a massive quark $Q$ bubble.
We adopt the approach of \rcites{Gritschacher:2013pha,Pietrulewicz:2014qza} where the corrections are calculated starting from the ${\cal O}(\alpha_s)$ graphs describing radiation of a gluon with mass $M$.
The corresponding one-loop diagrams are shown in \fig{secdia}.
In a second step a dispersion relation to account for the gluon splitting into a pair of a quark and antiquark with mass $m$ is applied. Using the dispersion relation the gluon propagator with the insertion of the bare massive quark vacuum polarization function in Feynman gauge and including the (infinitesimal) gluon mass regulator $\Lambda$ can be written as
\begin{align}
 \frac{-ig^{\mu\rho}}{p^2-\Lambda^2+i0}\;\Pi_{\rho\sigma}(p^2,m^2)\,\frac{-ig^{\sigma\nu}}{p^2-\Lambda^2+i0} ={}& \frac{1}{\pi}\int\limits_{4m^2}^\infty\!\frac{\dd {M}^2}{{M}^2}\,
 \frac{-i(g^{\mu\nu}-\frac{p^\mu p^\nu}{p^2})}{p^2-{M}^2+i 0}\;\mathrm{Im}\left[\Pi({M}^2,m^2)\right]\notag\\
 &-\frac{-i\left(g^{\mu\nu}-\frac{p^\mu p^\nu}{p^2}\right)}{p^2-\Lambda^2 + i0}\;\Pi(0,m^2)\,.
 \label{eq:dispersion1}
\end{align}
Here $\Pi(p^2,m^2)$ is the gluon vacuum polarization function arising from a massive quark bubble defined by
\begin{align}
\Pi_{\mu\nu}^{AB}(p^2,m^2) = -i(p^2 g_{\mu \nu}-p_\mu p_\nu)\Pi(p^2,m^2)\delta^{AB}\equiv\int\dd x^4\, \mathrm e^{ipx}\langle 0|\mathrm T J_\mu^A(x)J_\nu^B(0)|0\rangle \,,
\end{align}
with the vector current $J^A_\mu \equiv i g {\bar Q} T^A \gamma_\mu Q$.
The imaginary part of the one-loop coefficient $\Pi^{(1)}$ in $d=4-2\varepsilon$ dimensions evaluates to ($\tilde\mu \equiv\mu \, e^{\gamma_E/2}(4\pi)^{-1/2}$)
\begin{align}
 \mathrm{Im}\big[\Pi^{(1)}(p^2,m^2)\big]=\Theta \big(p^2\!-\!4m^2 \big) T_F \left(\frac{p^2}{\tilde \mu^2}\right)^{-\varepsilon}\frac{2^{4\varepsilon}\pi^{\frac32+\varepsilon}}{\Gamma(\frac52-\varepsilon)}
 \bigg(\frac{2m^2}{p^2}+1-\varepsilon\bigg)\bigg(1-\frac{4m^2}{p^2}\bigg)^{\!\!\frac12-\varepsilon}.
\end{align}
Note that the Heaviside $\Theta$-function restricts the gluon mass integration in the dispersion relation to $M>2m$. We could therefore safely set $\Lambda=0$ in the first term of \eq{dispersion1}.

Using \eq{dispersion1} for the gluon propagator with a heavy quark loop insertion in \fig{diagrams1} (o), (p), (q) and (r) we see that we effectively need to compute the ${\cal O}(\alpha_s)$ jet-function ME $\Jme_{f}^{(1)}$ with finite gluon mass $M$, i.e.\ the corresponding diagrams in \fig{secdia}. This is needed for the contribution from the first term on the RHS of \eq{dispersion1} (before integration over $M^2$). 
To include the contribution from the second term we also need the result expanded for $M\to\Lambda\ll M,p^2$. 
 
The $p^\mu p^\nu$ terms on the RHS of \eq{dispersion1} can be dropped in our computation
due to gauge invariance of quantum electrodynamics with a massive photon field~\cite{Stueckelberg:1957zz,Feldman:1963zz} and the fact that 
the secondary $\mathcal{O}(\alpha_s^2 C_F T_F)$ massive quark corrections are up to color factors identical in an Abelian gauge theory.
Dropping the $p^\mu p^\nu$ terms in \eq{dispersion1} we immediately see that diagrams (q$'$) and (r$'$) are proportional to $\bar n\cdot\bar n=0$ and thus vanish. 

The contribution of diagram (o$'$) to \eq{JmeQCD} before taking the imaginary part reads ($s\equiv p^2-m^2+i0$)
\begin{align}
\text{(o$'$)} &=-i\frac{4\,\alpha_s C_F\tilde \mu^{2\varepsilon}}{(2\pi)^d (\bn \cdot p)s^2} \int\dd^d k\,\frac{(\bar n\cdot k)(d-2) s + (\bar n\cdot p) (4 m^2 - (d-2) (2 p\cdot k + s))}{[-(p+k)^2+m^2][-k^2+M^2]} \\ 
&=\alpha_s C_F \,\frac{\tilde\mu^{2 \varepsilon}\, \Gamma(\varepsilon)}{2^{1 - 2 \varepsilon} \pi^{2 - \varepsilon}}
\frac{1}{s^2}\,
\int_0^1\dd u\,\frac{(\varepsilon - 1) s (u - 1) + 2 m^2 ((\varepsilon - 1) u - 1)}{({M}^2 (1-u) + u (s (u - 1) + m^2 u))^{\varepsilon}}\,, \nn
\end{align}
where the integration variable $u$ represents a Feynman parameter.
This expression is well-behaved in the large $M$ limit. 
It is convenient to separate the leading term for $M\to\infty$ in $d=4-2 \eps$ dimensions,
\begin{equation}
\label{eq:binf1}
(\text{o}')_{M\to\infty}=\alpha_s C_F \Gamma(\varepsilon)2^{-1 + 2 \varepsilon}\pi^{-2 + \varepsilon}\, \leri{\frac{{M}}{\tilde{\mu}}}^{-2\varepsilon}\frac{2(2 \varepsilon - 3) m^2 + (-1 + \varepsilon)^2 s}{(\varepsilon - 2) (\varepsilon - 1) s^2}\,.
\end{equation}
In the remaining contribution, which vanishes for $M^2\to\infty$, we can safely set $d=4$:
\begin{align}
\label{eq:be0}
(\text{o}')-(\text{o}')_{M\to\infty}&={}-\frac{C_F \alpha_s}{2\pi^2}\,  \frac{1}{s^2}
\int_0^1\dd u\,\left[s (u - 1) + 2 m^2 (1 + u)\right]\\
&\,\,\,\times\left[\log\Big({M}^2 (1-u)\Big) - \log\Big({M}^2 (1-u) + u (s (u - 1) + m^2 u)\Big)\right]+\mathcal O(\varepsilon)
\,.\nn
\end{align}
We now take the imaginary part according to \eq{ImDisc}.
For the non-logarithmic $1/s$ and $1/s^2$ terms we use $\mathrm{Im}[(s+i0)^{-1}]=-\pi\delta(s)$ and $\mathrm{Im}[(s+i0)^{-2}]=\pi\delta'(s)$, respectively.
In addition, we get a nonzero contribution from the second logarithm in \eq{be0} for $p^2>({M}+m)^2$ which also restricts the relevant integration range of the Feynman parameter $u$. 
Adding the $\ord{\alpha_s}$ counterterm from the renormalization of $m$ in the pole mass scheme
at tree level we obtain ($\xi\equiv\sqrt{1-4 m^2/M^2}$) 
\begin{align}
&\mathrm{Im}\big[\text{(o$'$)}\big] - 2 \,a_s m \, \delta m_M^{(1)} \, \delta^\prime(s)  ={} a_s C_F \nn\\
&\times\Bigg[-\Theta \left(p^2 - (m+M)^2\right)\frac{\sqrt{(M^2-s)^2-4 m^2 M^2} \left[4 m^4+2 m^2 (M^2+2 s)+s (M^2-s)\right]}{s^2 (m^2+s)^2}\notag\\
&\quad-2\,\delta(s)\,\mathrm e^{\gamma_E  \varepsilon} \leri{\frac{M}{\tilde{\mu}}}^{-2\varepsilon} \Gamma (\varepsilon)
\frac{\varepsilon-1}{\varepsilon-2}+\delta(s)\frac{3\, \xi  \left(2 m^4-M^4\right) \log \left(\frac{m^2}{M^2}\right)}{2 m^4 \xi }\notag\\
&\quad+\delta(s)\frac{3 \left[\left(-4 m^4-2 m^2 M^2+M^4\right) \log \left(\frac{1 - \xi}{1 + \xi}\right)-m^2 \xi  (3 m^2+2 M^2)\right]}{2 m^4 \xi }\Bigg]+\mathcal O(\varepsilon)\,.
\end{align}
Note that the $\delta'(s)$ terms from diagram (o$'$) and the contribution due to the quark pole mass counterterm $\delta m_M$ (where the subscript indicates the nontrivial dependence on the gauge boson mass $M$) cancel in the sum.

Diagram (p$'$) is rapidity divergent. With the symmetric $\eta$ regulator~\cite{Chiu:2011qc,Chiu:2012ir} we have
\begin{align}
\label{eq:massgluec}
\text{(p$'$)}& =-i\frac{8\,\alpha_s C_F\tilde \mu^{2\varepsilon}\nu^{\eta}}{(2\pi)^d} \frac{1}{s} \int\dd^d k\,\frac{\bar n\cdot(p+k)}{[-(p+k)^2+m^2][-k^2+M^2][-\bar n\cdot k]^{1+\eta}} \nn\\ 
&= \alpha _s C_F \frac{\tilde \mu^{2\varepsilon}\Gamma(\varepsilon)}{\pi^{2 - \varepsilon} 2^{1 - 2 \varepsilon}}
\left(\frac{\bar n\cdot p}{\nu}\right)^{-\eta}  \frac{1}{s}
\int_0^\infty\dd u\,\dd v\,\frac{v u^{-\eta -1} (u+v)^{2\varepsilon +\eta -2}\delta (1-u-v)}{ \left(m^2 u^2+v \left(M^2 (u+v)-s u\right)\right)^{\varepsilon }}\,. 
\end{align}
To compute the integral we use the same approach as for diagram (o$'$). The $M\to\infty$ piece is given by
\begin{equation}
\label{eq:binf2}
 (\text{p}')_{M\to\infty}= \alpha_s C_F 2^{-1 + 2 \varepsilon}\pi^{-2 + \varepsilon}\,\left(\frac{\bar n\cdot p}{\nu}\right)^{-\eta} \leri{\frac{{M}}{\tilde{\mu}}}^{-2\varepsilon}
 \frac{\Gamma(\varepsilon)\Gamma(2-\varepsilon)\Gamma(-\eta)}{\Gamma(2-\varepsilon-\eta)}
 \frac{1}{s}\,,
\end{equation}
and the remainder term reads
\begin{align}
\label{eq:binf3}
(\text{p}')-(\text{p}')_{M\to\infty}={}& \frac{C_F \alpha_s }{2 \pi^2}  \frac{1}{s} \int_0^1\dd u \frac{1-u}{u} \\
&\times 
\Big[\log \Big(M^2(1-u)\Big) - \log \Big(m^2 u^2+(1-u) \left(M^2-s u\right)\Big)\Big] + \ord{\varepsilon,\eta}\,.  \nn
\end{align}
At this point the rapidity regularization has done its job and is in particular not needed anymore for the dispersion integration. One may therefore expand in $\eta$ keeping only the divergent and finite terms.
To take the imaginary part we proceed in the same way as for diagram (o$'$). 

Accounting for the equivalent contribution from the mirror diagram of (p$'$) the complete result for the bare primary massive quark jet-function ME with gluon mass $M$ to ${\cal O}(\alpha_s)$ reads 
\begin{align}
&\Jme_{Q}^{\mathrm{bare}}(p^2,m^2,M^2) = \delta(s)\nn\\
& + a_s C_F \Bigg\{\Theta \left(p^2-(m+M)^2\right)\Bigg[\frac{4 \log \left(\frac{\sqrt{\left(M^2-s\right)^2-4 m^2 M^2}+M^2+s}{-\sqrt{\left(M^2-s\right)^2-4 m^2 M^2}+M^2+s}\right)}{s}\notag\\
&\hspace{5cm}-\frac{\left(2 m^2+s\right) \sqrt{\left(M^2-s\right)^2-4 m^2 M^2} \left(2 m^2+M^2+3 s\right)}{s^2 \left(m^2+s\right)^2}\Bigg]\notag\\
&+\delta(s) \Bigg[2 \mathrm e^{\gamma  \varepsilon } \Gamma (\varepsilon ) \left(\frac{M^2}{\tilde{\mu}^2}\right)^{-\varepsilon } \left(\frac{2}{\eta}-2 \log\left(\frac{Q}{\nu}\right)+2 H_{1-\varepsilon}-\frac{\varepsilon-1}{\varepsilon -2}\right)\notag\\
&+\frac{\left(4 m^4-10 m^2 M^2+3 M^4\right) \log \left(\frac{1 - \xi}{1 + \xi}\right)}{2 m^4 \xi }\notag\\
&-\frac{m^2 \left(-8 m^2 \log \left(\frac{1-\xi }{2}\right) \log \left(\frac{\xi +1}{2}\right)+m^2+6M^2\right)+\left(2 m^4-4 m^2 M^2+3 M^4\right) \log \left(\frac{m^2}{M^2}\right)}{2 m^4}\Bigg]\Bigg\}\nn\\
&+\mathcal O(\varepsilon,a_s^2)\label{eq:jmassgluebare1}
\,,
\end{align}
where $m$ is the quark pole mass and $H_n$ denotes the $n$-th harmonic number. For the expression shown here we have kept the exact dependence on $d=4-2\varepsilon$ for the terms that need regularization in the $M^2\to\infty$ limit of the dispersion integration. Otherwise the expansions in $\varepsilon$ and $\eta$ are carried out.

From \eq{jmassgluebare1} we can determine the ${\cal O}(\alpha_s)$ bare primary massive quark jet-function ME with the infinitesimal gluon mass regulator $\Lambda$, 
\begin{align}
&\Jme_{Q}^\mathrm{bare,(1)} (p^2,m^2,\Lambda^2) ={} C_F \Bigg\{\left[\frac{4}{\varepsilon}-4\log \left(\frac{\Lambda^2}{\mu ^2}\right)+\ord{\varepsilon}\right]\delta (s)\frac{1}{\eta}+\left[3-4 \log\left(\frac{Q}{\nu }\right)\right]\delta (s)\frac{1}{\varepsilon}\notag\\
&\quad+\delta (s) \left[2 L_m^2 +L_m+2 \log ^2\left(\frac{\Lambda^2}{\mu ^2}\right)+4 \log\left(\frac{\Lambda^2}{\mu ^2}\right) \log \left(\frac{Q}{\nu }\right)+8\right]\\
&\quad+\mathcal L_0(s) \left[-4 L_m-4 \log \left(\frac{\Lambda^2}{\mu^2}\right)-4\right]+8 \mathcal L_1(s)+\Theta (s)\left[\frac{s}{\left(m^2+s\right)^2}-\frac{4 \log\left(\frac{s}{m^2}+1\right)}{s}\right]\Bigg\}\,. \nn
\end{align}
We stress once again that a dimensionful regularization parameter such as the gluon mass is mandatory to fully separate the rapidity and IR singularities such that all remaining $1/\varepsilon$ terms are UV.

Using the dispersion identity \eq{dispersion1} the unrenormalized secondary massive quark ${\cal O}(\alpha_s^2 C_F T_F)$ correction to the jet-function ME $\Jme_Q(p^2, m^2)$ with an infinitesimal gluon mass regulator reads
\begin{align}
\label{eq:jetdisp1}
\left.\Jme_{Q}^\mathrm{bare,(2)}(p^2,m^2,\Lambda^2)\right|_{C_F T_F} &={}
\frac{1}{\pi}\int\limits_{4m^2}^\infty\frac{\dd {M}^2}{{M}^2}\Jme_{Q}^{\mathrm{bare},(1)}(p^2,m^2,{M}^2)\,\mathrm{Im}\left[\Pi^{(1)}({M}^2,m^2)\right]\notag\\
& \qquad
- \left(\Pi^{(1)}(m^2,0)-\frac{4 T_F}{3\varepsilon}\right)\Jme_{Q}^\mathrm{bare,(1)}(p^2,m^2,\Lambda^2)\\
& \hspace{-2cm}={} \left. Z_{\Jme}^{(2)}(s,m^2,\Lambda^2,\mu,Q/\nu)\right|_{C_F T_F} + \left. \Jme_{Q}^{(2)}(p^2,m^2,\Lambda^2,\mu,Q/\nu) \right|_{C_F T_F}\,.
\notag
\end{align}
The $4 T_F/(3 \eps)$ term in the second line of \eq{jetdisp1} is associated with the counterterm
coming from the heavy quark loop and is required to implement the $\msb$ strong coupling in the $(n_\ell+1)$ flavor scheme.
The dispersion integral in the first line of \eq{jetdisp1} can be solved analytically in terms of elementary functions and standard complete elliptic integrals using \texttt{Mathematica} with appropriate changes of integration variables, except for the logarithmic real radiation term shown in the second line of \eq{jmassgluebare1}. The latter can be integrated numerically very efficiently. It contributes to the corrections related to the three-particle $QQ\bar Q$-cut and corresponds to \eq{Itilde}. 
The final result for the finite (renormalized) contribution $\left. \Jme_{Q}^{(2)} \right|_{C_F T_F}$ is displayed in 
\eq{Jme1} and the divergent terms $\left. Z_{\Jme}^{(2)}\right|_{C_F T_F}$ are given in \eq{ZJme1}.

\section{Numerical Analysis}
\label{sec:analysis}

In this section we provide a brief numerical study concerning the structure and importance of the $\ord{\alpha_s^2}$ corrections to the primary massive quark SCET jet function. To be definite we use the universal jet function $J_Q^{\mathrm{uf}(n_\ell+1)}(p^2,m^2,\mu)$ of \eq{Jufdef} for the analysis since in practice it is sufficient to formulate factorization theorems for all possible scale hierarchies, see the discussion in section~\ref{sec:consistency}. We remind the reader that the universal SCET jet function agrees with the massless quark SCET jet function in the limit $m\to 0$, called ``massless limit'' in the following. However, its relation to the bHQET jet function in the limit $s=p^2-m^2\to 0$ involves the collinear-soft function $S_c$ that is singular by itself in this kinematic regime, see eqs.~\eqref{eq:jmmfdef}, \eqref{eq:Sc} and \eqref{eq:JBJmf}. Since the bHQET factorization theorem \eqref{eq:facttheorembHQET} and the factorization theorem for $p^2-m^2\sim p^2\lesssim m^2$ \eqref{eq:uffacth1} (which is identical to \eqref{eq:uffacth2}) can be smoothly connected, a comparison of the universal function $J_Q^{\mathrm{uf}(n_\ell+1)}$ with its leading singular limit for $s=p^2-m^2\to 0$, called ``heavy quark limit'' in the following, correctly visualizes the size of the mass corrections that are not captured by the leading power bHQET factorization theorem in the region $0< s=p^2-m^2\lesssim m^2$.

\begin{figure}
\centering
\includegraphics[width=0.48\textwidth]{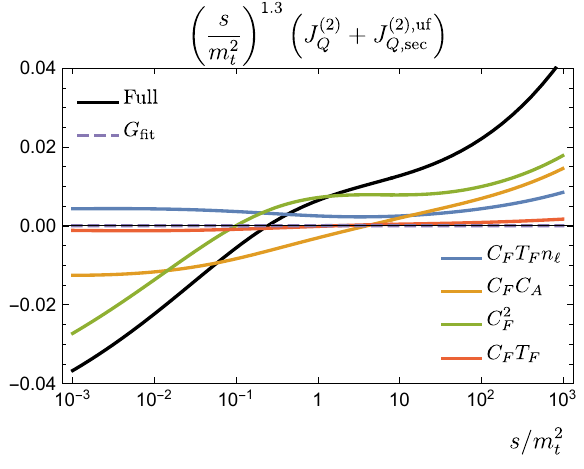}\quad
\includegraphics[width=0.48\textwidth]{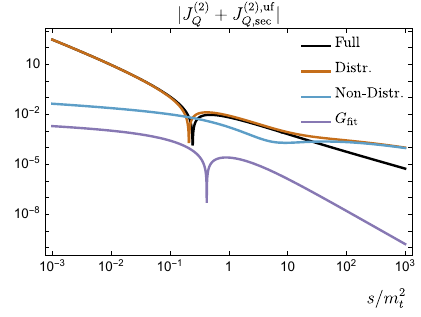}
\caption{\textit{Left:} Size of contributions to the $\mathcal O(\alpha_s^2)$ coefficient of the universal SCET massive primary quark jet function from different color structures as a function of $s/m_t^2$ for $\mu=\sqrt{s}+5\,\mathrm{GeV}$, $n_\ell=5$ and $m=m_t=173\,\mathrm{GeV}$ multiplied by $(s/m_t^2)^{1.3}$. \textit{Right:} Double logarithmic plot of the modulus of the full $\ord{\alpha_s^2}$ corrections and the corresponding contribution coming from $G_\mathrm{fit}$. We also show the sum of all distributional $\ord{\alpha_s^2}$ corrections and the non-distributional ones.}
\label{fig:ScetColor}
\end{figure}

\begin{figure}
\centering
\includegraphics[width=0.48\textwidth]{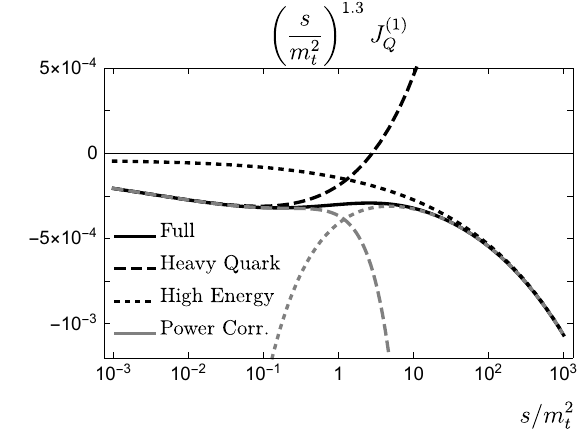}\quad
\includegraphics[width=0.48\textwidth]{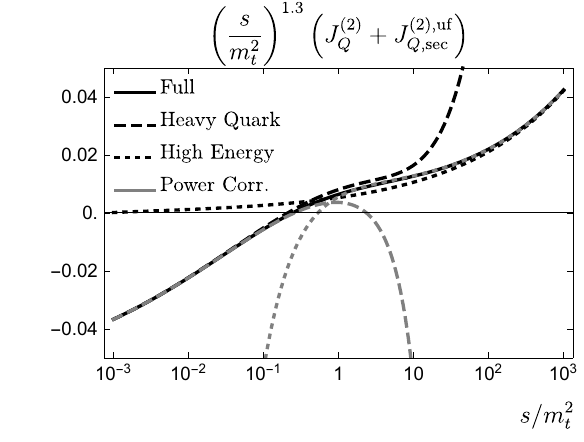}
\caption{\textit{Left:} $\ord{\alpha_s}$ coefficient of the SCET massive primary quark jet function (black solid) in comparison to the leading power results of the massless (black dotted) and bHQET (black dashed) limit. Also shown are the curves including the respective next terms in the small mass and bHQET expansions (gray dotted and dashed, respectively).
\textit{Right: } Analogous plot for the $\mathcal O(\alpha_s^2)$ coefficient of the universal SCET massive primary quark jet function.}
\label{fig:ScetPower}
\end{figure}

In the left panel of \fig{ScetColor} the size of the contributions of the two-loop coefficient $J_Q^{(2)}(p^2,m^2,\mu) + J_{Q,\mathrm{sec}}^{(2),\mathrm{uf}}(p^2,m^2,\mu)$ coming from the different color structures is displayed as a function of $s/m^2=(p^2-m^2)/m^2$ for $\mu=\sqrt{s}+5\,\mathrm{GeV}$. The terms proportional to $C_F^2$ (green), $C_F C_A$ (orange), $C_F T_F n_\ell$ (blue) and $C_F T_F$ (red) are displayed individually where we multiplied the factor $(s/m^2)^{1.3}$ to reduce the variation of the values over the whole kinematic regime.\footnote{
This convention is also adopted for \fig{ScetPower}.
For all numerical discussions we adopt the renormalization scale $\mu=\sqrt{s}+5\,\mathrm{GeV}$ and $m=m_t=173\,\mathrm{GeV}$.} 
The total coefficient is displayed in black. We see that the $C_F^2$, $C_F C_A$ and $C_F T_F n_\ell$ terms, not unexpectedly, provide the bulk of the $\ord{\alpha_s^2}$ corrections, and that the secondary heavy quark corrections are typically more than one order of magnitude smaller. Interestingly, due to the behavior of the $C_F^2$ and $C_F C_A$ terms, the $\ord{\alpha_s^2}$ correction changes sign at $\sqrt{p^2}\approx 1.1 m$ which is well within the bHQET regime.

In the left panel of \fig{ScetColor} we also show the total contribution of the fit function $G_\mathrm{fit}$ (purple dashed) which is part of the non-distributional terms $G_{C_F}^{(1Q)}$ and $G_{C_A}^{(1Q)}$ contained in the $C_F^2$ and $C_F C_A$ corrections, see \eqs{G1QCF}{G1QCA} respectively. Typically, it makes up for less than $1\%$ of the total $\ord{\alpha_s^2}$ correction, except close to the zero at $\sqrt{p^2}\approx 1.1m$. This is also shown in more detail in the right (double logarithmic) panel of \fig{ScetColor} where the modulus of the full $\ord{\alpha_s^2}$ corrections (black) and the corresponding contribution coming from the function $G_\mathrm{fit}$ (purple) are displayed. Since the $\ord{\alpha_s}$ corrections to the jet function do not have a zero (see below), the overall contribution of $G_\mathrm{fit}$ to the entire jet function never exceeds the per mille level.
In the right panel of \fig{ScetColor} we have also shown the sum of all distributional $\ord{\alpha_s^2}$ corrections (brown) and the non-distributional ones (light blue) which are constituted by the sum of all $G$-functions quoted in \sec{results}. We see that the non-distributional contributions are strongly suppressed compared to the distributional terms in the bHQET region and still at least an order of magnitude smaller as long as $s\lesssim 10 m^2$.
In the high energy limit, however, there are strong cancellations between them because the non-distributional corrections develop distributional terms in the limit $m\to 0$, see \app{Gmto0}. So the contributions of the non-distributional corrections are particularly important for $s/m^2 \gtrsim 10$ where they are not at all negligible.

We also would like to show the numerical impact of the genuine SCET massive primary quark jet function in comparison to the previously known jet functions for massless quarks (i.e.\ its high energy limit $s/m^2\to \infty$) and for the bHQET limit $s/m^2\to 0$. In the left panel of \fig{ScetPower} the full $\ord{\alpha_s}$ coefficient $J_Q^{(1)}(p^2,m^2,\mu)$ is displayed (black solid) in comparison to the singular results of the massless (black dotted) and bHQET (black dashed) limit.
We see that both expansions work well in the respective limits, but fail to provide reasonable approximations in the intermediate regime $0.3\lesssim s/m^2\lesssim 10$ which corresponds to $1.1\lesssim\sqrt{p^2}/m\lesssim 3$.
In the right panel of \fig{ScetPower} the analogous results are shown for the $\ord{\alpha_s^2}$ coefficient $J_Q^{(2)}(p^2,m^2,\mu)+J_{Q,\mathrm{sec}}^{(2),\mathrm{uf}}(p^2,m^2,\mu)$. Again, the massless as well as the bHQET expansion work well in the respective limits. But, interestingly, both also approximate the exact result within $15-20\%$ in the intermediate regime $0.3\lesssim s/m^2\lesssim 10$. That the $\ord{\alpha_s}$ and $\ord{\alpha_s^2}$ correction of the universal SCET massive primary quark jet function differ in this respect should be considered as an accidental feature. This can be seen by also including the respective next terms in the small mass and bHQET expansions (which represent terms belonging to subleading power expansions in massless quark SCET and bHQET, respectively). These are shown as the gray dotted and dashed lines, respectively, in \fig{ScetPower}. At $\ord{\alpha_s}$ as well as $\ord{\alpha_s^2}$ these subleading terms improve the approximation of the exact results substantially into the intermediate regime $0.3\lesssim s/m^2\lesssim 10$. But we see that, while at $\ord{\alpha_s}$ they also improve the approximation to $s/m^2\approx 1$, the same is not true at $\ord{\alpha_s^2}$.

\section{Conclusions}
\label{sec:conclusions}

In this work we have calculated the $\ord{\alpha_s^2}$ corrections to the SCET jet functions for primary massive quarks which are relevant for the inclusive description of jets initiated by massive quarks produced by hard interactions with high energy. The result provides important information for the intermediate region $p^2\sim p^2-m^2=s\sim m^2$, where the previously known $\ord{\alpha_s^2}$ jet function results in the massless limit $m^2/p^2\to 0$ and in the bHQET limit $s\to 0$ do not provide accurate approximations. Thus they represent the essential input to achieve a coherent description that smoothly interpolates these two extreme limits.

The interesting conceptual feature of the primary massive quark SCET jet function is that it can be defined in two different ways depending on whether - in addition to the zero-bin subtractions - one imposes soft mass mode bin subtractions or not for diagrams containing secondary massive quark effects, i.e.\ a massive quark-antiquark vacuum polarization subdiagram.
In the context of the view that jet functions require so-called zero-bin subtractions \cite{Manohar:2006nz} to constitute infrared finite factorization functions, the relevance of the soft mass mode bin subtraction~\cite{Chiu:2009yx,Gritschacher:2013pha} depends on the assumed power counting of the quark mass w.r.t.\ the jet invariant mass. An alternative view, that leads to equivalent results, but does not rely on a zero-bin expansion supplemented by power counting arguments, is to define the jet function with a subtraction defined from a collinear-soft matrix element~\cite{Bauer:2011uc,Procura:2014cba} where the effects of secondary massive quarks can be optionally included or not. The collinear-soft matrix elements provide unambiguous prescriptions of the subtractions eliminating the need to impose an additional power counting. The definitions and the results for the collinear-soft matrix elements are identical for massless and massive primary quarks and even apply for bHQET jet functions (to deal with corrections arising from secondary massive quarks that are lighter than the super-heavy primary quark).

If the secondary massive quark is included in the collinear-soft matrix element the resulting jet function is called \textit{universal} jet function. The universal jet function reduces to the massless quark jet function in the limit of zero quark mass and also has the same virtuality ($\mu$) anomalous dimension. It can be used for the kinematic region ranging from $m^2\sim p^2\sim p^2-m^2$ down to the massless limit $p^2\gg m^2$, where $p^2$ is the squared jet function invariant mass and $m$  the mass of the primary quark. If the secondary massive quarks are not included in the collinear-soft matrix element the jet function is called \textit{mass mode} jet function. It is infrared divergent for $m\to 0$ and has a virtuality and a rapidity anomalous dimension. It can be used for the kinematic region $m^2\sim p^2\sim p^2-m^2$. The two types of jet functions are related via the so-called collinear-soft function \cite{Pietrulewicz:2017gxc}, and we have provided the $\ord{\alpha_s^2}$ results for both. We have also demonstrated how both jet functions are used in the factorization approaches of \rcites{Gritschacher:2013pha,Pietrulewicz:2014qza,Hoang:2015vua} and \rcite{Pietrulewicz:2017gxc} which were developed 
to deal with the effects of secondary massive quarks in the context of factorization. 
While the approach of \rcite{Pietrulewicz:2017gxc} provides a more transparent view on the modular structure of the collinear, soft and collinear-soft mass modes, the approach of \rcites{Gritschacher:2013pha,Pietrulewicz:2014qza,Hoang:2015vua} provides a method with a smooth quark mass dependence
that allows to treat the cases $p^2-m^2\ll m^2\sim p^2$ and $ p^2-m^2\sim m^2\sim p^2$
in a single factorization theorem.
Taking a jet mass factorization theorem with the primary massive quark SCET jet function as an example we have shown how both approaches are related and in which way they are or can be rendered equivalent. We emphasize that, considered together, both approaches provide a thorough view on the field theoretic treatment of secondary massive quark effects within factorization approaches that separate soft and collinear quantum effects.

Regarding the calculation, we have treated the two-loop secondary mass corrections due to heavy quark loops separately from the corrections due to massless quark and gluon loops.
In fact, the correct treatment of the secondary mass corrections requires to carefully disentangle UV, IR and rapidity divergences present in this case.
This is achieved by employing a dimensionful IR regulator (in our case a small gluon mass). The loop integrations are than carried out using the dispersive method of \rcites{Gritschacher:2013pha,Pietrulewicz:2014qza,Hoang:2015vua}.

The remaining $\ord{\alpha_s^2}$ corrections (related to diagrams without a closed quark-antiquark loop) have been carried out in pure dimensional regularization.
We distinguished two types of relevant diagrams: planar and nonplanar.
While the contributions from the planar diagrams can be obtained using standard modern multi-loop techniques, we faced interesting technical issues in the calculation of the nonplanar diagrams.
In particular, for two nonplanar diagrams we have not been able to employ (standard automated) IBP reductions without generating spurious rapidity divergences that require an additional rapidity regulator. The latter made the solution of the corresponding master integrals unfeasible.
We therefore treated these two diagrams in a semi-numerical approach exploiting analytic information from the massless and bHQET limits. The corresponding numerical results are very accurate and only represent a very small contribution to the full $\ord{\alpha_s^2}$ corrections to the jet functions.
We have also provided precise numerical approximations of our analytic results (involving elliptic functions) for practical implementations.

\section*{Acknowledgments}
This work was supported by FWF Austrian Science Fund under the Project No.\ P28535-N27 and
in part by a GFK fellowship and by the clusters of {excellence} PRISMA (EXC 1098) and  PRISMA$^+$ (EXC 2118) at JGU Mainz.
We acknowledge partial support by the FWF Austrian Science Fund under the Doctoral Program ``Particles and Interactions'' No.\ W1252-N27. We also thank the Erwin-Schr\"odinger International Institute for Mathematics and Physics for partial support. 
CL thanks the JGU Mainz and the University of Salamanca for hospitality during parts of this work. 
We thank A.\ Carmona, D.\ Lechner, M.\ L\"oschner, P.\ Pietrulewicz, M.\ Preisser, and D.\ Samitz for helpful discussions.
\vspace*{0.3cm}

\begin{appendix}

\section{Other Jet Functions}
\label{app:OtherJet}

The mass mode jet function for massless primary quarks at $\mathcal O(\alpha_s^2)$ has the form
\begin{align}
\label{eq:Jmfmassless}
J_q^{\mathrm{mf},(n_\ell+1)}\left(p^2,m^2,\mu,\frac{Q}{\nu}\right) ={}& \delta(p^2) + a_s^{(n_\ell)} J_q^{(1)}(p^2,\mu)\\
& + \left(a_s^{(n_\ell)}\right)^2\left[J_q^{(2)}(p^2,\mu)+J_{q,\mathrm{sec}}^{(2),\mathrm{mf}}\left(p^2,m^2,\mu,\frac{Q}{\nu}\right)\right]+\mathcal O(\alpha_s^3)\,,\notag
\end{align}
with
\begin{align}
&J_q^{(1)}(p^2,\mu) ={} C_F \Bigg\{\left(7-\pi^2\right) \delta(p^2)-3 \mathcal L_{0}(p^2)+4 \mathcal L_{1}(p^2)\nonumber\\
&\quad+\varepsilon\Bigg[\left(-\frac{28 \zeta_3}{3}+14-\frac{3 \pi^2}{4}\right) \delta(p^2)+\left(\pi^2-7\right)\mathcal L_0(p^2)+3 \mathcal L_1(p^2)-2 \mathcal L_2(p^2)\Bigg]\nonumber\displaybreak[1]\\
&\quad+\varepsilon^2\Bigg[\left(-7 \zeta_3+28-\frac{7 \pi^2}{4}-\frac{\pi^4}{24}\right) \delta(p^2)+\left(\frac{28 \zeta_3}{3}-14+\frac{3 \pi^2}{4}\right) \mathcal L_0(p^2)\notag\\
&\qquad+\left(7-\pi^2\right) \mathcal L_1(p^2)-\frac{3}{2}\mathcal L_2(p^2)+\frac{2}{3} \mathcal L_3(p^2)\Bigg]\Bigg\}\,, \label{eq:Jmf1massless}\\
&J_q^{(2)}(p^2,\mu) ={} C_F^2 \Bigg[\left(-18 \zeta_3+\frac{205}{8}-\frac{67 \pi^2}{6}+\frac{14 \pi^4}{15}\right) \delta(p^2)+\left(-8 \zeta_3-\frac{45}{2}+7 \pi^2\right) \mathcal L_{0}(p^2)\nonumber\\
&\quad+\left(37-\frac{20 \pi^2}{3}\right) \mathcal L_{1}(p^2)-18 \mathcal L_{2}(p^2)+8 \mathcal L_{3}(p^2)\Bigg]\nonumber\displaybreak[1]\\
&\quad+C_F C_A \Bigg[\left(-\frac{206 \zeta_3}{9}+\frac{53129}{648}-\frac{208 \pi^2}{27}-\frac{17 \pi^4}{180}\right) \delta(p^2) + \left(40 \zeta_3-\frac{3155}{54}+\frac{22 \pi^2}{9}\right) \mathcal L_{0}(p^2)\nonumber\\
&\qquad+\left(\frac{367}{9}-\frac{4 \pi^2}{3}\right) \mathcal L_{1}(p^2)-\frac{22}{3} \mathcal L_{2}(p^2)\Bigg]\nonumber\displaybreak[1]\\
&\quad+C_F T_F n_\ell \Bigg[\left(\frac{494}{27}-\frac{8 \pi^2}{9}\right) \mathcal L_{0}(p^2)+\left(\frac{16 \zeta_3}{9}-\frac{4057}{162}+\frac{68 \pi^2}{27}\right) \delta(p^2)\notag\\
&\qquad-\frac{116}{9} \mathcal L_{1}(p^2)+\frac{8}{3} \mathcal L_{2}(p^2)\Bigg]\,, \label{eq:Jmf2massless}\displaybreak[1]\\
&J_{q,\mathrm{sec}}^{(2),\mathrm{mf}}\left(p^2,m^2,\mu,\frac{Q}{\nu}\right) ={} C_F T_F\Bigg\{\Bigg[\left(-\frac{8}{3}L_m^2-\frac{80}{9}L_m-\frac{224}{27}\right) \log \left(\frac{Q}{\nu}\right)+2 L_m^2\notag\\
&\qquad+\frac{2}{9}\left(3+4\pi^2\right)L_m-\frac{8 \zeta_3}{3}+\frac{20 \pi ^2}{27}+\frac{73}{18}\Bigg]\delta(p^2) + G_\mathrm{sec}\Theta\left(p^2-(2m)^2\right)\Bigg\}, \label{eq:Jmf2secmassless}
\end{align}
\begin{align}
&G_\mathrm{sec} ={}\frac{1}{p^2}\Bigg[-\frac{32}{3}\mathrm{Li}_2\!\left(\frac{b-1}{1+b}\right)+\frac{16}{3}\log\left(\frac{1-b^2}{4}\right)\log\left(\frac{1-b}{1+b}\right)-\frac{8}{3}\log^2\left(\frac{1-b}{1+b}\right)\nonumber\\
&\quad+\left(\frac{1}{2}b^4-b^2+\frac{241}{18}\right)\log\left(\frac{1-b}{1+b}\right)-\frac{5}{27}b^3+\frac{241}{9}b-\frac{8\pi^2}{9}\Bigg]\notag\\
&\hphantom{G_\mathrm{sec}}= \frac{(p^2-(2m)^2)^{7/2}}{560 m^9}+\mathcal O\!\left(\frac{(p^2-(2m)^2)^{9/2}}{m^{11}}\right),
\end{align}
where $b\equiv\sqrt{1-4m^2/p^2}$.

The universal jet function for massless primary quarks at $\mathcal O(\alpha_s^2)$ has the form \cite{Pietrulewicz:2014qza}
\begin{align}
J_q^{\mathrm{uf},(n_\ell+1)}(p^2,m^2,\mu) ={}& \delta(p^2) + a_s^{(n_\ell + 1)} J_q^{(1)}(p,\mu)\notag\\
& + \left(a_s^{(n_\ell+1)}\right)^2 \left[J_q^{(2)}(p,\mu)+J_{q,\mathrm{sec}}^{(2),\mathrm{uf}}(p^2,m^2,\mu)\right] + \mathcal O(\alpha_s^3)\,,
\label{eq:Jufmassless}
\end{align}
with
\begin{align}
&J_{q,\mathrm{sec}}^{(2),\mathrm{uf}}(p^2,m^2,\mu) = C_F T_F\Bigg\{\Bigg[\frac{8}{9}L_m^3+\frac{58}{9}L_m^2+\left(\frac{718}{27}-\frac{8 \pi ^2}{9}\right) L_m\nn\\
&\qquad+\frac{1531}{54}+\frac{10\pi^2}{27}-\frac{80\zeta_3}{9}\Bigg]\delta(p^2) + \left[-\frac{8}{3}L_m^2-\frac{116}{9}L_m-\frac{224}{27}\right] \mathcal L_0(p^2)+\frac{16}{3}L_m\mathcal L_1(p^2)\notag\\
&\quad + G_\mathrm{sec} \Theta\left(p^2-(2m)^2\right)\Bigg\}\,.\label{eq:Juf2secmassless}
\end{align}
Both jet functions satisfy \eq{Scdef} in the same way as the corresponding primary massive jet functions.

The renormalized primary massless jet-function ME including contributions related to 1-loop corrections and secondary massive quark production at 2-loop in the computational scheme with an infinitesimal gluon mass $\Lambda$ is given by
\begin{align}
&\left.\Jme\!\!\left(p^2,0,\Lambda^2,\mu,\frac{Q}{\nu}\right)\right|_{C_F T_F} = \delta(p^2)\\
&\quad + a_s^{(n_\ell+1)} C_F\Bigg\{\left[2 \log ^2\left(\frac{\Lambda ^2}{\mu^2}\right)+4 \log \left(\frac{\Lambda ^2}{\mu ^2}\right) \log \left(\frac{Q}{\nu}\right)-\frac{2 \pi ^2}{3}+7\right]\delta(p^2)\notag\\
&\qquad + \left[-4 \log \left(\frac{\Lambda ^2}{\mu^2}\right)-3\right]\mathcal L_0(p^2)+4 \mathcal L_1(p^2)\Bigg\}\notag\\
&\quad + \left(a_s^{(n_\ell+1)}\right)^2 C_F T_F \Bigg\{\Bigg[\left(\frac{8}{3} \log ^2\left(\frac{\Lambda ^2}{\mu^2}\right)+10\right) L_m+2 L_m^2\notag\\
&\qquad\quad+\log \left(\frac{Q}{\nu }\right) \left(\frac{16}{3} \log\left(\frac{\Lambda ^2}{\mu ^2}\right) L_m-\frac{80}{9} L_m-\frac{8}{3} L_m^2-\frac{224}{27}\right)-\frac{8\zeta_3}{3}+\frac{20 \pi ^2}{27}+\frac{73}{18}\Bigg]\delta(p^2)\notag\\
&\qquad + \left(-\frac{16}{3} \log\left(\frac{\Lambda ^2}{\mu ^2}\right)-4\right)L_m \mathcal L_0(p^2) +\frac{16}{3} L_m \mathcal L_1(p^2) + G_\mathrm{sec}\Theta(p^2-(2m)^2)\Bigg\} + \ord{\alpha_s^3}\,. \nn
\end{align}
The ${\cal O}(\alpha_s)$ and ${\cal O}(\alpha_s^2 C_F^2, \alpha_s^2 C_F C_A, \alpha_s^2 C_F T_F n_\ell)$ corrections in the computational scheme with a massless gluon in dimensional regularization analytically agree with those of the massless quark SCET jet function, see \cite{Becher:2006qw,Pietrulewicz:2014qza}.

\section{\texorpdfstring{$G$}{G}-Functions in the \texorpdfstring{Limit $m\to 0$}{Massless Limit}}
\label{app:Gmto0}

In the limit that the massive primary quark $Q$ becomes massless the singular limits of the non-distributional $G$-functions have the form
\begin{align}
G_1 ={}& \left(- 2 L_m^2- L_m - 1 - \frac{2 \pi^2}{3}\right) \delta(p^2) + (1 + 4 L_m) \mathcal L_0(p^2) - 4 \mathcal L_1(p^2) + \mathcal O\!\left(\frac{m^2}{p^4}\right),\displaybreak[1]\\
G_1^{(\varepsilon)} ={}& \left(- \frac{3}{2} L_m^2 + \frac{2\pi^2}{3} L_m - 2 - \frac{\pi^2}{3} - 4 \zeta_3\right) \delta(p^2)\nn\\
&+ (1 + 4 L_m + 2 L_m^2) \mathcal L_0(p^2) + (-5 - 8 L_m) \mathcal L_1(p^2) + 6 \mathcal L_2(p^2) + \mathcal O\!\left(\frac{m^2}{p^4}\right),\displaybreak[1]\\
G_1^{(\varepsilon^2)} ={}& \Bigg[- \frac{1}{6} L_m^4 - \frac{1}{6} L_m^3 + \left(-4 + \frac{\pi^2}{6}\right)L_m^2 + \left(\frac{7 \pi^2}{12} + 4 \zeta_3\right)L_m\displaybreak[1]\\
&\quad - 4 - \frac{13 \pi^2}{12} - \frac{\pi^4}{30} - 2 \zeta_3\Bigg] \delta(p^2)\notag\displaybreak[1]\\
&+ \left[\frac{2}{3} L_m^3 + 2 L_m^2 + (8 - \pi^2)L_m + 2 - \frac{\pi^2}{4}\right] \mathcal L_0(p^2)\notag\\
& + [- 4 L_m^2 - 8 L_m - 9 + \pi^2] \mathcal L_1(p^2) + \Big[8 L_m + \frac{13}{2}\Big] \mathcal L_2(p^2) - \frac{14}{3} \mathcal L_3(p^2) + \mathcal O\!\left(\frac{m^2}{p^4}\right)\notag,
\end{align}
\begin{align}
&G_{C_F}^{(1Q)} ={} \Biggl[-2 L_m^4-2 L_m^3+\left(2 \pi^2-\frac{35}{2}\right) L_m^2+L_m \left(56 \zeta_3-\frac{27}{2}+\pi^2\right)\\
&\qquad-68 \zeta_3+\frac{43 \pi^4}{30}+\frac{125}{2}+\pi^2 \left(24 \log 2-\frac{131}{6}\right)\Biggr] \delta(p^2)\notag\\
&\quad+\left[8 L_m^3+12 L_m^2+\left(38-\frac{20 \pi^2}{3}\right) L_m-72 \zeta_3+\frac{\pi^2}{3}+\frac{33}{2}\right] \mathcal L_{0}(p^2)\notag\\
&\quad+\left[-32 L_m^2-40 L_m+\frac{20 \pi^2}{3}-45\right] \mathcal L_{1}(p^2)+6 [8 L_m+5] \mathcal L_{2}(p^2)-24 \mathcal L_{3}(p^2) + \mathcal O\!\left(\frac{m^2}{p^4}\right),\nn\displaybreak[1]\\
&G_{C_A}^{(1Q)} ={} \Bigg[\frac{1}{9} \left(6 \pi^2-179\right) L_m^2+\frac{1}{9} L_m \left(-180 \zeta_3-139+56 \pi^2\right)+42 \zeta_3 \nn\\
&\qquad+\frac{7 \pi^4}{180}-\frac{508}{9}+\pi^2 \left(\frac{25}{27}-12 \log 2\right)\Bigg] \delta(p^2)\nn\\
&\quad+\frac{1}{9} \left[66 L_m^2+\left(391-12 \pi^2\right) L_m+180 \zeta_3-34 \pi^2+172\right] \mathcal L_{0}(p^2)\nn\\
&\quad-\frac{4}{9} \left[66 L_m-3 \pi^2+106\right] \mathcal L_{1}(p^2)+22 \mathcal L_{2}(p^2) + \mathcal O\!\left(\frac{m^2}{p^4}\right), \displaybreak[1]\\
&G_{T_F} ={} \left[\frac{58}{9} L_m^2 + \left(\frac{74}{9} - \frac{16 \pi^2}{9}\right) L_m + \frac{110}{9} + \frac{46 \pi^2}{27}\right] \delta(p^2) \nn\\
&\quad + \left[- \frac{8}{3} L_m^2 - \frac{128}{9} L_m - \frac{86}{9} + \frac{8 \pi^2}{9}\right] \mathcal L_0(p^2) + \left[\frac{32}{3} L_m + \frac{140}{9}\right] \mathcal L_1(p^2)\notag\\
&\quad - 8 \mathcal L_2(p^2) + \mathcal O\!\left(\frac{m^2}{p^4}\right), \displaybreak[1]\\
&G^{(3Q)} ={} \left[L_m^2+7 L_m+88 \zeta_3-91+\pi^2 (11-16 \log 2)\right] \delta(p^2)+[-2 L_m-7] \mathcal L_{0}(p^2)\nn\\
&\quad+2 \mathcal L_{1}(p^2)+\mathcal O\!\left(\frac{m^2}{p^4}\right),\displaybreak[1]\displaybreak[1]\\
&G_{\mathrm{sec}}^{(3Q)} ={} \frac{2}{81} \left[-36 L_m^3-261 L_m^2+3 \left(12 \pi^2-359\right) L_m+216 \zeta_3+171 \pi^2-2783\right] \delta(p^2)\nn\\
&\quad+\frac{2}{27} \left[36 L_m^2+174 L_m-12 \pi^2+359\right] \mathcal L_{0}(p^2)-\frac{4}{9} [12 L_m+29] \mathcal L_{1}(p^2)\label{eq:GTFsec3Qml} \nn\\
&\quad+\frac{8}{3} \mathcal L_{2}(p^2) + \mathcal O\!\left(\frac{m^2}{p^4}\right).
\end{align}

\section{Anomalous Dimensions}
\label{app:AnoDim}

The $\mu$-anomalous dimensions as defined in \eq{MuAnomDims} are given by
\begin{align}
&\gamma_c^{(n_f)}(Q,\mu) ={} -\gamma_Q^{(n_f)}(Q,\mu)\,,\\
&\gamma_Q^{(n_f)}(Q,\mu) ={} \Gamma_\mathrm{cusp}^{(n_f)}[\alpha_s]\log\Bigl(\frac{-Q^2-i0}{\mu^2}\Bigr) + \gamma_Q^{(n_f)}[\alpha_s]\,, \displaybreak[1]\\
&\gamma_Q^{(n_f)}[\alpha_s] ={} -6\, C_F a_s^{(n_f)} + \left(a_s^{(n_f)}\right)^2\Bigg[C_F^2\left(-3+4\pi^2-48\zeta_3\right)\notag\\
&\qquad + C_F C_A\left(-\frac{961}{27}-\frac{11\pi^2}{3}+52\zeta_3\right)+C_F T_F n_f\left(\frac{260}{27}+\frac{4\pi^2}{3}\right)\Bigg]+\mathcal O(\alpha_s^3)\,,\displaybreak[1]\\
&\gamma_{c_B}^{(n_f)}(Q,m,\mu) ={} -\Gamma_\mathrm{cusp}^{(n_f)}[\alpha_s] \log\Bigl(\frac{-Q^2-i0}{m^2}\Bigr) + \gamma_{c_B}^{(n_f)}[\alpha_s]\,,\displaybreak[1]\\
&\gamma_{c_B}^{(n_f)}[\alpha_s] ={} 4\, C_F a_s^{(n_f)} + \left(a_s^{(n_f)}\right)^2\Bigg[C_F C_A\left(\frac{196}{9}-\frac{4\pi^2}{3}+8\zeta_3\right)+C_F T_F n_f\left(-\frac{80}{9}\right)\Bigg]\notag\\
&\quad+\mathcal O(\alpha_s^3)\,,\displaybreak[1] \\
\label{eq:gammaJdef}
&\gamma_J^{(n_f)}(p^2,\mu) ={} -2\Gamma_\mathrm{cusp}^{(n_f)}[\alpha_s]\mathcal L_0(p^2) + \gamma_J^{(n_f)}[\alpha_s]\delta(p^2)\,,\displaybreak[1]\\
&\gamma_J^{(n_f)}[\alpha_s] ={} 6\, C_F a_s^{(n_f)} + \left(a_s^{(n_f)}\right)^2 \Bigg[C_F^2\left(3-4\pi^2+48\zeta_3\right) \nn \\
&\qquad+C_F C_A\left(\frac{1769}{27}+\frac{22\pi^2}{9}-80\zeta_3\right) + C_F T_F n_f\left(-\frac{484}{27}-\frac{8\pi^2}{9}\right)\Bigg]+\mathcal O(\alpha_s^3)\,, \displaybreak[1]\\
&\gamma_{J_B}^{(n_f)}(\hat s,\mu) ={} -2\Gamma_\mathrm{cusp}^{(n_f)}[\alpha_s] \mathcal L_0(\hat s)+\gamma_{J_B}^{(n_f)}[\alpha_s]\delta(\hat s)\,, \displaybreak[1]\\
&\gamma_{J_B}^{(n_f)}[\alpha_s] ={} 4\, C_F a_s^{(n_f)} + \left(a_s^{(n_f)}\right)^2\Bigg[C_F C_A\left(\frac{1396}{27}-\frac{23\pi^2}{9}-20\zeta_3\right)\notag\\
&\qquad + C_F T_F n_f\left(\frac{4\pi^2}{9}-\frac{464}{27}\right)\Bigg]+\mathcal O(\alpha_s^3)\,,\displaybreak[1]\\
&\gamma_S^{(n_f)}(\ell^+,\ell^-,\mu) ={} \delta(\ell^+)\gamma_s^{(n_f)}(\ell^-,\mu) + \delta(\ell^-)\gamma_s^{(n_f)}(\ell^+,\mu)\,, \\
&\gamma_s^{(n_f)}(\ell,\mu) ={} 2\Gamma_\mathrm{cusp}^{(n_f)}[\alpha_s]\mathcal L_0(\ell) + \gamma_s^{(n_f)}[\alpha_s] \delta(\ell)\,, \displaybreak[1]\\
&\gamma_s^{(n_f)}[\alpha_s] ={} \left(a_s^{(n_f)}\right)^2\Bigg[C_F C_A \left(-\frac{808}{27}+\frac{11 \pi^2}{9}+28\zeta_3\right) + C_F T_F n_f\left(\frac{224}{27}-\frac{4\pi^2}{9}\right)\Bigg]\nn\\
&\quad + \mathcal O(\alpha_s^3)\,,\displaybreak[1]\\
&\Gamma_\mathrm{cusp}^{(n_f)} ={} 4\, C_F a_s^{(n_f)} + \left(a_s^{(n_f)}\right)^2\left[C_F C_A\left(\frac{268}{9}-\frac{4\pi^2}{3}\right)+C_F T_F n_f\left(-\frac{80}{9}\right)\right]+\mathcal O(\alpha_s^3).
\end{align}

The $\mu$- and $\nu$-anomalous dimensions of the jet-function ME $\Jme_f^{(n_\ell+1)}(p^2, m^2)$ and the collinear-soft MEs $ \Scme^{(n_\ell+1)}(\ell,m)$ and $ \Scme^{(n_\ell)}(\ell)$, defined in analogy to \eqs{MuAnomDims}{NuAnomDims}, read
\begin{align}
&\left. \gamma_{\Jme}(p^2,\mu,Q/\nu) \right|_{C_F T_F} = a_s^{(n_\ell+1)} C_F \left[6-8\log\left(\frac{Q}{\nu}\right)\right]\delta(p^2) \nn\\
&\quad + \left(a_s^{(n_\ell+1)}\right)^2 C_F T_F\left[\frac{160}{9} \log \left(\frac{Q}{\nu }\right)-\frac{16 \pi ^2}{9}-\frac{4}{3}\right]\delta(p^2) + \ord{\alpha_s^3}\,, \displaybreak[1]\\
&\gamma^{(n_\ell)}_{\Scme}(\ell,\mu,\nu) = a_s^{(n_\ell)}C_F\left[8\log \left(\frac{\nu}{\mu }\right)\delta (\ell) + 8\mathcal L_0(\ell)\right] + \ord{\alpha_s^2}\,,\displaybreak[1]\\
&\left. \gamma^{(n_\ell+1)}_{\Scme}(\ell,\mu,\nu) \right|_{C_F T_F} = a_s^{(n_\ell+1)}C_F\left[8 \log \left(\frac{\nu}{\mu }\right)\delta(\ell) + 8 \mathcal L_0(\ell)\right] \notag\\
&\quad+\left(a_s^{(n_\ell+1)}\right)^2 C_F T_F\Bigg[\Bigg(-\frac{160}{9} \log \left(\frac{\nu}{\mu }\right)-\frac{8 \pi ^2}{9}+\frac{448}{27}\Bigg)\delta(\ell) - \frac{160}{9}\mathcal L_0(\ell)\Bigg] + \ord{\alpha_s^3}\,,\displaybreak[1]\\
&\left. \gamma_{\nu,\Jme}(m,\Lambda,\mu) \right|_{C_F T_F} = a_s^{(n_\ell+1)} C_F \left[-4\log\left(\frac{\Lambda^2}{\mu^2}\right)\right]\notag\\
&\quad + \left(a_s^{(n_\ell+1)}\right)^2 C_F T_F \Bigg[-\frac{16}{3} \log \left(\frac{\Lambda^2}{\mu^2}\right) L_m+\frac{8}{3} L_m^2+\frac{80}{9} L_m+\frac{224}{27}\Bigg] + \ord{\alpha_s^3}\,,\displaybreak[1]\\
&\gamma^{(n_\ell)}_{\nu,\Scme}(\Lambda,\mu) = a_s^{(n_\ell)}C_F\left[-4\log\left(\frac{\Lambda^2}{\mu^2}\right)\right] + \ord{\alpha_s^2}\,,\displaybreak[1]\\
&\left. \gamma^{(n_\ell+1)}_{\nu,\Scme}(m,\Lambda,\mu)\right|_{C_F T_F} = a_s^{(n_\ell+1)}C_F\left[-4\log\left(\frac{\Lambda^2}{\mu^2}\right)\right] \notag\\
&\quad + \left(a_s^{(n_\ell+1)}\right)^2 C_F T_F \Bigg[-\frac{16}{3} \log \left(\frac{\Lambda^2}{\mu^2}\right) L_m+\frac{8}{3} L_m^2+\frac{80}{9} L_m+\frac{224}{27}\Bigg] + \ord{\alpha_s^3}\,.
\end{align}
The ${\cal O}(\alpha_s^2 C_F^2, \alpha_s^2 C_A C_F, \alpha_s^2 C_F T_F n_\ell)$ contributions to the $\mu$- and $\nu$-anomalous dimensions of the jet-function and
collinear-soft MEs are unknown. We therefore only displayed the ${\cal O}(\alpha_s^2 C_F T_F)$ corrections coming from the massive quark $Q$, which is indicated by subscript ``$C_F T_F$''.

The virtuality $\mu$-anomalous dimensions for the collinear-soft function $S_c$ and the mass mode jet function $J_f^{\mathrm{mf},(n_\ell+1)}$ are
\begin{align}
&\gamma_{S_c}(\ell,m,\mu,\nu) ={} a_s^2 C_F T_F \Bigg[\left(-\frac{32}{3} L_m-\frac{160}{9}\right)\mathcal L_0(\ell)\nn\\
&\qquad + \left(-\frac{32}{3} \log\left(\frac{\nu}{\mu }\right)L_m-\frac{160}{9} \log\left(\frac{\nu}{\mu }\right)-\frac{8 \pi ^2}{9}+\frac{448}{27}\right)\delta(\ell)\Bigg] + \ord{\alpha_s^3},\displaybreak[1]\\[2mm]
&\left. \gamma_{J^\mathrm{mf}}(p^2,m^2,\mu,Q/\nu) \right|_{C_F T_F} ={} a_s^{(n_\ell+1)}C_F\left[-8\mathcal L_0(p^2)+6\delta(p^2)\right]\nn\\
&\quad + \left(a_s^{(n_\ell+1)}\right)^2 C_F T_F \Bigg[\left(\frac{32}{3} L_m \log \left(\frac{Q}{\nu}\right)+\frac{160}{9} \log \left(\frac{Q}{\nu }\right)-\frac{16 \pi^2}{9}-\frac{4}{3}\right) \delta (p^2)\nn\\
&\qquad - \frac{32}{3} L_m \mathcal L_0(p^2) \Bigg] + \ord{\alpha_s^3}\,,
\end{align}
where the remaining color factor contributions to $\gamma_{J^\mathrm{mf}}$  at ${\cal O}(\alpha_s^2)$
(which do not arise from diagrams containing the massive quark $Q$ loop)
coincide with the ${\cal O}(\alpha_s^2)$ contributions of $\gamma_{J}^{(n_\ell)}$ given in \eq{gammaJdef}.

\section{Master Integrals}
\label{app:MI}

The family A master integrals depicted in \fig{mis1} evaluate to ($y\equiv m^2/p^2$, $\bar y\equiv 1-y$)
\begin{align}
 &M_1={}\mathcal I(-y;0,0,0,1,1,0,0,0)=(-y/\bar y)^{2 - 2 \varepsilon}\, \Gamma(\varepsilon - 1)^2\,, \\
 &M_2={}\mathcal I(-y;1,0,0,1,1,0,0,0)=-(-y/\bar y)^{2 - 2 \varepsilon}\, \Gamma(\varepsilon - 1)^2\, {}_2F_1(1, 2 - 2 \varepsilon; 2 - \varepsilon;1/\bar y)\,,\displaybreak[1]\\
 &M_3={}\mathcal I(-y;1,1,0,1,1,0,0,0)=(-y/\bar y)^{2 - 2 \varepsilon}\,
  \Gamma(\varepsilon-1)^2 \,{}_2F_1(1, 2 - 2 \varepsilon; 2 - \varepsilon; 1/\bar y)^2\,,\displaybreak[1]\\
 &M_4={}\mathcal I(-y;1,0,1,0,1,0,0,0)= -2 \Gamma (1-\varepsilon ) \Gamma (\varepsilon ) \Gamma (2 \varepsilon -2) \, _2F_1\left(2-2 \varepsilon ,2 \varepsilon -1;2-\varepsilon ;1/\bar{y}\right),\displaybreak[1]\\
 &M_5={}\mathcal I(-y;1,0,1,0,2,0,0,0)=-\Gamma (1-\varepsilon ) \Gamma (\varepsilon -1) \Gamma (2 \varepsilon ) \, {}_2F_1\left(2-2 \varepsilon ,2 \varepsilon ;2-\varepsilon;1/\bar{y}\right),\displaybreak[1]\\
 &M_6={}\mathcal I(-y;1,0,1,1,1,0,0,1)=\mathrm{e}^{-2\varepsilon\gamma_E}\bigg[\frac{\pi^2}{12 \varepsilon^2}\nonumber\\
 &\quad+\frac{1}{\varepsilon}\left(\frac{1}{6} \pi^2 H(-1;-y)-H(-2,-1;-y)+H(-2,0;-y)+\frac{7 \zeta_3}{2}\right)\nonumber\\
 &\quad + 7 \zeta_3 H(-1;-y)-\frac{2}{3} \pi^2 H(-2;-y)+H(-3,-1;-y)-H(-3,0;-y)+3 H(-2,-2;-y)\nonumber\\
 &\quad+\frac{1}{3}\pi^2 H(-1,-1;-y)-3 H(-2,0,0;-y)-2 H(-1,-2,-1;-y)+2 H(-1,-2,0;-y)\nonumber\\
 &\quad+\frac{11 \pi^4}{72}+\mathcal O(\varepsilon)\bigg]\,, \displaybreak[1]\\
 &M_7={}\mathcal I(-y;1,0,1,1,1,0,0,2)=\mathrm{e}^{-2\varepsilon\gamma_E}\bigg[-\frac{1}{4 \varepsilon^3}\nonumber\\
 &\quad+\frac{1}{\varepsilon}\left(-H(-1,-1;-y)+H(-1,0;-y)+\frac{5 \pi^2}{24}\right)-\frac{1}{6} \pi^2 H(-1;-y)-2 H(-2,-1;-y)\nn\\
 &\quad+2 H(-2,0;-y)+3 H(-1,-2;-y)-3 H(-1,0,0;-y)+\frac{29 \zeta_3}{3}\nn\\
 &\quad+\varepsilon\bigg(-2 \zeta_3 H(-1;-y)-\pi ^2 H(-2;-y)+2 H(-3,-1;-y)-2 H(-3,0;-y)\nn\\
 &\qquad+4 H(-2,-2;-y)-7 H(-1,-3;-y)+\frac{5\pi^2}{6} H(-1,-1;-y)+\frac{\pi^2}{3} H(-1,0;-y)\nn\\
 &\qquad-4 H(-2,0,0;-y)+4 H(-1,-2,-1;-y)-4 H(-1,-2,0;-y)\nn\\
 &\qquad-4 H(-1,-1,-1,-1;-y)+4 H(-1,-1,-1,0;-y)+7 H(-1,0,0,0;-y)+\frac{497\pi^4}{1440}\bigg)\nn\\
 &\quad+\mathcal O(\varepsilon^2)\bigg]\,,
 \label{eq:mitopoa}
\end{align}
where the $H(\vec n;-y)$ denote HPLs according to the conventions of~\rcite{Maitre:2005uu}.

The relevant imaginary parts and $\bar y$-expansions related to the family B master integrals depicted in \fig{topoBmis} were determined or directly taken from \rcites{Laporta:2004rb,Bauberger:1994by,Remiddi:2016gno} and read
\begin{align}
&M_8={}\mathcal I(-y;0,0,0,1,1,1,0,0)=\mathrm e^{-2\varepsilon\gamma_E}\left(\frac{-y}{\bar y}\right)^{1-2 \varepsilon} \biggl[-\frac{3}{2 \varepsilon^2}-\frac{\frac{1}{-y}+18}{4 \varepsilon}\nn\\
&\quad-\frac{1}{8} \left(59+2 \pi^2\right)-\frac{3 \bar y}{8} - \frac{1}{8} \left(\pi^2-7\right) \bar y^2 - \frac{1}{96} \left(9 \pi^2-56\right) \bar y^3 + \mathcal O\!\left(\varepsilon,\bar y^4\right)\biggr], \displaybreak[1]\\
&\mathrm{Im}\big[(-s-i0)^{-2\varepsilon}M_8\big]\overset{s>0}{=}{}\Theta(1/9-y)\frac{\pi}{2\bar y} \sqrt{1-\sqrt{y}} \sqrt{3 \sqrt{y}+1}\nn\\
&\quad\times\Biggl[8 y\, \mathrm K\left(\frac{\left(\sqrt{y}+1\right)^3 \left(3\sqrt{y}-1\right)}{\left(\sqrt{y}-1\right)^3 \left(3 \sqrt{y}+1\right)}\right) + \left(\sqrt{y}-1\right) (1+3 y) \mathrm \, \mathrm E\left(\frac{\left(\sqrt{y}+1\right)^3 \left(3 \sqrt{y}-1\right)}{\left(\sqrt{y}-1\right)^3 \left(3 \sqrt{y}+1\right)}\right)\Biggr]
\nn\\&\quad
+\mathcal O(\varepsilon), \displaybreak[1]\\
&M_9={}\mathcal I(-y;0,0,0,2,1,1,0,0)=\mathrm e^{-2\varepsilon\gamma_E}\left(\frac{-y}{\bar y}\right)^{-2 \varepsilon}\biggl[\frac{1}{2 \varepsilon^2} + \frac{1}{2 \varepsilon} - \frac{1}{12} \left(6-\pi^2\right)\nonumber\\
&\quad-\left(\frac{\pi^2}{12}-1\right) \bar y - \frac{1}{32} \left(4-\pi^2\right) \bar y^2+\mathcal O\!\left(\varepsilon,\bar y^3\right)\biggr],\displaybreak[1]\\
&\mathrm{Im}\big[(-s-i0)^{-2\varepsilon}M_9\big]\overset{s>0}{=}{}-\Theta(1/9-y)\frac{\pi  \sqrt{3 \sqrt{y}+1}}{3 \sqrt{1-\sqrt{y}}}\nn\\
&\quad\times \Biggl[4 \left(2 \sqrt{y}-1\right)\, \mathrm K\left(\frac{\left(\sqrt{y}+1\right)^3 \left(3 \sqrt{y}-1\right)}{\left(\sqrt{y}-1\right)^3\left(3 \sqrt{y}+1\right)}\right)+3 \left(\sqrt{y}-1\right)^2\, \mathrm E\left(\frac{\left(\sqrt{y}+1\right)^3 \left(3 \sqrt{y}-1\right)}{\left(\sqrt{y}-1\right)^3 \left(3 \sqrt{y}+1\right)}\right)\Biggr]\nn\\
&\quad+\mathcal O(\varepsilon)\,,\displaybreak[1]\\
&M_{10}={}\mathcal I(-y;1,1,0,1,1,1,0,0)=\bar y\Bigg[-\frac{1}{6} \pi^2 H(1;1/y) - 2 H(0, 1, 1; 1/y)\nonumber\\
&\quad+ H(1, 0, 1; 1/y) + \frac{2}{3} \int_9^\infty\!\dd t\, \frac{t-9}{t - 1} \log\left(1 - \frac{1}{y t}\right)\frac{2\, \mathrm K\left(\frac{\left(\sqrt{t}-3\right) \left(\sqrt{t}+1\right)^3}{\left(\sqrt{t}-1\right)^3 \left(\sqrt{t}+3\right)}\right)}{\sqrt{\left(\sqrt{t}-1\right)^3 \left(\sqrt{t}+3\right)}}\,\Bigg]+\mathcal O(\varepsilon)\label{eq:mi10full}\\
&\,={}\bar y\Bigg[\bigg(\frac{17 \zeta_3}{4}-\frac{3}{2} \pi^2 \log 2\bigg) + \bar{y} \left(3 \log \left(-\bar{y}\right)-\log ^2\left(-\bar{y}\right)-\frac{3\pi ^2}{16}-\frac{5}{2}\right)\Bigg]+\mathcal O(\varepsilon,\bar y^3)\,,\displaybreak[1] \label{eq:m10exp}\\
&\mathrm{Im}\big[(-s-i0)^{-2\varepsilon}M_{10}\big]\overset{s>0}{=}{}\bar y\, \pi\Bigg[\Theta(1-y)\left[2 \log (y) \log (\bar{y}/y)-3\text{Li}_2(-\bar{y}/y)\right] \nn\\
&\quad - \Theta(1/9 - y) \frac{4}{3} \int_9^{1/y}\dd t\, \frac{t-9}{t - 1} \frac{\mathrm K\left(\frac{\left(\sqrt{t}-3\right) \left(\sqrt{t}+1\right)^3}{\left(\sqrt{t}-1\right)^3 \left(\sqrt{t}+3\right)}\right)}{\sqrt{\left(\sqrt{t}-1\right)^3 \left(\sqrt{t}+3\right)}}\,\Bigg]+\mathcal O(\varepsilon)\,.\label{eq:mi10Im}
\end{align}
The remaining integral in \eq{mi10Im} can be evaluated numerically very efficiently. 
It originates from the branch cut of the logarithm in the explicit integral in the full $\eps$ expanded result for $M_{10}$ given in \eq{mi10full} when taking the imaginary part.
The expansion of $M_{10}$ around $s\sim 0$, i.e.\ $y\sim 1$, in \eq{m10exp} was computed by expanding the integrand and solving the resulting integrals using a PSLQ-type algorithm implemented in \texttt{Mathematica}.

All results stated in this appendix were checked numerically using sector decomposition.

\end{appendix}

\bibliography{./sources}
\bibliographystyle{JHEP}

\end{document}